\documentclass[journal,twoside]{cls/ieeecolornologo}
\definecolor{subsectioncolor}{rgb}{0,0,0}
\usepackage{cite}
\usepackage{amsmath,amssymb,amsfonts}
\usepackage{graphicx}
\usepackage{hyperref}
\usepackage{textcomp}
\usepackage{bbm}
\usepackage{mathtools}
\mathtoolsset{centercolon}

\usepackage{algorithmic}
\usepackage[ruled,vlined,linesnumbered]{algorithm2e}

\usepackage{acronym}
\usepackage[percent]{overpic}

\newcommand{\calA}{{\cal A}}

\newcommand{\calC}{{\cal C}}
\newcommand{\calD}{{\cal D}}

\newcommand{\calF}{{\cal F}}

\newcommand{\calH}{{\cal H}}

\newcommand{\calL}{{\cal L}}
\newcommand{\calM}{{\cal M}}

\newcommand{\calO}{{\cal O}}
\newcommand{\calP}{{\cal P}}

\newcommand{\calR}{{\cal R}}

\newcommand{\calT}{{\cal T}}

\newcommand{\calV}{{\cal V}}

\newcommand{\calX}{{\cal X}}
\newcommand{\calY}{{\cal Y}}

\newcommand{\bfa}{\mathbf{a}}
\newcommand{\bfb}{\mathbf{b}}

\newcommand{\bfd}{\mathbf{d}}
\newcommand{\bfe}{\mathbf{e}}
\newcommand{\bff}{\mathbf{f}}
\newcommand{\bfg}{\mathbf{g}}
\newcommand{\bfh}{\mathbf{h}}

\newcommand{\bfk}{\mathbf{k}}

\newcommand{\bfu}{\mathbf{u}}
\newcommand{\bfv}{\mathbf{v}}
\newcommand{\bfw}{\mathbf{w}}
\newcommand{\bfx}{\mathbf{x}}
\newcommand{\bfy}{\mathbf{y}}

\newcommand{\bfalpha}{\boldsymbol{\alpha}}

\newcommand{\bftau}{\boldsymbol{\tau}}

\newcommand{\bfxi}{\boldsymbol{\xi}}

\newcommand{\bfB}{\mathbf{B}}

\newcommand{\bfW}{\mathbf{W}}

\newcommand{\bbE}{\mathbb{E}}

\newcommand{\bbR}{\mathbb{R}}

\newif\ifarxiv
\arxivtrue

\ifarxiv
   {}
\else
   \usepackage{xr}
\fi

\definecolor{darkgray}{gray}{0.4}
\definecolor{darkgreen}{rgb}{0,0.5,0}
\definecolor{darkyellow}{rgb}{0.9,0.8,0}
\definecolor{pink}{rgb}{0.9,0.0,0.9}
\definecolor{orange}{rgb}{1.0,0.7,0}

\newcommand{\teq}{\coloneqq}
\newcommand{\bfone}{\mathbf{1}}%
\DeclareMathOperator{\sgn}{sign}
\DeclareMathOperator{\silu}{SiLU}
\newcommand{\KAN}{\mbox{KAN}}

\newcommand{\defeq}{\coloneqq}
\DeclareMathOperator*{\argmin}{arg\,min} 

\newtheorem{theorem}{Theorem}
\newtheorem{definition}{Definition}
\newtheorem{assumption}{Assumption}

\newtheorem{lemma}{Lemma}
\newtheorem{remark}{Remark}

\usepackage{color}
\usepackage{colortbl}

\usepackage{booktabs}
\usepackage{multirow, float, boldline, hhline}

\definecolor{headerblue}{rgb}{0.784, 0.784, 1.0} 
\definecolor{gray}{gray}{0.9}  
\definecolor{lightgray}{gray}{0.95}
\definecolor{lightblue}{rgb}{0.9,0.9,1.0}
\newcommand{\TwoRow}[2]{
    #1 \\ 
    \rowcolor{lightgray}
    #2 \\
}
\newcommand{\ARow}[4]{
\cellLeft{#1} & #2 & & \cellLeft{#3} & #4 &
}

\renewcommand{\arraystretch}{1.5}
\newcolumntype{C}{>{\centering\arraybackslash}p{0.5cm}}
\newcolumntype{T}{>{\columncolor{lightblue}}c}
\newcolumntype{U}{>{\columncolor{lightgray}}c}

\usepackage{comment}
\newif\ifspaceconstrained
\spaceconstrainedtrue 
\ifspaceconstrained
    \excludecomment{cutforspace} 
\else
    
\fi

\title{Distance-Aware Error for Spline Networks: 
\\A Bottom-Up Approach to Uncertainty
}

\author{Masoud Ataei, Mohammad Javad Khojasteh, \IEEEmembership{Member, IEEE}, and Vikas Dhiman, \IEEEmembership{Member, IEEE}
\thanks{
This work is supported by the National Science Foundation under Grant No. 2218063 and the Gleason Endowment at RIT.
Preliminary results of this paper appeared in the Proceedings of the International Conference on Acoustics, Speech, and Signal Processing~\cite{ataei2025darek}.\\
Masoud Ataei and Vikas Dhiman are with the Department of Electrical and Computer Engineering of University of Maine, Orono, ME 04469, USA. \texttt{\{masoud.ataei,vikas.dhiman\}@maine.edu}. \\
Mohammad Javad Khojasteh is with Electrical and Microelectronic Engineering Department of Rochester Institute of Technology, Rochester, NY 14623, USA. \texttt{mjkeme@rit.edu}.}}

\newacro{NN}{neural network}
\newacroplural{NN}[NNs]{neural networks}
\newacro{KAN}{Kolmogorov-Arnold network}
\newacroplural{KAN}[KANs]{Kolmogorov-Arnold networks}
\newacro{KAT}{Kolmogorov-Arnold theorem}
\newacro{KKAN}{\Kurkova{} Kolmogorov-Arnold network}
\newacro{GP}{Gaussian process}
\newacroplural{GP}[GPs]{Gaussian processes}
\newacro{MLP}{multi-layer perceptron}
\newacro{SNR-MLP}{spectral-normalized ReLU activated MLP}
\acrodefindefinite{MLP}{an}{a}
\newacro{DAREK}{distance-aware uncertainty for Kolmogorov network}
\newacroplural{DAREK}[DAREKs]{distance-aware uncertainty for Kolmogorov networks}
\newacro{K-DAREK}{Distance-aware uncertainty for \Kurkova{} Kolmogorov networks}
\newacro{DKL}{deep kernel learning}
\newacro{SM}{spectral mixer}
\newacro{ReLU}{rectified linear unit}
\newacro{MSE}{mean square error}
\newacro{MPC}{model predictive control}
\newacro{CBF}{control barrier functions}
\newacro{SNN} {spline neural network}
\newacroplural{SNN}[SNNs]{spline neural networks}
\newacro{BNN} {Bayesian neural network}      
\newacro{SGLD}{stochastic gradient Langevin dynamics}    
\newacro{MC}  {Monte Carlo}      
\newacro{ML}  {machine learning}      
\newacro{MAP} {maximum a posteriori}   
\newacro{RBF} {radial basis function}     
\newacro{SNGP}{spectral-normalized Gaussian processes}    
\newacro{DUQ} {deterministic uncertainty
quantification}   
\newacro{DDU} {deep deterministic uncertainty}     
\newacro{DUE} {deterministic uncertainty estimation}     
\newacro{ReLU}{regularized linear unit}
\newacro{SiLU}{sigmoid linear unit}
\newacro{KNN}{k-nearest neighbors}
\newacro{PPE}{piecewise polynomial error}
\newacroplural{PPE}[PPEs]{piecewise polynomial errors}

\begin{document}
\maketitle

\begin{abstract}
We develop a new class of error bounds that are distance-aware and tightly characterize the approximation error induced by \acp{SNN}.
Our bottom-up approach starts by analyzing the error bounds of each neuron (a spline), and then extends to the \ac{NN}.
We begin with a review of error bounds for Newton's polynomial, which is then generalized to an arbitrary spline, under higher-order Lipschitz continuity assumptions.
We then extend these bounds to function compositions, which are the core of deep networks such as \acp{KAN}.
By the analysis of the propagation of approximation error through composed spline layers, we provide error bounds for the entire spline-based network.
These bounds are deterministic, do not rely on sampling or probabilistic assumptions, and hold under mild regularity conditions.
We evaluate our method on a variety of experiments, including estimating the object's shape from sparse laser scan points and safe navigation in an unstructured environment
to show its effectiveness.
We find that our method is faster than \ac{GP} and \ac{MC} approaches, and that our error bounds reliably enclose the true error. 
We also develop a metric for the distance-awareness of an uncertainty estimator, and show that \ac{DAREK} is distance-aware in more regions of the space than the baselines.
For ease of adoption, we provide a library that can be integrated into existing KAN models to estimate error bounds, available at: \url{https://masoud-ataei.github.io/DAREK/} 

\end{abstract}


\begin{table*}[ht]
\caption{Notations and definitions of important quantities. 
} 
\label{tbl:symbol_notation}
\setlength{\tabcolsep}{1pt} 
\renewcommand{\arraystretch}{1.3}
\providecommand {\cellLeft} [1] {\hspace{0.2cm}#1}
\centering
\begin{tabular}{  m{2.5cm}  m{6.0cm} m {0cm} |  m{2.5cm} m{6.0cm} m{0cm}}
\hline 
\rowcolor{headerblue}  
{\textbf{Notation}} & \textbf{Definition} & &
{\textbf{Notation}} & \textbf{Definition}  & \\[0.25ex]
\hline

\TwoRow{
    \ARow{$f$}{True function}
         {$[.]f$}{Divided differences of a function $f$ (defined in Eq~\ref{eq:divided-differences})}}
{   \ARow{$\bftau_{1:m}$}{Vector or set of $m$ knots, $\{\tau_1,\dots,\tau_m\}$}
         {$\tau_i$}{The knot $i$ in the set $\bftau_{1:m}$}}
\TwoRow{\ARow{$\calT$}{Matrix of knots}
             {$\calT_{i,j}$}{The $j$th knot in $i$th spline}}
       {\ARow{$\bftau_{1:k}^{j(x)}(\eta_{1:m})$}{k-nearby points to $x$ in set $\eta_{1:m}$ (Def.~\ref{def:k-nearby_knots})}
             {$\tau_{i}^{j(x)}$}{$i$th element in the k-nearby knots $\bftau_{1:k}^{j(x)}$}}
\TwoRow{\ARow{$\calD_f(\calX)$ }{
Operator that creates a dataset of pairs of data $\calD_f(\calX) \defeq \{(\bfx, f(\bfx))\}_{\bfx \in \calX}$}
             {$\calP_{k,n}[\calD_f(\bftau_{1:m})](x)$}{The $n$th piece of 
             $k$th-order Newton's piecewise polynomial approximation of 
             $f$ over set of knots $\bftau_{1:m}$  for $x \in [\tau_n, \tau_{n+1})$}}
       {\ARow{$b_{k,i}(x)$}{Spline-basis of order $k$ non-zero between $[\tau_i, \tau_{i+1})$ (Def.~\ref{def:newton-form-poly})}
             {$w_{k,i}$}{The weight of basis $b_{k,i}$}}
\TwoRow{\ARow{$\phi_k[\bftau_{1:m}]$}{A spline of order $k$ on the set of knots $\bftau_{1:m}$}
             {$\calL_f^k$}{$k$th order Lipschitz constant of $f$ (defined  in  Eq~\ref{eq:lipschitz_def})}}
       {\ARow{$\bar{u}_f(x;\bftau_{1:m})$}{The error bound of $f$ over set of knots $\bftau_{1:m}$ evaluated at $x$ with zero error at knots}
             {$u_f(\bfx;\bftau_{1:m})$}{The uncertainty estimator that calculates the error bound of function $f$ over set of knots $\bftau_{1:m}$ evaluated at $x$ with non-zero error at knots.}}
\TwoRow{\ARow{$\calX$}{Input space}
             {$\calX_{\calD}$}{Set of training inputs}}
       {\ARow{$\hat{f}$}{A piecewise polynomial approximation of $f$ on set of knots $\bftau_{1:m}$}
             {$\hat{f}_j$}{The $j$th piece of $\hat{f}$ defined on $x \in [\tau_j , \tau_{j+1})$}}
\ARow{$e_j^f$}{True error which is difference between the true function $f$ and the approximation $\hat{f}$ at interval $(\tau_j,\tau_{j+1})$}
             {$\mathbf{1}_n$ }{A vector of $n$ ones} \\
\hline
\end{tabular}
\end{table*}

\acresetall 
\section{Introduction}
\IEEEPARstart{S}{plines} are smooth, piecewise polynomials commonly used for approximation and modeling, particularly in control engineering and signal processing~\cite{egerstedt2009control,unser2002splines}. The use of spline functions in \acp{NN} dates back to the early 1990s, where techniques such as cubic splines and B-splines (Basis splines) were employed to improve generalization and enable more efficient training~\cite{friedman1990adaptive, uncini1997adaptive}. 
These early efforts aimed to leverage the smoothness and flexibility of splines to build data-efficient models. 
Recent advancements in computational power and the growing demand for more expressive and interpretable\footnote{Interpretability in machine learning is widely valued but inconsistently defined~\cite{lipton2018mythos}, with prescriptions ranging from mechanistic~\cite{nanda2023progress} to symbolic explanations~\cite{liu2025kan,tohme2024isr}. In this paper, we pursue a form of interpretability based on ``explanation by examples''~\cite{lipton2018mythos}, which emphasizes a human-understandable mapping between model inputs and outputs---by relating predictions to nearby training instances.
} models have sparked a renewed interest in \acp{SNN}~\cite{igelnik2003kolmogorov,bohra2020learning,aziznejad2020deep,diamant2024conformalized,keskin2018splinenets,gunther2021spline,fakhoury2022exsplinet}. In particular, \ac{KAN}~\cite{liu2025kan} and its extensions~\cite{bozorgasl2024wav,bresson2024kagnns,abueidda2025deepokan,vaca2024kolmogorov,genet2024tkan,koenig2024kan} have adopted B-spline as activation functions, while drawing inspiration from the  \ac{KAT}~\cite{kolmogorov1957representation}. 
There is also growing interest in applying these models to control systems, physics-based problems, and trajectory planning~\cite{beltran2024b, qian2024investigating}, though challenges related to uncertainty quantification and reliability remain unresolved. 

Uncertainty estimation for \ac{ML} models can be categorized into two classes:
probabilistic methods and worst-case methods.
Probabilistic methods estimate a probability distribution on the output of the \ac{ML} model, while worst-case methods provide error bounds on the output.
A popular class of probabilistic methods~\cite{lakshminarayanan2017simple, gal2016dropout, maddox2019simple, jantre2024learning} uses \ac{MC} approaches to estimate uncertainty. These approaches are architecture invariant and can be applied to any \ac{NN}, including \Acp{SNN}, regardless of the number of layers and activation functions. 
However, being architecture invariant comes at a price. These methods ignore the inductive biases embedded into an architecture, resulting from the choice of layers and activation functions.
Additionally, \ac{MC} uncertainty estimation requires training wider networks~\cite{gal2016dropout} or training multiple networks~\cite{lakshminarayanan2017simple}, which increases computational cost.

\Acp{GP}~\cite{rasmussen2006gaussian} offer a probabilistic alternative approach that, unlike \ac{MC} sampling methods, provides architecture-specific uncertainty quantification. 
In \acp{GP}, the architecture of the covariance function~\cite{rasmussen2006gaussian} determines how the predictions and uncertainty change with the input.
Over the past decade, Deep \acp{GP}~\cite{damianou2013deep,song2024novel,chakraborty2024scalable} have blurred the boundaries between \acp{NN} and \acp{GP} by facilitating the learning of covariance functions as well as interpreting \acp{NN} as \acp{GP}. 
Moreover, while infinitely wide \acp{NN} converge to \acp{GP}, this limit is impractical and necessitates approximation~\cite{pleiss2021limitations,lee2018deep,jacot2018neural}.
\acp{GP}' limitations are high computational cost and the need to store a large amount of training data for inference.
However, scalability can be improved by selecting inducing points~\cite{chakraborty2024scalable}. 
Nevertheless, kernel choice significantly impacts the performance of Deep \acp{GP}, and an ensemble of Deep \acp{GP} with diverse kernels can enhance flexibility and performance~\cite{song2024novel}.

Probabilistic uncertainty approaches are often avoided in safety-critical applications due to their computational complexity, 
large sample size requirements, reliance on hyperparameters~\cite{cheng2021limits}, and hazards of poorly specified prior~\cite{seaman2012hidden}. 
We instead focus on worst-case analysis~\cite{jaulin2001interval}, which is easier to understand, implement, and debug. 
Moreover, they
are robust to bounded input noise and adversarial attacks, and effective where probabilistic assumptions might not hold due to insufficient data.
Furthermore, worst-case analysis is appropriate for safety-critical systems, where the costs of ignoring low-probability events are very high. 
Combined with suitable robust control algorithms, worst-case analysis is akin to placing a safety cage around dangerous machinery, ensuring a system's robustness, reliability, and failure resilience, even when a single failure can lead to disastrous consequences. 

Unlike \acp{BNN}, we develop a \emph{distance-aware} error analysis for \acp{SNN},
meaning the error bound at a given input is a monotonically increasing function of its distance from the training data~\cite{liu2020simple,mukhoti2023deep,van2020uncertainty, van2021feature}.
Intuitively, the predictions of a well-trained model are more accurate within the neighborhood of the training data, and the model should avoid being overly confident when confronted with out-of-distribution inputs, which is numerically demonstrated in~\cite{liu2020simple}. 
Distance-awareness is crucial in safety-critical applications, such as autonomous driving and human-robot interaction, where decisions must be made with caution based on how novel the input is compared to the training data. 
Liu et al.~\cite{liu2025kan} provide error bounds for KANs, but these are potentially loose, not distance-aware, and only defined up to an unknown constant. In this work, we focus on \acp{KAN} and study distance-aware uncertainty estimation for \acp{SNN}, and its implications for sensing and control.

Our approach relies on a trained \ac{KAN} network, except that instead of randomly sampling the knots\footnote{“Spline” originates from flexible wooden strips used by shipbuilders, constrained by control points (or knots).}, we restrict them to be from the training data.
We build an uncertainty estimate for a KAN network bottom-up. We first estimate the uncertainty of a single neuron in the network, which is then composed layer-wise to estimate the uncertainty of a complex network.
To estimate uncertainty at a neuron level, we use Newton's polynomial error bound~\cite{de1978practical}, which assumes the function to be higher-order Lipschitz continuous and function approximation to be a piecewise continuous polynomial (Section~\ref{sec:PPE}).
We get the required Lipschitz constant for Newton's polynomial by assuming higher-order continuity on the network-level true function, which is then attributed to the continuity at the neuron level (Section~\ref{sec:Lip-division}).
Similarly, errors made at the knots in the network level function approximation are attributed to the neuron level errors (Section~\ref{sec:Error-division}). 
Given the neuron-level error bounds, we compose the errors to a two-layer network for ease of exposition and then generalize it to a multi-layer network (Section~\ref{sec:multilayer-spline-error-analysis}).
Finally, we extend error bounds to layers that were so far restricted to be piecewise continuous polynomials, which allow for an additive smooth function (e.g., SiLU,  Section~\ref{sec:residual-layer}).

In summary, we introduce \acp{DAREK}, a bottom-up approach that is more \textit{efficient} than traditional \acp{GP}, and provides \textit{distance-aware} uncertainty bounds that are more \textit{interpretable} and \textit{tighter} than those provided by \acp{BNN}.
We specifically make the following \textbf{contributions}: 
1) establish analytical worst-case \textit{distance-aware} error bounds for a multi-layer spline neural network; 
2) propose and evaluate approaches for Lipschitz attribution and error attribution between network layers and neurons; 
3) extend error bounds to layers with a sufficiently smooth function added to the spline neurons; 
4) develop a metric for distance-awareness and compare commonly used uncertainty estimators for distance-awareness; 
and
5) validate the scalability and efficacy of our method, as compared to \acp{GP} and Ensembles, on various examples, including object shape estimation and safe navigation in an unstructured environment.

\begin{table}[!t]
    \centering
    \caption{Expansions of 
    important acronyms.
    }
    \label{tbl:acronyms}
    \rowcolors{2}{lightgray}{white}  
    \resizebox{\columnwidth}{!}{%
    \begin{tabular}{ll} 
        \hline
        \rowcolor{headerblue}             
            \textbf{Acronym} & \textbf{Expansion}\\      
        \hline
            \acs{NN}      & \Acl{NN} \\
            \acs{SNN}     & \Acl{SNN}  \\
            \acs{BNN}     & \Acl{BNN}  \\    
            \acs{GP}      & \Acl{GP} \\
            \acs{MC}     & \Acl{MC}  \\    
            \acs{RBF}     & \Acl{RBF}  \\    
            \acs{ReLU}     & \Acl{ReLU}  \\             
            \acs{SNGP}     & \Acl{SNGP}  \\    
            \acs{KAT}     & \Acl{KAT} \\
            \acs{KAN}     & \Acl{KAN} \\
            \acs{PPE}     & \Acl{PPE} \\
            \acs{DAREK}   & \Acl{DAREK} \\
            \acs{MPC}     & \Acl{MPC}  \\
            \acs{CBF}     & \Acl{CBF}  \\            
        \hline
    \end{tabular}
    }
\end{table}

\section{Related Work}
\textbf{\acp{BNN}} quantify uncertainty by treating network parameters as random variables and learning a posterior distribution over them, rather than relying on fixed weights that yield a single-point prediction~\cite{blundell2015weight,gawlikowski2023survey}.
Computing the exact posterior distribution is intractable; therefore, several approximate inference methods have been proposed. 
\ac{MC} Dropout~\cite{gal2016dropout} assumes a Bernoulli distribution over the weights.
Bayes By Backprop (BBB)  and other variational bayes approaches~\cite{wu2018deterministic} use variational inference to learn weight distribution~\cite{blundell2015weight}, and \ac{SGLD} applies stochastic gradients to sample from the posterior~\cite{welling2011bayesian}.  
Probabilistic backpropagation (PBP) developed an approximation technique that directly updates weight distribution in the network, without relying on global variational inference~\cite{hernandez2015probabilistic}.
Deep ensembles train multiple models with different initializations, which have been shown to improve generalization~\cite{lakshminarayanan2017simple}. 
Other approaches include randomized \ac{MAP} sampling~\cite{lu2017ensemble}, which injects noise into the optimization process to mimic posterior sampling, and the Laplace approximation for estimating unimodal posteriors~\cite{ritter2018scalable}.
While \acp{BNN} can provide confidence levels, they often require careful calibration to reflect the true posterior~\cite{ guo2017calibration}. 
Ataei et al.~\cite{ataei2024dadee} propose an approach to effectively estimate well-calibrated~\cite{berkenkamp2017safe,berkenkamp2019no} uncertainty in both in-domain and out-of-domain settings~\cite{liu2020robust}. 
While these approaches estimate a probabilistic  \emph{distance-unaware} uncertainty, we adopt a worst-case \emph{distance-aware} analysis instead. 

\textbf{Distance-awareness} is a desirable property of uncertainty estimators, which ensures higher uncertainty in regions far from the training data~\cite{liu2020simple}.
\Acp{GP} are non-parametric models that interpolate between training data, providing mean predictions and uncertainty estimates using a given kernel function.
These models are inherently distance aware because the \Ac{GP} uncertainty increases with the decrease in kernel value, which acts as a proxy for inverse-distance~\cite{rasmussen2006gaussian}. 
The computational complexity of traditional \acp{GP} is high, and several approximations have been developed for specific applications; however, each of these methods introduces its trade-offs~\cite{liu2020gaussian, nguyen2009model, snelson2005sparse, hensman2015scalable}. 

\textbf{\Ac{DKL}} maps high-dimensional inputs to a feature space using a \ac{NN} and applies a \ac{GP} in feature space. Several researchers~\cite{liu2020simple,van2020uncertainty,van2021feature,mukhoti2023deep} explore \ac{DKL} for distance-aware uncertainty estimation.
\Ac{SNGP}~\cite{liu2020simple} applies spectral normalization to make the network Lipschitz continuous and replaces the final layer with a scalable \ac{GP} approximation using random Fourier features, which causes the predictive uncertainty to converge to zero when the number of data points goes to infinity, even for inputs far from the training data~\cite{van2021feature}.
\Ac{DUQ}~\cite{van2020uncertainty} focuses on stabilizing distance-aware uncertainty using \acs{RBF} in feature space, with constraints to prevent feature collapse.
Feature collapse happens when substantially different model inputs have identical representations in feature space. 
\Ac{DUQ} requires modeling the centroid of each class; hence, it does not generalize to regression problems.
\Ac{DUE}~\cite{van2021feature} improves on \Ac{DUQ} by enforcing bi-Lipschitz constraints, similar to \ac{SNGP}~\cite{liu2020simple}, and uses inducing point \acp{GP} to retain non-parametric behavior,
resulting in uncertainty estimates that are more robust, sensitive to prior choice and architectural details. 
\Ac{DDU}~\cite{mukhoti2023deep} fits a class-conditional Gaussian density in the feature space of a spectrally-normalized network to detect out-of-distribution inputs, making it computationally efficient but primarily empirical in nature.
One drawback of \Ac{DKL} methods is that they estimate uncertainty using \ac{GP} only after feature mapping and ignore the uncertainty induced by the learned mapping.
We propose a \emph{bottom-up} approach that computes the uncertainty of \acp{SNN} and hence can be combined with \ac{DKL} approaches.
Moreover, these methods are probabilistic, whereas we do a worst-case analysis for the Spline-based network.
Our method uses Newton's polynomial definition to bring distance awareness in error calculation and uses the Lipschitz constant to determine the bounds. 

\textbf{Error estimation of splines} has been extensively studied, with global error bounds derived for cubic splines~\cite{fakhoury2022exsplinet,de1978practical,takacs2016approximation}. 
Takacs et al.~\cite{takacs2016approximation} found the error bound of B-splines with maximum smoothness and fixed knot interval, and showed that this error is independent of the polynomial degree, later extended to arbitrary smoothness~\cite{sande2020explicit}. 
Although these works focused on \textit{distance-unaware} error bounds, \cite{de1978practical} provides distance-aware error bounds and shows that placing knots at Chebyshev points can significantly reduce error. However, this analysis assumes zero error at knots.
Our work introduces a \textit{distance-aware} error bound for spline approximation with non-zero error at knots and tracks it in a hierarchical spline network, which has not been previously explored.

\textbf{Spline neural networks} use splines as activation functions to enhance \ac{NN} architectures.
Early works, such as~\cite{igelnik2003kolmogorov}, replaced the univariate functions in the Kolmogorov-Arnold theorem~\cite{kolmogorov1957representation} with cubic splines. 
At the same time, Compolucci~et~al.~\cite{campolucci1996neural} and Harris~et~al.~\cite{harris1993intelligent} introduced learnable spline and B-spline activations to improve flexibility.
Hong et al.\cite{hong2011modeling} introduce a complex-valued B-spline network for Wiener systems.
According to~\cite{fakhoury2022exsplinet}, using at least second-degree splines in \acp{SNN} ensures differentiability for gradient-based training, while dimensionality reduction and tensor-product B-splines help manage model complexity.
Additionally, a \textit{distance-unaware} upper bound for deep \ac{ReLU} networks is derived~\cite{montanelli2020error}, which can be viewed as first-order spline networks. Also, \cite{balestriero2018spline} demonstrates that any deep \ac{ReLU} network can be equivalently represented as a multivariate spline network.
Moreover, \ac{KAN}~\cite{liu2025kan} introduces \textit{distance-unaware} error bounds for hierarchical spline networks, with an unknown constant of proportionality.
Recently, \acp{SNN} have been extended to convolution \acp{NN} as well~\cite{fey2018splinecnn}.
Our algorithm, \ac{DAREK}, provides a novel distance-aware error bound for spline networks, a more precise characterization of approximation errors that enhances reliability and interpretability. 
 
\begin{figure*}[!ht]
    \begin{overpic}[width=0.98\textwidth,trim=0pt 10pt 0pt 0pt, clip]{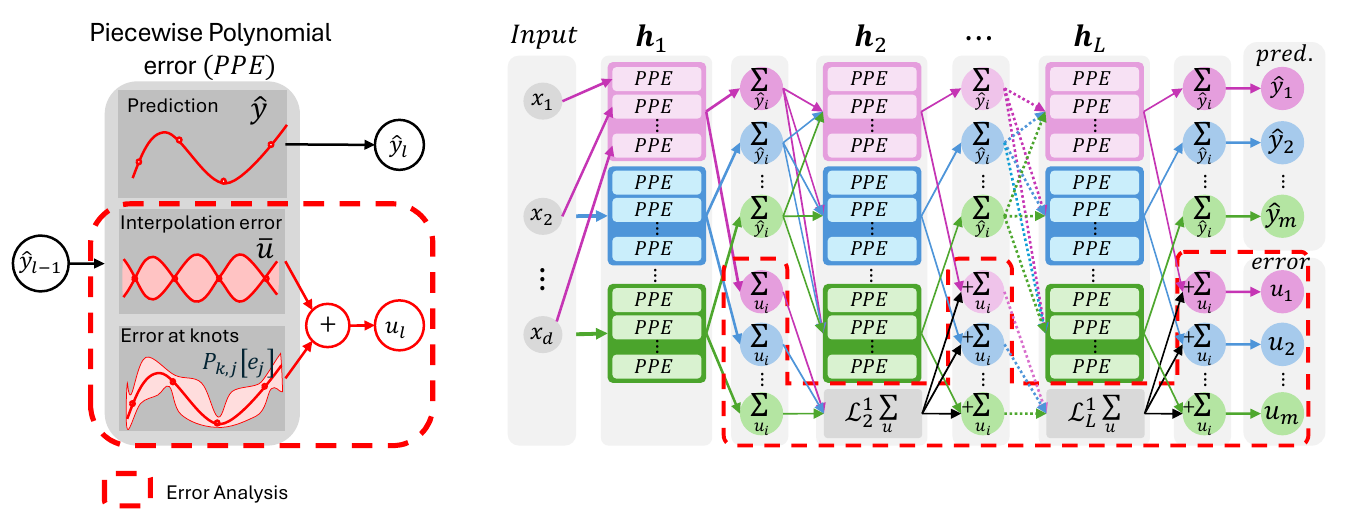}
        \put(-1,35){\textbf{(a)}} 
        \put(33,35){\textbf{(b)}} 
    \end{overpic}
    \caption{ 
    Block diagram of DAREK. \textbf{(a)} 
    \Ac{PPE} takes a scalar $\hat{y}_{l-1}$ as input and computes three values: 1) $\hat{y}_l$ - the spline value at given input, 2) $\bar{u}_l$ - the interpolation error using Theorem~\ref{thm:poly-interp-bound}, and 3) $\calP_{k,j}[e_j]$ - the error at knot using Lemma~\ref{thm:error-at-knots}. 
    The output of the \ac{PPE} consists of the mean prediction $\hat{y}_l$ and error bound term $u=\bar{u} + \calP_{k,j}[e_j]$. 
    \textbf{(b)} DAREK architecture consists of $L$ layers, where each layer $h_l$ has $m_l$ block of \acp{PPE} and each has $m_{l-1}$ \acp{PPE} in it. The cumulative error bound up to layer $l$ is computed as: $\bfu_l = \bfu_{h_l} + \calL_{l}^{1} \sum{\bfu_{l-1}}$ Theorem~\ref{thm:two-layer-poly-bound}.
    }
    \label{fig:DAREK}
\end{figure*}

\section{Overview of DAREK}
\label{sec:overview}
\Ac{DAREK} is a bottom-up approach to worst-case uncertainty (error bound) estimation in \acp{SNN}. 
We begin by finding error bounds of each neuron and then combine them to estimate the error bounds of the full network.
Fig.~\ref{fig:DAREK} provides an overview of the \ac{DAREK} algorithm.
Fig.~\ref{fig:DAREK}~{(a)} illustrates different components of the error bounds for a single neuron, denoted as \ac{PPE} block, and {(b)} shows how predictions and error bounds from each \ac{PPE} block contribute to produce the model's prediction, as well as the associated error bounds estimate. 

\Ac{DAREK} estimates the uncertainty of a trained \ac{SNN} with input $\bfx \in \bbR^d$, output $\hat{\bfy} \in \bbR^m$, and corresponding error bound $\bfu \in \bbR^m$, under reasonable assumptions of continuity that are enumerated later.
Predictions and error bounds are computed layer by layer and combined to obtain the \ac{NN}'s error bound.
The core component of \ac{DAREK} is \ac{PPE} block, which maps each input $\hat{y}_{l-1}$ to a prediction $\hat{y}_l$ and associated error $u_l$.
This error estimation is driven by aggregating two sources of error: interpolation error $\bar{u}$ (Theorem~\ref{thm:poly-interp-bound}) and error at knot $\calP_{k,j}[e_j]$ (Lemma~\ref{thm:error-at-knots}).
Each layer, $\bfh_l : \bbR^{m_{l-1}} \rightarrow \bbR^{m_l}$ for $l \in [1,\dots,L]$, forms output as linear combination of \acp{PPE} predictions and errors as linear combination of \acp{PPE} errors. 
The propagated error is computed by scaling the sum of errors from the previous layer by the Lipschitz constant of the current layer and adding the current layer's error, $\calL_{l-1}^{1} \sum {\bfu_{l-1}} + \bfu_l$. 
Further details are provided in Section~\ref{sec:method}.

By selecting spline knots from training data, our worst-case error analysis for \ac{SNN} leverages a semi-parametric model that combines the accuracy and efficiency of parametric methods (e.g., \acp{NN}) with the interpretability and analyzability of non-parametric models (e.g., \acp{GP}). As discussed below, this design also improves computational efficiency compared to state-of-the-art methods.

\textbf{Computational Cost:}
Consider an \acl{MC} (MC) method that estimates uncertainty by sampling $n$ different sets of \ac{NN} weights. Suppose each \ac{NN} evaluation has a computational cost of $\calO(L H^2)$, where $L$ is the number of layers and $H$ is the average width of the hidden layer.  
In that case, \ac{MC} sampling results in a computational complexity of $\calO(n LH^2)$. 
This method is more efficient than exact \acp{GP}, which requires matrix inversion with a cost of $\calO(N^3)$, where $N$ is the number of data samples and is typically greater than $L$, $H$, and $n$.
In contrast, \ac{DAREK} identifies the $k+1$ closest knots to a given test point at each layer among a set of $m$ knots (see Algorithm~\ref{alg:DAREK} for more details). 
If the $k+1$-consecutive-closest-knots are identified using a binary tree, then the test-time complexity is $\calO(\log_2(m)LH)+\calO(LH^2)$.
The binary tree is constructed once as a post-training step. 
Typically, \ac{MC} methods use $n \in \{5, \dots, 20\}$ sampled networks, while we use $H \in \{2, \dots, 20\}$ and $m\in \{10,\dots, 100\}$ for \ac{DAREK}.
When $m < 2^nH $, which is a common scenario, \ac{DAREK} is more efficient than \ac{MC} methods. 
A numerical comparison of computational costs is also demonstrated in  Fig.~\ref{fig:Err-comparison}.

\section{Background}
\label{sec:back}
\noindent\textit{\underline{Notations:}}
Let $\bftau_{i:j} \teq \{\tau_{i},\tau_{i+1},\dots,\tau_{j}\}$ be a sorted sequence of distinct points, and $f: \calX \subseteq \bbR \to \bbR$ be a scalar function.
Given $\bftau_{1:m}$ with $\tau_i \in \calX$, the set of input-output pairs is $\calD_f(\bftau_{1:m}) \teq \{(\tau_i, f(\tau_i))\}_{i=1}^m$.
Separately, let $\calD \teq \{(\bfx_i, \bfy_i)\}_{i=1}^n$ denote the training dataset, and define its input set as 
$\calX_\calD \teq \{\bfx_i\}_{i=1}^n$.
We use $\bbR^+$ to denote the set of non-negative real numbers.
We use capital letters for matrices, lowercase boldface for vectors, and lowercase regular font for scalars.
The notation $[.]f$ represents the divided difference of function $f$ on a non-repeating set $\bftau_{i:j}$, which is defined recursively as follows~\cite{de1978practical}:
\begin{align}
    [\tau_1]f &\teq f(\tau_1), \qquad
    [\tau_1,\tau_2]f \teq \frac{[\tau_2]f-[\tau_1]f}{\tau_2 - \tau_1},\notag\\
    [\bftau_{1:k}]f &\teq \frac{[\bftau_{2:k}]f-[\bftau_{1:k-1}]f}{\tau_k - \tau_1}.
   \label{eq:divided-differences}
\end{align}
Intuitively, divided differences are closely related to numerical derivatives. 
Important notations and definitions used in the paper are summarized in Table~\ref{tbl:symbol_notation}. 

Our objective is to determine efficient worst-case distance-aware uncertainty bounds for hierarchical spline architectures. 
To achieve this, we introduce our definitions of knot selection, distance awareness, piecewise polynomial, and Lipschitz constant. We then draw on Newton’s polynomial, which offers a tight bound for piecewise polynomials~\cite{de1978practical},
and review relevant background, including B-splines, \ac{KAT}~\cite{kolmogorov1957representation}, and KAN~\cite{liu2025kan}.

\begin{definition}[$k$-nearby knots]
\label{def:k-nearby_knots}
Let $\eta_{1:m} = \{\eta_1, \dots \eta_m\}$ be a sorted sequence of $m$ unique real knots.
Define $k$-nearby knots $\bftau^{j(x)}_{1:k}(\eta_{1:m})$, with $k\ge2$, given a number $x \in [\eta_1, \eta_m)$ to be a set of $k$ knots nearby $x$ chosen such that:
\begin{enumerate}
    \item the nearby knots are \emph{consecutive},
    \item the nearby knots includes both nearest knots $\{\eta_{j(x)}, \eta_{j(x)+1}\}$ with nearest-pair index $j(x)$ defined such that $\eta_{j(x)} \le x < \eta_{j(x)+1}$, and
    \item the nearby knots are a \emph{deterministic} function of the nearest-pair index $j(x)$.
\end{enumerate}
We shorten the notation to $\bftau^{j(x)}_{1:k}$ when the set of knots $\eta_{1:m}$ are clear from the context.
\end{definition}
There are several options for choosing the $k$-nearby knots.
For example, one can choose the right-window as $\bftau_{1:k}^{j(x),\text{right}}(\eta_{1:m}) \defeq \{\eta_{j(x)}, \dots, \eta_{j(x)+k-1}\}$,  
or the left window $\bftau_{1:k}^{j(x),\text{left}}(\eta_{1:m}) \defeq \{\eta_{j(x)-k+2}, \dots, \eta_{j(x)+1}\}$, 
or the left $k$-nearest neighbors $\bftau_{1:k}^{j(x),\text{lNN}}(\eta_{1:m}) = \text{KNN}_{k-1}(\eta_{j(x)}, \eta_{1:m}\setminus \{\eta_{j(x)+1}\}) \cup \{ \eta_{j(x)+1} \}$ where  $\text{KNN}_{k-1}(x, \calX)$ finds the $k-1$ nearest neighbors of $x$ in the set $\calX$.
One counter example that does not agree with our definition is if one takes the nearest neighbors of $x$ directly, $\bftau_{1:k}^{j(x),\text{NNx}}(\eta_{1:m}) = \text{KNN}_{k-1}(x, \eta_{1:m}) \cup \{ \eta_{j(x)}, \eta_{j(x)+1}\}$, as this may result in a different set of knots depending on $x$ even when $j(x)$ is fixed.
Our theory in the rest of the paper is independent of the choice of this method as long as the same method is used consistently. 
In our experiments, we use the $k$-nearest neighbor approach, $\bftau_{1:k}^{j(x),\text{NN}}(x)$. We use $\tau_i^{j(x)}$ to denote $i$th element in $\bftau_{1:k}^{j(x)}$.

\begin{definition}[Newton's polynomial operator]
Given data points $\calD_f(\bftau_{1:m})$, we can fit a $k$th-order Newton's polynomial using $k+1$ knots, ensuring it passes through all the selected knots. For each interval $x \in [\tau_n, \tau_{n+1})$, let $\bftau_{1:k+1}^{(j)}$ to be $k+1$ nearby knots (Definition~\ref{def:k-nearby_knots}). 
Then the piecewise polynomial fit in the $n$th interval $x\in [\tau_n, \tau_{n+1})$ is given by~\cite[p7]{de1978practical}:
\begin{align}
    \calP_{k,n(x)}[\calD_f(&\bftau_{1:m})](x) 
    \teq 
    \left[\tau_1^{n(x)}\right]f 
    \notag\\
    &+ 
    \sum_{i=2}^{k+1} \left[\bftau_{1:i}^{n(x)}\right]f \prod_{j=1}^{i-1} \left(x-\tau_j^{n(x)}\right).
    \label{eq:newton-form-poly}
\end{align}
    \label{def:newton-form-poly}
\end{definition}

Any $k+1$ distinct points sampled from a $k$th-order polynomial determine a unique Newton's polynomial 
with the same coefficients, regardless of the chosen points. 
This implies that for a polynomial $f(x)$ of order $k$, the Newton's polynomial satisfies $\calP_{k,n}[\calD_f(\bftau_{n:n+k})](x) = f(x)$~\cite{de1978practical}.

\begin{definition}[Piecewise polynomial] 
\label{def:piecewise-poly}
A piecewise polynomial $\hat{f}(x)$ of order $k$ on a given set of knots $\bftau_{1:m}$ consists of $m-1$ polynomials $\hat{f}_{[j]}(x)$ of order $k$, as follows:
\begin{align}
    \hat{f}(x) &=  \sum_{i=0}^k c_{i,j}x^k \eqcolon \hat{f}_{[j]}(x) \qquad \forall x \in [\tau_j, \tau_{j+1}), \notag\\
    \text{s. t.} \quad
    \hat{f}_{[j]}(& \tau_{j+1})
    = \hat{f}_{[j+1]}(\tau_{j+1}).      
    \label{eq:piecewise-poly}
\end{align}
The polynomial $\hat{f}_{[j]}(x)$ represents the $j$th piece of $\hat{f}(x)$. Piecewise polynomials are continuous but not necessarily smooth.
When we refer to $\calP_{k,j}$, we also mean the $j$th piece of the piecewise polynomial.
\end{definition}

\begin{figure}
\centering
\includegraphics[width=0.24\textwidth,trim=0pt 3pt 0pt 0pt, clip]{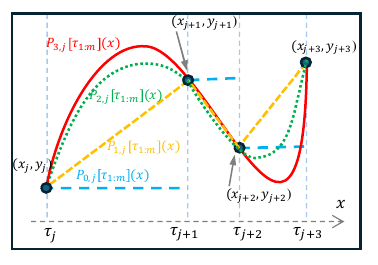}%
\includegraphics[width=0.24\textwidth,trim=0pt 0pt 0pt 0pt, clip]{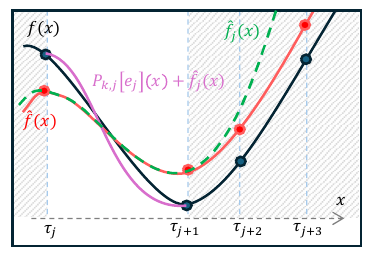}
\caption{\textbf{left)} Newton's piecewise polynomial of different orders is shown, with \textbf{black circles} representing selected knots at real values. The {\color{cyan}\textbf{blue line}} shows a zero-order piecewise polynomial, the {\color{darkyellow}\textbf{yellow line}} is the first-order, the {\color{darkgreen}\textbf{green line}} represents a quadratic Newton's polynomial, and the {\color{red}\textbf{red line}} is the cubic polynomial on the selected knots. 
\textbf{right)} The plot illustrates the notations for error as defined in Lemma 1. Black, {\color{red}red}, {\color{darkgreen}green}, and {\color{pink}pink} lines are true functions, spline approximation with non-zero error at knots, Newton's polynomial of interval $j$, and the adjusted polynomial of interval $j$ to compensate for non-zero error, respectively. }
\label{fig:Spline-Err-at-knot}
\label{fig:NewtonPoly}
\end{figure}

An illustration of this operator for different $k$ values is shown in Fig.~\ref{fig:NewtonPoly} (left).
Newton's polynomial is rarely used for spline fitting due to non-smooth transitions at knots. Smooth splines are typically constructed using B-splines. We base our spline error analysis on Newton's polynomial, as they might provide tight distance-aware error bounds (cf. Theorem~\ref{thm:poly-interp-bound}).

\textbf{B-splines:}
A B-spline of order $k$ is a smooth piecewise polynomial of the same order, defined in terms of basis functions called B-spline basis. B-splines are defined on $m$ knots $\bftau_{1:m}$ for $k$s greater than one as: 
\begin{align}
b_{k,i}(x) &\teq (t_{i+k} - t_{i})[t_i,\dots,t_{i+k}](. - x)_{+}^{k-1},
\end{align}
where $(x)_{+} =\max\{0, x\}$~\cite{wahba2006spline}.
Any smooth spline can be written as a linear combination of B-spline basis functions~\cite[p97]{de1978practical}, 
$\phi_k[\bftau_{1:m}](x) = \sum_{i=1}^{m-k} \alpha_i b_{k,i}(x) = \bfalpha^\top \bfb_{k,\bftau_{1:m}}(x)$
where $\alpha_i \in \bbR$ are the coefficients and $\bfalpha = (\alpha_1,\dots,\alpha_{\beta})^\top$, $\bfb_{k,\bftau_{1:m}}(x) = (b_{k,i}(x))_{i=1}^{\beta}$ are corresponding \mbox{basis} vectors. We denote ${\beta \teq m-k}$ as the size of the $k$th-order B-spline basis for $m$ knots and emphasize the dependence of the basis functions on the knots $\bftau_{1:m}$ by including it in the subscript.
Note that, in total, we have $m$ knots, $m-k$ B-splines, and $m-1$ polynomials defined between two consecutive knots.

\textbf{Kolmogorov-Arnold Theorem:} For any given continuous multivariate function $f(\bfx):\bbR^d \to \bbR; f \in \calC^0$; \acf{KAT}~\cite{kolmogorov1957representation} states that the following two-layer representation exists with a set of continuous univariate activation functions, $\{h_{l,i,j} \in \calC^0 \}_{l,i,j}$,  $f(\bfx) = \sum_{i=1}^{2d+1} h_{2,1,i}(\sum_{j=1}^d h_{1,i,j}(x_j))$. 

A natural question is whether the continuity assumptions for the multivariate function $f$ and univariate activations functions $h_{l,i,j}$ can be strengthened (higher order continuity, $f, h_{l,i,j} \in \calC^k: k \ge 1$)  or simplified (discontinuous $f, h_{l,i,j} \notin \calC^0$).
Notably, the theorem does not hold for continuously differentiable multivariate ($f \in \calC^1$) and univariate activation functions ($h_{l,i,j} \in \calC^1$)~\cite{morris2021hilbert}.
By increasing the number of layers in the KAT~\cite{liu2025kan} and relaxing equality to approximation~\cite{kuurkova1992kolmogorov}, one can extend KAT representation results with weaker or alternative assumptions~\cite{schmidt2021kolmogorov,hecht1987kolmogorov,ismailov2023three}, and one of the extensions is discussed next. 

\textbf{Kolmogorov-Arnold Networks:}
\ac{KAN}~\cite{liu2025kan} replaces the univariate activation functions in \ac{KAT} with splines. 
A two-layer model $\KAN_2: \bbR^{m_1} \to \bbR$ can be written as,
\[
    \KAN_2(\bfx) = \sum_{i=1}^{m_2} \phi_{2,1,i}\big(\sum_{j=1}^{m_1} \phi_{1,i,j}(x_j)\big), \notag
\]
where $\phi_{l,i,j}$ is a $k$th-order spline in layer $l$ that connects input $j$ to output $i$.
Consider the case when $\phi_{l,i,j}(x_j) = \bfalpha_{l,i,j}^\top \bfb_{k,\bftau_{l,j}}(x_j)$. Then KAN can be written as:
\begin{align}
\KAN_2(\bfx) 
&= \bfw_2^\top \bfB_{k,\bftau_{2}}(\bfW_1 \bfB_{k,\bftau_{1}}(\bfx)),
\label{eq:mlp-eq-kan}
\end{align}%
where $\bfW_1 = ([(\bfalpha_{1,i,j})_{j=1}^{m_1}]^\top)_{i=1}^{m_2} \in \bbR^{m_2\times m_1\beta}$, $\bfw_2 = (\bfalpha_{2,1,i})_{i=1}^{m_2} \in \bbR^{m_2\beta}$, $\bfB_{k,\bftau_1}(\bfx) = (\bfb_{k,\bftau_{1,j}}(x_j))_{j=1}^{m_1}$ and $\bfB_{k,\bftau_2}(\bfy) = (\bfb_{k,\bftau_{2,i}}(y_i))_{i=1}^{m_2}$. Here the notation $(\bfv_i)_{i=1}^{m_2}$ denotes vertical stacking of the column vectors $\bfv_i$ into a single column vector. Equation~\eqref{eq:mlp-eq-kan} is a \ac{MLP} with $\bfB_{k,\bftau_2}$ as its activation function and $\bfB_{k,\bftau_1}$ as a feature map of the inputs $\bfx$. 
Since the basis functions depend on the knots, and the model weights are learned from the training data, \ac{KAN} is a \textit{semi-parametric} model that brings the computational efficiency and accuracy of parametric models (e.g. \acp{NN}) with the interpretability of non-parametric models (e.g. splines, k-nearest neighbors). We use this property of \ac{KAN} to develop a distance-aware uncertainty estimator.

For ease of notation, let define $i$th output of layer $l$, which has $m_l$ components, as $y_{l,i}=\sum_{j=1}^{m_{l-1}}\phi_{l,i,j}(y_{l-1,j})$ and layer $l$ of network predicts the output $\bfy_l = \bfh_l(\bfy_{l-1}) = [y_{l,i}]_{i=1}^{m_l}=[\sum_{j=1}^{m_{l-1}}\phi_{l,i,j}(y_{l-1,j})]_{i=1}^{m_l}$, and $\bfy_0 = \bfx$. 
With this, in general, a \acf{SNN} consists of multiple \ac{KAN} layers each represented by $\bfh_l:\bbR^{m_l} \rightarrow \bbR^{m_{l+1}}$, which maps a $m_l$ dimensional input to $m_{l+1}$ dimensional output. Therefore, an $L$ layer multi-input multi-output \ac{KAN} can be written as:
\begin{align}
    \KAN_L(\bfx) = \bfh_L \circ \dots \circ \bfh_l \circ \dots \circ \bfh_1(\bfx).
    \label{eq:multilayer_kan_eq}
\end{align}

\textbf{Distance awareness:}
Distance-aware uncertainty is driven by the intuition that a model’s prediction uncertainty increases with distance from the training data. We define distance awareness for worst-case analysis inspired by~\cite{liu2020simple}.
\begin{definition}[Input distance awareness]
    \label{def:distance_awareness}
    Consider a predictive model $\hat{y} = \hat{f}(\bfx)$ that estimates a true function $y = f(\bfx)$ using $x \in \calX_\calD$. 
    Let $u_{\hat{f}}(\bfx) : \calX \to \bbR^+$
    be an uncertainty estimator such that for all $\bfx \in \calX$, the true output is bounded $y \in [\hat{f}(\bfx)-u_{\hat{f}}(\bfx), \hat{f}(\bfx)+u_{\hat{f}}(\bfx)]$.
    The uncertainty estimator $u_{\hat{f}}(\bfx)$ is said to be distance-aware if it monotonically increases with the test point's distance from the training data with respect to some distance function $d(\bfx, \calX_\calD)$.
\end{definition}

Unlike Liu~et~al.~\cite{liu2020simple}, who define the set distance $d(., \calX_\calD)$ as the expected distance of a test point to all the training data, we define it to be the distance of the test point from the nearest training point (or knot) $\tau^*$,
\begin{align}
    d(\bfx, \calX_\calD) = \min_{\tau \in \calT} d_u(\bfx, \tau) = d_u(\bfx, \tau^*),
\end{align}
where $\calT \subset \calX_\calD$ is the set of training points (or knots).
We call this distance function $d_u(\bfx, \tau^*)$ to be the \emph{inducing distance} for the uncertainty estimator $u_{\hat{f}}$.
The inducing distance does not need to be Euclidean, or $\ell_p$, 
it can be a geodesic distance over the data manifold. 
If both the uncertainty estimator and the inducing distance are differentiable, the distance-awareness property can be compactly stated as,
\begin{align}
    [\nabla_\bfx u_{\hat{f}}(\bfx)]^\top \nabla_\bfx d_u(\bfx, \tau^*) \ge 0 \quad \forall \bfx \in \calX.
    \label{eq:dist-aware-cond}
\end{align}
The above condition enforces that the uncertainty monotonically increases with distance.

\textbf{Measuring Distance-Awareness:} 
Evaluating \eqref{eq:dist-aware-cond} over the entire input space is intractable, as it requires global monotonicity verification over $\calX$.
Therefore, we adopt a sampling-based metric.
For computational efficiency and consistency across uncertainty estimators, we assume the inducing distance to be Euclidean.

\begin{definition}[Sampled Distance-Awareness (SDA)]
    \label{eq:sampled-distance-awareness}
    Let $\bfx_t \sim \calX_{\text{test}}$ be sampled uniformly from the test data.
    Then the \emph{sampled distance-awareness} (SDA) of $u$ is defined as,
    \begin{align}
        \label{eq:probability_distance_aware}
        \text{SDA} = \frac{1}{N}\sum_{\bfx_t \sim \calX_{\text{test}}}
        \mathbbm{1}
        \left[
        \nabla_{\bfx} u_{\hat{f}}(\bfx_t)^\top (\bfx_t - \bftau^*) \ge 0 \right],
    \end{align}
\end{definition}%
where $\mathbbm{1}[.]$ is the indicator function.
A higher SDA value indicates that, with high probability, uncertainty increases as one moves farther from the nearest knot, which captures the distance-awareness property of an uncertainty estimator.

We assume Lipschitz continuity to obtain distance-awareness bounds, as it is a common assumption in control literature, and many real-world functions obey this criterion.
Since we are working with $k$th-order splines, we define an extended notion of $k$th-order Lipschitz continuity as follows:

\begin{assumption}[$k$th-order Lipschitz continuity]
\label{assm:k-Lips}
For a ${k-1}$ time differentiable function $f: [a,b] \to \bbR$, the
$k$th-order Lipschitz constant $\calL^{k}_f$ is the maximum rate of change in the $(k-1)$th derivative of the function $f$ to the change of input,
\begin{align}
    \frac{|f^{(k-1)}(x)-f^{(k-1)}(y)|}{d(x,y)} \le \calL^{k}_f \quad \forall x \neq y.
    \label{eq:lipschitz_def}
\end{align}%

If such a constant exists, the function is $k$th-order Lipschitz continuous. The inequality holds for all $x,y \in [a,b]$.
As $y$ approaches $x$, the left-hand side converges to the derivative, $\lim_{y \rightarrow x} \frac{|f^{(k-1)}(x)-f^{(k-1)}(y)|}{|x-y|} = |f^{(k)}(x)|$. Thus, the inequality is valid for all $x$ and $y$, and it is applied to the maximum of the $k$th derivative as well.
When the distance function $d(x,y) = |x-y|$ is the absolute difference and the function $f$ is $k$ times differentiable, then $\max_{x\in[a,b]} |f^{(k)}(x)|$$ \le \calL^{k}_f$. Note that for a multivariate function $f$, the Lipschitz constant of its derivatives is the supremum of changes in $f$ along any direction; see Equation~\eqref{eq:mul-Lips}, similar to Lemma 1 in~\cite{dhiman2021control}.
\end{assumption}

If we assume a function $f$ and its approximator $\hat{f}$ to be first-order Lipschitz continuous with constant $\calL^{1}_f$, then a trivial distance-aware uncertainty estimator is 
\begin{align}
    \bar{u}_{\hat{f}}^\ell(\bfx) = (2+\epsilon_{\text{safe}}) \calL^{1}_f d(\bfx,\tau^*).
    \label{eq:trivial-bound}
\end{align}
Here, we considered the true function and approximation to have the same Lipschitz constant and $\epsilon_{\text{safe}}$ as a safety margin.
In the next section, we extend this notion to tighter bounds for higher-order splines and later multi-layer \ac{KAN} networks. 

\section{Problem formulation}
We formalize the problem of worst-case uncertainty estimation from the perspective of safe control.
Suppose the function being learned (e.g., system dynamics or cost-to-go) is $f: \bfx \mapsto y \in \bbR$.
Also, we know a set of assumptions about the true functions so that we can constrain it to a function class $f \in \calF$.
In this paper, we consider the class $\calF$ to be a class of all $k+1$-order Lipschitz functions with known constants $\calL^{k+1}_f$.
We also have a dataset of noisy input-output pairs available to learn the desired function $f$, $\calD \defeq \{ (\bfx_1, y_1), (\bfx_2, y_2), \dots, (\bfx_n, y_n)\}$, where $y_i = f(\bfx_i) + \epsilon_i \in \bbR$ are the corresponding observed (possibly noisy) outputs, with bounded noise $\epsilon_i \le \epsilon_{\text{max}}$. 
The first step is to learn the function approximation $\hat{f}(\bfx; \Theta) \in \calH$ from the hypothesis space $\calH$, where $\Theta$ is the combination of spline parameters and knots locations required for the function representation.
In this paper, our hypothesis space is the class of multi-layer KAN networks with a fixed architecture.
Let a learning algorithm provide us the optimal parameters $\Theta^*$ according to \eqref{eq:training},
with $l: \calY \times \calY \to \bbR^+$ as a given loss function.
Our objective is to find the worst-case bound for the learned function approximation.
With slight abuse of notation, let $\calF \cap_\epsilon \calD$ denote the space of all functions in $\calF$ that may generate the dataset $\calD$, $\calF \cap_\epsilon \calD \defeq \{ g \in \calF \mid l(g(\bfx_i), y_i) \le \epsilon_{\text{max}}, \, \forall (\bfx_i, y_i) \in \calD \}$\footnote{This bounded feasible set can be extended to probabilitic set as well~\cite{bertsimas2021probabilistic}, $\{ g \in \calF \mid P( l(g(\bfx_i), y_i) \le \epsilon_{\text{th}}) \geq 1-\delta, \, \forall (\bfx_i, y_i) \in \calD \}$.}. 
Then the uncertainty bound $u_f(\bfx)$ at a given test point $\bfx$ is the loss if the true function is chosen to be the worst possible in the given range of possibilities,
\begin{align}
    \Theta^* &= \argmin_\Theta \sum_{(\bfx_i,\bfy_i)\in\calD} l(\hat{f}(\bfx_i, \Theta), \bfy_i), \label{eq:training} \\
    u_f(\bfx) &\ge  \sup_{f \in \calF \cap_\epsilon \calD} l(f(\bfx), \hat{f}(\bfx; \Theta^*)).
    \label{eq:prob-form}
\end{align}%
In this paper, we do not modify the training process $\eqref{eq:training}$, but focus on estimating the tight worst-case bound $u_f(\bfx)$. 
Optimizing over $\calF \cap_\epsilon \calD$ is in general intractable.
In this paper, we focus on a class of $k$th-order Lipschitz smooth functions.

\section{Method}
\label{sec:method}
In this section, we derive an error bound for the \ac{KAN} model in three steps.
First, we introduce the error bound for Newton's piecewise polynomial approximation. 
Second, we extend this result to arbitrary splines where the error at knots may be non-zero. 
Finally, we extend the resulting error bound to \acp{KAN} through the addition and composition of spline functions. 
We then present the error bound for multilayer networks and a method for handling residual connections, which is used in structures such as \ac{KAN}.
Additionally, we explored various approaches to dividing Lipschitz constants across layers and effectively managing error propagation.

\subsection{Piecewise Polynomial Error (PPE)}
\label{sec:PPE}
We begin by establishing an error bound for Newton's piecewise polynomial approximation.
\begin{theorem}[Newton's polynomial error bound]
\label{thm:poly-interp-bound}
Let $f \in \calC^{k+1}$ which is defined over $[a,b]$ and has $k+1$ continuous derivatives and is $k+1$th-order Lipschitz continuous with constant $\calL^{k+1}_f$. The approximation error for $k$th-order Newton's piecewise polynomial~\cite{de1978practical} fit, $\calP_{k,j}[\calD_f(\bftau_{1:m})]$, that passes through the knots $\calD_f(\bftau_{1:m})$, at the test point $x \in [\tau_j, \tau_{j+1})$ for all $j \in \{1, \dots, m - 1 \}$ is bounded as follows:

\begin{align}
\hspace{-3mm}
    | f(x) - \calP_{k,j}[\calD_f(\bftau_{1:m})](x) | &\le  
    \underbrace{
        \frac{\calL_f^{k+1}}{(k+1)!} \left|\prod_{i=1}^{k+1} (x-\tau_i^{(j)})\right|} 
        _{\eqqcolon \bar{u}_f^{k+1}(x; \bftau_{1:m})},
    \label{eq:int-error}
\end{align}
where $\tau_i^{(j)}$ is a knot in the set $\bftau_{1:k+1}^{j(x)}$ of $k + 1$ nearby knots to $x$ in $\bftau_{1:m}$ as defined in Definition~\ref{def:k-nearby_knots}. 
The above bound is tight when the $k+1$th derivative is constant at the maximum amount, $f^{(k+1)}(x) = \calL^{k+1}_f$.
\end{theorem}
\begin{proof}
    A function $f$ can be written as a $k$th-order polynomial  approximation along with the remainder term~\cite{phillips2003interpolation}:    
    \begin{align}
        f(x) &= \calP_{k,j}[\calD_f(\bftau_{1:k+1}^{j(x)}) ](x) \notag \\
        &\quad + (x-\tau_{1}^{(j)}) \hdots (x-\tau_{k+1}^{(j)}) [\tau_1^{(j)},\hdots,\tau_{k+1}^{(j)},x]f.
    \end{align}%
    Rearranging the above expression and using the mean value theorem for divided differences (Theorem~\ref{thm:divided-diff-mean-value}), we obtain
    \begin{align}                
        |f(x) - &\calP_{k,j}[\calD_f(\bftau_{1:k+1}^{j(x)})](x)| = \frac{f^{(k+1)}(\zeta)}{{(k+1)}!}   \prod_{i=1}^{k+1} (x-\tau_i^{(j)}) \notag\\
        &\le \frac{\calL_f^{k+1}}{{(k+1)}!} \left|\prod_{i=1}^{k+1} (x-\tau_i^{(j)})\right|
        , \quad \exists \zeta \in [a,b].
    \end{align}
\end{proof}

\begin{remark}
    Newton's polynomial bound $\bar{u}_f^{k+1}(x; \bftau_{1:m})$, defined in \eqref{eq:int-error}, provides the worst-case bound as defined in $\eqref{eq:prob-form}$, when the true function class $\calF$ consist of all $k+1$-order Lipschitz continuous function, the hypothesis class $\calH$ is the class of splines and $\epsilon_\text{max}=0$.
    Additionally, the bound is distance-aware with a distance that is proportional to the geometric mean of the distances to the (k+1)-nearby knots, $d(x, \calX_\calD) \defeq \prod_{i=1}^{k+1}\left|x-\tau_i^{(j)}\right|$.
    In Appendix~\ref{ptf:thm-error-kan-layer}, we provide the proof that Newton's polynomial is distance-aware.
\end{remark}

\begin{remark}
    This higher-order error bound can be loose when the test point is far away from the knots. To address this, we consider the minimum of the trivial bound \eqref{eq:trivial-bound} and Newton’s polynomial error bound as interpolation error, $\bar{u}_f(x; \bftau_{1:m}) = \min \, \{\bar{u}_f^{\ell}(x; \bftau_{1:m}), \bar{u}_f^{k+1}(x; \bftau_{1:m})\}$. Also, if the local Lipschitz constant is known rather than a global Lipschitz constant, it can provide a tighter bound. 
\end{remark}

As discussed in Section~\ref{sec:back}, we have selected to use the error bound for Newton's polynomial as the basis of our analysis, because it offers a tight and distance-aware bound for piecewise polynomials, the building block of \acp{SNN}.
The interpolation error in Theorem~\ref{thm:poly-interp-bound} assumes zero error at the knots, which is not typically possible in many conditions, such as when data is noisy, when there are fewer knots than data points, or when we require the piecewise polynomial to be smooth. 
In the next Lemma, we provide an error bound for any spline approximation that has non-zero error at the knots.  

\begin{lemma}[Piecewise polynomial interpolation at knots]
\label{thm:error-at-knots}
With the same assumptions as Theorem~\ref{thm:poly-interp-bound}, consider a piecewise polynomial approximation $\hat{f}(x)$ that does not necessarily pass through all the knots $\calD_f(\bftau_{1:m})$.
Let $\hat{f}_{[j]}(x)$ be the $j$th polynomial piece of $\hat{f}(x)$ (Definition~\ref{def:piecewise-poly}).
Because $\hat{f}_{[j]}(x)$ is a polynomial, we can extend its domain to the domain of $f(x)$, $[a, b]$.
Also, define the corresponding error function as $e_j^f(x) \teq f(x) - \hat{f}_{[j]}(x)$. Note that $e_j^f(x) : [a,b] \to \bbR$ is defined on the entire domain of $f(x)$, not just $x \in [\tau_j, \tau_{j+1})$ and has known values at the knots, $\calD_{e_j^f}(\bftau_{1:m})$. Then the error for the interval $x\in [\tau_j, \tau_{j+1})$ for all $j \in \{1, \dots, m-1\}$ is bounded by,
\begin{align}
    \label{eq:error-at-knots-ebs}
    |f(x)-\hat{f}(x)| \le &\bar{u}_f(x; \bftau_{1:m}) + |\calP_{k,j} [\calD_{e_{j}^f}](x)| \notag\\ 
    &\eqqcolon u_f(x; \bftau_{1:m}) \quad \text{(EBS)},
\end{align}
where $\bar{u}_f(x)$ is the error bound from Theorem~\ref{thm:poly-interp-bound}. We refer to~\eqref{eq:error-at-knots-ebs} as the \textit{error bound with spline fit} (EBS).
The above bound is tight when the inequality in Theorem~\ref{thm:poly-interp-bound} is tight and $\sgn(f(x) -\calP_{k,j}[\calD_f](x)) = \sgn(\calP_{k,j}[\calD_{e^f_j}](x))$ for ${x \in [\tau_1,\tau_{m})}$. 
This Lemma holds for any $k$th-order piecewise polynomial, regardless of the learning algorithm for $\hat{f}$.
\end{lemma}

Fig.~\ref{fig:Spline-Err-at-knot} (right) depicts the actual function $f$ in black, the non-zero spline approximation $\hat{f}$ in red, the $j$th piece of spline approximation $\hat{f}_{[j]}$ in green, and the adjusted $j$th piece of $k$th-order Newton's polynomial in pink. 
The proof of Lemma~\ref{thm:error-at-knots} is provided in Appendix~\ref{ptf:thm-error-at-knot} and the proof of distance awareness of Theorem~\ref{thm:poly-interp-bound} and Lemma~\ref{thm:error-at-knots} are provided in Appendices~\ref{sec:background-proofs-DA_of_IntErr} and~\ref{sec:background-proofs-DA_of_Spline}, respectively.
Although the bound in Lemma~\ref{thm:error-at-knots} is tight, in practice the knot errors $\calD_{e^f_j}$ may contain measurement noise and modeling error. 

When knots are densely spaced, fitting a high-order polynomial $\calP_{k,j}[\calD_{e^f_j}](x)$ can introduce oscillations between knots, similar to Runge's phenomenon~\cite{epperson1987runge}.
These oscillations may lead to numerical instability, are distance-unaware, and are undesirable because the error at the knots is expected to primarily capture measurement noise and modeling error, which should not contribute more to the interpolation error than the error at nearby knots itself.
To avoid these oscillations, we fit a piecewise linear function $|\calP_{1,j}[\calD_{e_j^f}](x)|=[\tau_{j}, \tau_{j+1}]|e_j^f| (x-\tau_j) + |e_j^f|$, where the slope is the divided difference $[\tau_{j}, \tau_{j+1}]|e_j^f|= ( |e_{j+1}^f| - |e_j^f| ) / (\tau_{j+1} - \tau_j)$ and refer to this approach as \textit{error bound with linear fit} (EBL).
\begin{align}
    \label{eq:error-at-knots-ebl}
    u_f(x; \bftau_{1:m}) \coloneqq &\bar{u}_f(x; \bftau_{1:m}) + |\calP_{1,j} [\calD_{e_{j}^f}](x)| \quad \text{(EBL)}.
\end{align}
This choice smooths the contribution of the error at knots, reduces sensitivity to noise, and lowers computational complexity. 
If the worst-case measurement error is known, that can be added to the error at knots. For simplicity, we use the same notation $u_f$ for both EBS~\eqref{eq:error-at-knots-ebs} and EBL~\eqref{eq:error-at-knots-ebl}, considering EBS as the default error bound and explicitly mentioning EBL when it is used.
\subsection{Multilayer Spline Error Analysis}
\label{sec:multilayer-spline-error-analysis}

Next, we discuss how the error propagates when composing spline functions.
To extend layer-wise error bound to a two-layer case, consider $f(x): [a, b] \to \bbR$ approximated by a two-layer architecture $\hat{f}(x)= \hat{h}(\sum_{i=1}^n\hat{g}_i(x)) = \hat{h}(\bfone_n^\top \hat{\bfg}(x))$, where $\mathbf{1}_n$ is a vector of $n$ ones.
Any true 1-D function $f$ can always be decomposed into the same network structure as the approximation, $f(x) = h(\bfone_n^\top \bfg(x))$, with infinitely many valid decompositions.
For example, when $n=2$, one such decomposition is $g_1(x) = f(x)$, $g_2(x) = 0$, and $h(x) = x$.
Moreover, given any decomposition $f(x) = h(g_1(x)+g_2(x))$ and an invertible function $v(x)$, we can get a new decomposition $f(x) = h'(g'_1(x) + g_2(x))$ where $h'(y) \teq h(v^{-1}(y))$ and $g'_1(x) \teq v(g_1(x)+g_2(x))-g_2(x)$.
The tightest worst-case error can be achieved by picking a true decomposition that minimizes layer-wise error among all possible decompositions.
We revisit the discussion of picking a true decomposition in subsection~\ref{sec:Error-division}.
The next theorem provides error bounds given knowledge of the true decomposition values at the knots.

\begin{theorem}[Two-layer KAN error bound]
\label{thm:two-layer-poly-bound}
    Consider a true two-layer function $f(x): [a, b] \to \bbR$, approximated by a composition of $k$th-order piecewise polynomials $\hat{f}(x) = \hat{h}(\bfone_n^\top \hat{\bfg}(x))$. 
    Let the true decomposition at the knots be 
    given as
    $f(\tau_i)= h(\bfone_n^\top \bfg(\tau_i))$ for all $\tau_i \in \bftau_{1:m}$.
    Define the errors at knots as $e^h(.) \teq h(.)-\hat{h}(.)$ and $e^\bfg(.) \teq \bfg(.) - \hat{\bfg}(.)$, and assume $h$ is first-order Lipschitz continuous with constant $\calL^1_h$.
    Let $f,h,\bfg \in \calC^{k+1}$, with known Lipschitz constants, and let $\hat{h}$, $\hat{\bfg}$ be piecewise polynomials.
    Then the error bound for the two-layer approximation is given by:
    \begin{align}
    \label{eq:two-layer-thrm}
    \hspace{-2.5mm}
    |f(x) -\hat{f}(x)|
    \le u_h(\bfone_n^\top \hat{\bfg}(x);\bfxi_{\bfg_m}) + \calL_h^1 \bfone_n^\top \boldsymbol{u}_{\bfg}(x;\bftau_{1:m}),
    \end{align}
    where {$\bfxi_{\bfg_m} \teq \{\bfone_n^\top \hat{\bfg}(\tau_i)\}_{i=1}^m$ is input for $\hat{h}$ layer} at knots, and $\boldsymbol{u}_\bfg(x) \teq (u_{g_i})_{i=1}^n$ is the vector of error bound in $\hat{g}_i$ spline approximations as estimated in Lemma~\ref{thm:error-at-knots}.
    The above bound is tight when $h^{(1)}(t) = \sgn(h(\bfone^\top_n\hat{\bfg}(x))-\hat{h}(\bfone_n^\top \hat{\bfg}(x)))\calL^1_h$, for all $t \in [\bfone_n^\top \hat{\bfg}(x),\bfone_n^\top \bfg(x)]$.
\end{theorem}
\begin{proof}
For ease of exposition, we use a hidden layer of size two $\hat{\bfg}(x) = [g_1(x), g_2(x)]^\top$.
Consider the left-hand side,
\begin{align}
    |f(x)-\hat{f}(x)| &= |h(g_1(x)+g_2(x)) - \hat{h}(\hat{g}_1(x)+\hat{g}_2(x))| \notag\\
    &= |h(y) - \hat{h}(\hat{y})|,
\end{align}
where $\hat{y} \triangleq \hat{g}_1(x)+\hat{g}_2(x)$ and $y\triangleq g_1(x)+g_2(x)$. To use the bound on the approximation of $h$, we need to consider the change in $h$ due to $y$. We use the Taylor series with integral remainder on function $f(y)$~\cite{apostol1967calculus}:
\begin{align}
    |h(y) - \hat{h}(\hat{y})| &=  |h(\hat{y}) + \int_{\hat{y}}^{y} h^{(1)}(t) dt - \hat{h}(\hat{y})| \notag \\
    &\le |h(\hat{y}) - \hat{h}(\hat{y})| + |\int_{\hat{y}}^{y} h^{(1)}(t) dt | \notag \\
    &\le |h(\hat{y}) - \hat{h}(\hat{y})| + \calL^1_h |y-\hat{y}|. \label{eq:two-layer-proof1}
\end{align}
Substituting Theorem~\ref{thm:poly-interp-bound} into Equation~\eqref{eq:two-layer-proof1} results in Equation~\eqref{eq:two-layer-thrm}.
The first inequality is tight when the sign of integral term is the same as $h(\hat{y})-\hat{h}(\hat{y})$ and second inequality is tight when $h^{(1)}(t) = \calL^1_h$ for the integral range, $t \in [\hat{y},y]$.
\end{proof}

\textbf{Error of Multi-Input Function:}
Consider a function $f(\bfx)$ approximated by $\hat{f}(\bfx)=\hat{h}(\sum_{i=1}^{m_i}\hat{g}_i(x_i))$. Let the set of input knots be represented by a matrix $\calT \in \bbR^{m_i \times m}$, where $m_i$ is the number of inner spline functions $g_i$ and $m$ is the number of knots per spline. 
Define $y_1=\sum_{i=1}^{m_i}g_i(x_i)$ and $ \bfxi_{1:m} = \{\sum_{i=1}^{m_i}g_i({\calT_{i,j}})\}_{j=1}^{m}$, where $\calT_{i,j}$ is $j$th knot of $i$th spline. Similar to Theorem~\ref{thm:two-layer-poly-bound}, the error is bounded by:
\begin{align}
    u_{f}(\bfx)&= u_h(y_1;\bfxi_{1:m}) +\sum_{i=1}^{m_i} \calL_{\partial h,i}^1 u_{g_i}(\bfx_i;\calT_i), 
\end{align}
where $\calL_{\partial h,i}^1$ is first-order component-wise Lipschitz constant of $h(.)$ with respect to its $i$th input. 
Formally for a multi-input output function $\bfh: \bbR^{m_i} \to \bbR^{m_o}$,
(similar to~\cite{lederer2019uniform, dhiman2021control}),
\begin{align}
    \calL^{(k)}_{\partial \bfh_o,i} &\teq \sup_{\bfx \in \bbR^{m_i},\lambda \in \bbR \setminus \{0\}} \left|
    \frac{\bfh^{(k-1)}_{o}(\bfx + \lambda \bfe_i) - \bfh^{(k-1)}_o(\bfx)}{\lambda} \right|, \notag\\ 
    &\forall 
        i \in \{1,\dots,m_i\},\quad
        o \in \{1,\dots,m_o\},
    \label{eq:mul-Lips}
\end{align}
where $\bfe_i = (0, 0, \dots, 1, \dots, 0)$ is the standard basis vector that is all zeros except at the $i$th coordinate. 
This definition is different from that in Assumption~\ref{assm:k-Lips}, as the change in input is allowed only in one dimension. However, the two definitions are equivalent with different constants~\cite{lederer2019uniform, dhiman2021control}. Henceforth, we use $\calL^{(k)}_\bfh$ for both when clear from context.

\begin{theorem}[Multi-layer KAN error]
\label{thm:multi-layer-poly-bound}
For an $L$ layers network (Equation~\eqref{eq:multilayer_kan_eq}), let the output of layer $l$ at knots be $\Xi_l = \bfh_l(\Xi_{l-1})$ and the output evaluated at the input be $\hat{\bfy}_l = \bfh_l(\hat{\bfy}_{l-1})$, where $\Xi_0 = \calT$ and $\hat{\bfy}_0 = \bfx$ (the input). 
Then the error of layer $l$ is $\bfu_l(\hat{\bfy}_{l-1};\Xi_{l-1})+ \calL_{h_{l}}^{1} \bfu_{l-1}(\hat{\bfy}_{l-2};\Xi_{l-2})$.
Note that $\bfu_0 = \mathbf{0}_{m_0}$ is a zero vector of size $m_0$. 
Then the overall error, with all the same assumptions at Theorem~\ref{thm:two-layer-poly-bound} for all the $L$ layers, will be:
\begin{align}
    |\bff(\bfx) &- \KAN_L(\bfx)| \le \bfu_L(\hat{\bfy}_{L-1};\Xi_{L-1}) \notag \\
                &\quad + \sum_{l=1}^{L-1} ( \bfu_l(\hat{\bfy}_{l-1};\Xi_{l-1}) \prod_{j=l+1}^{L} \calL^1_{h_j}  ).
    \label{eq:multi-layer-poly-bound}
\end{align}
\end{theorem}

The above theorem shows that the error has two components: the error introduced by each layer and the error propagated from previous layers. Depending on the Lipschitz constant of the outer layers, any error in the inner layers can quickly scale up in the outer layers. 

\begin{algorithm}
\small
\DontPrintSemicolon
\KwData{Input-output pairs $\bff(\calT)$, trained model
$\hat{\bff}(\bfx)=\hat{\bfh}_L(\dots\hat{\bfh}_2(\hat{\bfh}_1(\bfx)))$, test point $\bfx^*$, order $k$, Lipschitz constants $\calL^{(k+1)}_\bff$, $\calL^{(1)}_\bff$.}
Precompute errors $e^\bff = |\bff(\calT) - \hat{\bff}(\calT)|$. \;
Divide $e^\bff$ among layers, $e^{\bfh_1}, \dots, e^{\bfh_L}$ at knots.\; 
Divide $\calL_\bff^1, \calL_\bff^{k+1}$ among layers.\;
\For{$l,n,p \in (\{0,\dots, L-1\} \times \{1, \dots, m_l\} \times \{1, \dots, m_{l+1}\})$}{
    \tcc{$\calO(\log_2(m))$ binary search.}
    Find $j$ such that  $\hat{\bfy}_{l,n}(\calT_{1:m_0,j}) \le \hat{\bfy}_{l,n}(\bfx^*) < \hat{\bfy}_{l,n}(\calT_{1:m_0,j+1}).$\;
    Fit Newton's Polynomials $\calP_{k,j}[\calD(e^{h_{l,n}})]$ on the knots $\{(\hat{\bfy}_{1:k+1,l,n}^{(j)}(\calT_{1:m_0}), e^{{h}_{{l+1},p}}(\hat{\bfy}_{1:k+1,l,n}^{(j)}(\calT_{1:m_0})))\}_{i=1}^{k+1}$ \;
    Find 
    $\bfu_{{l+1},n}(\hat{\bfy}_{l,n}(\bfx^*); \hat{\bfy}_{l,n}(\calT))$ [Lemma~\ref{thm:error-at-knots}].\;    
 }

Compute error bound over $f$ [Thm.~\ref{thm:two-layer-poly-bound}]
\caption{DAREK \label{alg:DAREK}}
\end{algorithm}

\textbf{DAREK Algorithm:}
We summarize \ac{DAREK} in Algorithm~\ref{alg:DAREK}, which computes error bounds for an L-layer spline network with $m_l$ hidden units in layer $l$.
The intermediate layer values are denoted as $\hat{\bfy}_0(\bfx) \teq \bfx$, $\hat{\bfy}_l(\bfx) \teq \hat{\bfh}_{l}(\dots  \hat{\bfh}_2(\hat{\bfh}_1(\bfx)))$, and $\hat{\bfy}_L(\bfx) \teq \hat{\bff}(\bfx)$.
Here, subscript $n$ denotes an element of a vector, and $j$ denotes the $j$th piece of a piecewise polynomial.
To understand the algorithm, consider a two-layer model $\hat{f}(x)=\hat{h}_{2,1}(\hat{h}_{1,1}(x)+\hat{h}_{1,2}(x))$. 
After dividing the prediction error $e^f$ and Lipschitz constant budgets, $\calL_f^{(1)}$ and $\calL_f^{(k+1)}$, among neurons ($\hat{h}_{2,1}$, $\hat{h}_{1,1}$, and $\hat{h}_{1,2}$), the algorithm calculates each neuron's error by substituting its knots into Equations~\eqref{eq:int-error} and~\eqref{eq:error-at-knots-ebs}.
It then computes the error of each layer ($u_{h_2}=u_{h_{2,1}}$ and $u_{h_1}=u_{h_{1,1}}+u_{h_{1,2}}$), and the total error using Equation~\eqref{eq:two-layer-thrm}, $u_f=u_{h_2} + \calL_{h_2}^{(1)} u_{h_1}$.

The algorithm requires Lipschitz division and error at knots division in the second and third lines of Algorithm~\ref{alg:DAREK}, which is discussed in detail in Subsections~\ref{sec:Lip-division} and~\ref{sec:Error-division}. 

\subsection{Residual Connections in KAN Layer}
\label{sec:residual-layer}
In \ac{KAN}~\cite{liu2025kan}, the authors use the residual connection in the form of $\hat{f}(x) =: r(x) + \phi_k(x)$ to improve the optimization stability and convergence of the model. 
The $r(x)$ in \ac{KAN} implementation is a \ac{SiLU} function defined as $\silu(x) = {x}/{(1+\exp(-x))}$.
The following theorem shows how to compute the neuron's error when it is added to an arbitrary function, assuming the residuals are errorless. 

\begin{theorem}[Error bound of spline + any function]
\label{thm:error-kan-layer}
The error bound of the general structure for function defined on the interval $[\tau_j,\tau_{j+1})$ as $\hat{f}_{[j]}(x) = \hat{r}(x) + \phi_{k}(x)$ where $\phi_{k}$ is a polynomial of order $k$, and $\hat{r} \in \calF$ has the same assumptions as $f(x) \in \calF$, then is given by: 
\begin{align}
    |f- \hat{f}_{[j]}| &\le 
    \bar{u}_{f-\hat{r}}(x; \bftau_{1:m}) +  |\calP_{k,j}[\calD_{e_j^{f-\hat{r}}}]| = u_{f-\hat{r}}.
\end{align}
\end{theorem}
Intuitively, if we assume the true function is $f-\hat{r}$, then all error arises from the $\phi_{k}$. 
Proof is provided in Appendix~\ref{ptf:thm-error-kan-layer}.

\subsection{Lipschitz Division among Neurons}
\label{sec:Lip-division}

While the Lipschitz constant of the overall true function $f$ might be known from the problem definition $\calF$, determining the Lipschitz constants for the intermediate layers remains an open problem.
In this section, we explore empirical methods for dividing the Lipschitz constant of the true function among the neurons in each layer $\bfh_l$, possibly informed by the Lipschitz constant of the approximated neurons in each layer $\bfh_l$.
As discussed in Section~\ref{sec:multilayer-spline-error-analysis}, our goal is to find the closest true function decomposition that respects the known constraints of the true function $f$. 

The Lipschitz constant of each layer of the approximate function can be estimated directly from the data, representing the sharpest change in the layer's output relative to its input.
Alternatively, more efficient methods, such as those proposed by Shi et al., can be used~\cite{shi2022efficiently}.
Other works aim to bound the function approximation during training~\cite {aziznejad2020deep, liu2020simple}; however, we aim to assign a Lipschitz constant to each neuron of the trained network.
Before introducing our proposed approaches, we first review two properties of Lipschitz constants.

\begin{lemma}[Lipschitz constant of a sum]
\label{def:Lip-sum}
Let $g(x) = \sum_{i=1}^{d}g_i(x)$, where each $g_i$ has Lipschitz constant $\calL_{g_i}$. 
Recalling from Equation~\eqref{eq:lipschitz_def}, $|g(x)-g(y)| \le \calL_g |x-y|$, the Lipschitz constant of $g$ would be:
\begin{align}
    |g(x)-g(y)| &= 
    \left|\sum_{i=1}^{d}(g_i(x)-g_i(y))\right| \le \sum_{i=1}^{d}|g_i(x)-g_i(y)| \notag \\
                &\le \sum_{i=1}^{d} \calL_{g_i} |x-y| = (\sum_{i=1}^{d}\calL_{g_i}) |x-y|.
\end{align}
\end{lemma}

\begin{lemma}[Lipschitz constant of composite function]
\label{eq:LipschitzComposite}
Let the composite function $f(x)=h(g(x))$, where $g$ and $h$ have Lipschitz constants $\calL_g$ and $\calL_h$, respectively. 
Then:
\begin{align}
    |f(\bfx)-f(\bfy)|&=|h(g(\bfx))-h(g(\bfy))| \notag \\
    &\le \calL_h |g(\bfx)- g(\bfy)| \le \calL_h \calL_g |\bfx - \bfy|.
\end{align}
\end{lemma}
Based on Definition~\ref{def:Lip-sum} and Equation~\eqref{eq:LipschitzComposite}, for a $\KAN_L$ model (Equation~\eqref{eq:multilayer_kan_eq}) with $L$ layers and $m_l$ nodes per layer contributing to each output dimension, the total Lipschitz constant is bounded by:
$\calL_f = \prod_{l=1}^{L}(m_l \calL_l)$, where $\calL_l$ is the Lipschitz constant of each node in layer $l$.
We define $\hat{\calL}_l$ as the empirical Lipschitz constant of each neuron in layer $l$, computed numerically using Equation~\eqref{eq:lipschitz_def}.
Our goal is to determine, layer-wise Lipschitz constants $\calL_L$ for true function $f$ so that they are proportional to the layer-wise Lipschitz constants $\hat{\calL}_l$ for the approximated function $\hat{f}$, while satisfying the total Lipschitz budget constraint of
$\calL_f = \prod_{l=1}^{L}(m_l \calL_l)$.
Furthermore, we aim to balance the computational complexity of the approaches against their impact on uncertainty estimation. 
We proposed five different methods varying in computational cost and level of detail in modeling the problem: 
optimization-based division,
logarithmic division, 
linear division,
equal division, and
worst-case division.
In the following, we explain each one and then compare them in toy examples on a trained model.

\subsubsection{Optimization Based Division}
We start with the most detailed formulation, consequently the most computationally costly, which seeks the tightest possible division by explicitly accounting for higher-order Lipschitz constants.
Consider a two-layer model $f=\KAN_2(\bfx)= \sum_{i=1}^{m_2} \phi_{2,1,i}\big(\sum_{j=1}^{m_1} \phi_{1,i,j}(x_j)\big)$.
Our goal is to determine the Lipschitz constants of each layer given the Lipschitz constant of $f$. 
Assuming $\calL^{(k)}_{\phi_{2,1,i}} = \calL^{(k)}_{\phi_2}$ and $\calL^{(k)}_{\phi_{1,i,j}} = \calL^{(k)}_{\phi_1}$ for all $i$ and $j$, the first-order Lipschitz constant can be written as
\begin{align}
    \calL_f^{(1)} = m_2 m_1 {\calL_{\phi_2}^{(1)}}{\calL_{\phi_1}^{(1)}}, 
    \label{eq:Lf1_kan2}
\end{align}
and, for cubic splines, the $4$th-order Lipschitz constant as
\begin{align}
    \calL_f^{(4)} = m_2 m_1^4\calL_{\phi_2}^{(4)} \calL_{\phi_1}^{(1)} (\calL_{\phi_1}^{(2)})^3 &+ 3 m_2 m_1^3 \calL_{\phi_2}^{(3)} \calL_{\phi_1}^{(1)} \calL_{\phi_1}^{(2)} \calL_{\phi_1}^{(3)} \notag \\
    &+ m_2 m_1^2 \calL_{\phi_2}^{(2)} \calL_{\phi_1}^{(1)} \calL_{\phi_1}^{(4)},
    \label{eq:Lfk_kan2}
\end{align}
which reflects how the Lipschitz bounds are divided across layers.
We know $\calL^{(1)}_f$ and $\calL^{(4)}_f$, then we need to satisfy constraints~\eqref{eq:Lf1_kan2} and~\eqref{eq:Lfk_kan2} while maximizing the approximation error to account for the worst-case Lipschitz distribution among the layers.
We ignore the contribution from error at knots because Lipschitz constants only affect interpolation error.
The error is formulated as $u_f(\bfx^*) = \sum_{i=1}^{m_2}(\bar{u}_{\phi_{2,1,i}}(\bfy_1^*,\Xi_1) + \calL_{\phi_2}^{(1)} \sum_{j=1}^{m_1}  \bar{u}_{\phi_{1,i,j}}(\bfx^*,\calT)) $, where $\bfx^*$ is the test point that function approximation makes the most error. 
For ease of notation, consider Equation~\eqref{eq:int-error} as $\bar{u}_f(\bfx,\bftau) = \calL_f^{(k+1)} \psi(\bfx, \bftau)$, where $\psi(\bfx, \bftau) =  \prod_{i=1}^{k+1} |x-\tau_i^{(j)} |/ (k+1)!$. 
Furthermore, 
since worst-case test point is unknown, we consider the expectation of error as, $\bar{\psi}(\bftau) = \bbE_{\bfx \in \calX} [\psi(\bfx, \bftau)] \approx \sum_{\bfx \in \calX_\calD} \psi(\bfx, \bftau)$ .
We then formulate the following optimization problem:
\begin{align}
    \max_{\mathbf{\calL}} &\quad \calL_{\phi_2}^{(k+1)} \sum_{i=1}^{m_2}(\bar{\psi}(\Xi_1)) + \calL_{\phi_2}^{(1)}\calL_{\phi_1}^{(k+1)} \sum_{j=1}^{m_1}  \bar{\psi}(\bftau)) \notag \\
    s.t.  &\quad \calL_f^{(4)} = m_2 m_1^4\calL_{\phi_2}^{(4)} \calL_{\phi_1}^{(1)} (\calL_{\phi_1}^{(2)})^3 \notag\\
    &\quad \quad + 3 m_2 m_1^3 \calL_{\phi_2}^{(3)} \calL_{\phi_1}^{(1)} \calL_{\phi_1}^{(2)} \calL_{\phi_1}^{(3)} \notag 
     + m_2 m_1^2 \calL_{\phi_2}^{(2)} \calL_{\phi_1}^{(1)} \calL_{\phi_1}^{(4)} \notag\\
    &\quad \calL_f^{(1)} = m_2 m_1 {\calL_{\phi_2}^{(1)}}{\calL_{\phi_1}^{(1)}} \notag\\
    &\quad 0 \le \calL^{(k)}_{\phi_2},\calL^{(k)}_{\phi_1} \le \calL^{(k)}_{f} \quad \forall k \in \{ 1, 4\}.
\end{align}
This is a non-convex optimization problem that can be solved using the Lagrangian method~\cite{boyd2004convex}, but it has a high computational cost.

\subsubsection{Logarithmic Division} 
To avoid solving the computationally costly non-convex optimization, we propose a simpler approach.
Since the Lipschitz constant accumulates by product over the layers, we can linearize it in the log domain, 
$\log(\calL_f) = \sum_{l=1}^{L} \log(m_l \calL_l)$.
With this, we can distribute the total Lipschitz proportional to the sampled Lipschitz constant of the approximated function as:
\begin{align}
    \calL_{l} &= \frac{1}{m_l}\exp\left( \frac{\log(m_l\hat{\calL}_l)}{\sum_{i=1}^{L}\log(m_i\hat{\calL}_i)}\log(\calL_f)\right).
\end{align}
This approach factorizes the network-level Lipschitz constant into layers in proportion to the learned function's layer-wise Lipschitz constant.

\subsubsection{Linear Division}
An even simpler alternative is to distribute the Lipschitz budget linearly based on empirical estimates:
$w_l = \hat{\calL}_l / (\sum_{i=1}^{L}m_i\hat{\calL}_i)$, where the portion assigned to each neuron in layer $l$ is computed as 
$\calL_l = w_l \calL_f$.

\subsubsection{Equal Division}
The simplest approach assumes uniform distribution across all nodes, inspired by~\cite{liu2020simple}.
Let assume the Lipschitz constant of each node is $\calL_h$, then we have,  $\calL_f = \prod_{l=1}^{L} m_l \calL_h = ({\calL_h})^L \prod_{l=1}^{L} m_l$.
Solving for the node's Lipschitz constant, we obtain, $\calL_h = \sqrt[L]{{\calL_f}/{\prod_{l=1}^{L} m_l}}$.

\subsubsection{Worst-Case Division}
Another simplest, yet conservative, approach is to assign the full Lipschitz budget to each component $\calL_l = \calL_f$.
This guarantees a valid bound, but it is overly conservative and produces loose uncertainty estimates. 
 
\textbf{Case Studies:}
\label{lbl:test_lipdivision}
To evaluate the effectiveness of the Lipschitz division methods, we tested them on a set of benchmark functions, including $f_1(x)=\cos(x)$, $f_2(x,y)=\exp(\sin(\pi x)+y^2)$, a discontinuous function,
\begin{align}
    h(t) &=  
    \begin{cases}
        \frac{5}{4} \sum\limits_{k=1}^{4} \sin(k t), & \text{for } t < 0 \\[8pt]
        5 \cos(5 t), & \text{for } t \geq 0
    \end{cases} \notag\\[8pt]
    f_3(x,y) &= h(x) h(y)  
    \label{eq:discont_eq}
\end{align}
and $f_4(x,y)$: the Two Moon dataset~\cite{scikit-learn}, defined as:
\begin{align}
    &\begin{aligned}
        x_1 &= r \cos(\theta) + \epsilon_x,  & y_1 &= r \sin(\theta) + \epsilon_y, \\
        x_2 &= r - r \cos(\theta) + \epsilon_x,  & y_2 &= -r \sin(\theta) + \epsilon_y, 
    \end{aligned} \notag\\
    &f_4(x,y) =  
    \begin{cases}
        1, & \quad \text{if } (x,y) = (x_1, y_1) \\
        -1, & \quad \text{if } (x,y) = (x_2,y_2)
    \end{cases},
\end{align}
where \( \theta \) is sampled from \( [0, \pi] \), \( r \) is the radius, and noise terms \( \epsilon_x, \epsilon_y \sim 
\mathcal{N}(0, \sigma^2) \).

\begin{figure}
    \begin{overpic}[width=0.24\textwidth,trim=0pt 0pt 0pt 0pt, clip]{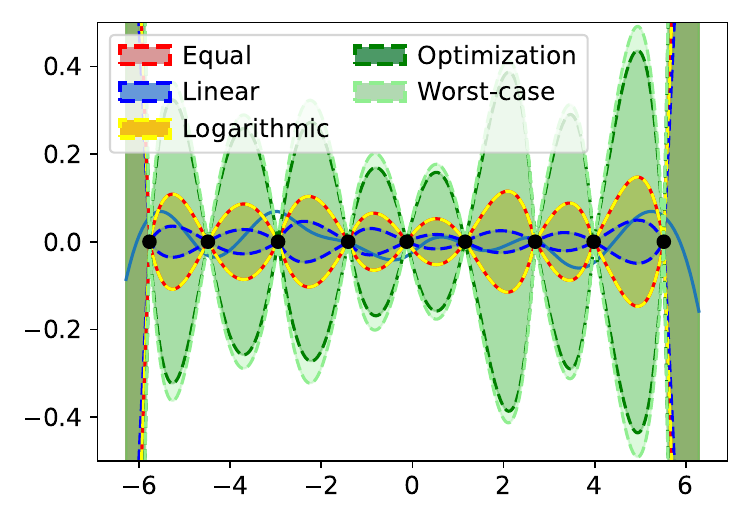}
        \put(-1,63){\textbf{(a)}} 
    \end{overpic}%
    \begin{overpic}[width=0.24\textwidth,trim=0pt 0pt 0pt 0pt, clip]{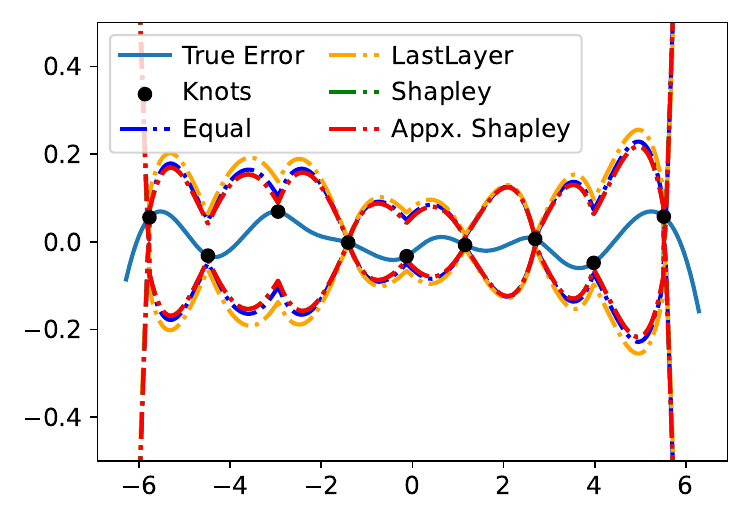}
        \put(-1,63){\textbf{(b)}} 
    \end{overpic}
    \caption{
    \textbf{a)} The comparison of error bounds of different Lipschitz division methods. All the algorithms can recover the true error.
    \textbf{b)} Error bound using different error division approaches on $\cos$ function.
    }
    \label{fig:Lipschitz-division-and-Error-division}
\end{figure}

We trained two-layer KANs on collected data on these functions, assumed zero error at knots to isolate the effect of Lipschitz division, used EBL~\eqref{eq:error-at-knots-ebl} for spline error, and computed the number of violations of the interpolation error bound to assess each method.
The results in Table~\ref{tbl:Lipschitz-division-summary} indicate that the \textbf{non-optimized worst-case} method achieves the fewest violations, the lowest computational complexity, and the most relaxed error bound among the compared approaches. 
The \textbf{worst-case} method yields the second-lowest violation rate but requires costly optimization.
The \textbf{equal division} method ranks third in terms of violation rate, yet it remains a reasonable choice due to its simplicity and minimal computation overhead, making it our \textbf{preferred choice}. 
Fig.~\ref{fig:Lipschitz-division-and-Error-division}~{(a)} visualizes 
that all the approaches successfully recover the true error bounds on the cosine function,  with both the non-optimized worst-case and worst-case error bounds appearing overly conservative. In contrast, the linear and logarithmic heuristic methods provide a tighter error bound.
Please check Appendix~\ref{sec:appdx_Lipschitz_division_results} for more information on network architecture, calculated Lipschitz, and additional details. 

\begin{table}[!ht]
    \centering
    \caption{Comparison of Lipschitz division methods on estimating different functions.}
    \label{tbl:Lipschitz-division-summary}
    \rowcolors{4}{lightgray}{white}  
    \begin{tabular}{lcccc} 
        \hline
        \rowcolor{headerblue}     
        & \multicolumn{4}{c}{\textbf{Violation rate}}\\ 
        \hhline{>{\arrayrulecolor{headerblue}}->{\arrayrulecolor{black}}----}
        \rowcolor{headerblue}     
        \textbf{Methods} & {$f_1$} & {$f_2$} & {$f_3$} & {$f_4$}\\              
        \hline
        Equal             & 0.171 & 0.132 & 0.472 & 0.150 \\
        Linear           & 0.487 & 0.057 & 0.733 & 0.777 \\
        Logarithmic      & 0.171 & 0.176 & 0.558 & 0.206 \\
        Worst-case   & 0.051 & 0.004 & 0.074 & 0.014 \\
        Optimization                 & 0.058 & 0.005 & 0.082 & 0.014 \\
        \hline
    \end{tabular}
\end{table}

\subsection{Division of Error at Knots}
\label{sec:Error-division}
Since only the total approximation error $e^f(\tau_i) \teq f(\tau_i) - \hat{f}(\tau_i)$ is observable and available at knots, rather than the layer-wise errors required for Theorem~\ref{thm:two-layer-poly-bound}, we divide the total error across layers proportionally to their contributions.
For an $L$-layer network similar to Equation~\eqref{eq:two-layer-thrm}, this obtains:
\begin{align}
    e^f(\bftau) \le e^{\bfh_L}(\hat{\bfy}_{L-1})  + \sum_{l=1}^{L-1} \big( \hat{e}^{\bfh_l}(\hat{\bfy}_{l-1}) \prod_{j=l+1}^{L} \calL^1_{h_j}  \big),
    \label{eq:multi-layer-poly-bound2}
\end{align}
where $\hat{e}^{\bfh_l}(\hat{\bfy}_{l-1}) = \sum_{j=1}^{m_l}e^{\bfh_{l,j}}(\hat{\bfy}_{l-1,j})$ represents the error contribution from layer $l$, and $\hat{\bfy}_{l-1,j}$ is the input to the $j$th node of layer $l$.
To evaluate interpolation error $\bar{u}_h$ (Theorem~\ref{thm:poly-interp-bound}), only the spline's knot locations are needed; however, the error at knots $\calP[e_j^h]$ requires the true function values at inner layers, which are unavailable.
To address this, we propose three alternative approaches for estimating the inner-layer error: 1- assigning all error to the last layer, 2- equally dividing the error across knots, and 3- distributing the error based on Shapley values. Once the error at the knots is estimated, Theorem~\ref{thm:two-layer-poly-bound} can be applied to obtain the error bound for the full approximation. 
Finally, we compare these approaches using 2D examples.

\subsubsection{Last Layer Error}

Our first approach depends on the following assumption:

\begin{assumption}[No early information loss (NEIL)]
\label{assum:NEIL}
In approximating a function $f(x)$ using function compositions, for instance, $\hat{f}(x) = \hat{h}(\hat{\bfg}(x))$, the early layers $\hat{\bfg}$ do not throw away vital information of $f$.
This means that for any two distinguishable inputs $x_1 \neq x_2$, which map into $f(x_1) \neq f(x_2)$, all the early layer approximations should be distinguishable $\hat{\bfg}(x_1) \neq \hat{\bfg}(x_2)$:
\begin{align}
    \text{if } f(x_1) \neq f(x_2) \implies  \hat{\bfg}(x_1) \neq \hat{\bfg}(x_2) \quad \forall x_1 \neq x_2.
    \label{eq:neil}
\end{align}
This assumption is equivalent to the invertibility of $\hat{\bfg}$, for all $x_1, x_2 \in \calX$ for which $f(x_1) \ne f(x_2)$.
\end{assumption}

If Assumption~\ref{assum:NEIL} is satisfied, that is there is no loss of information in the earlier layers, then all the error can be compensated in the last layer.
Moreover this assumption can be enforced by using bi-Lipschitz continuity techniques as used in~\cite{liu2020simple,van2021feature}.
So our first method for dividing error at knots assumes zero error in all the layers except the last layer.
Then the last layer error at knots is same as the total error at knots.

\subsubsection{Equally Dividing Error at Knot}
We divide this error across neurons in proportion to their Lipschitz constants. 
The error assigned to neuron $\phi_{l,j,i}$ in layer $l$ with $m_l$ nodes is
\begin{align}
    e^{\phi_{l,j,i}} = \frac{e^f }{C_e}, \quad C_e = \sum_{l=1}^{L-1} m_l \prod_{l'=l+1}^{L}\calL_{h_{l'}}.
\end{align}
Note that, although $e^f$ is defined on the entire domain, we only divide it at the knots, $x \in \bftau_{1:m}$.

\subsubsection{Shapley Value}
While the aforementioned methods are effective, they ignore the interactions between neurons. 
To address this, we adapt \textit{Shapley value} from cooperative game theory, which fairly distributes the total payoff among players based on each player's contribution to all possible coalitions~\cite{shapley1953value}. 
The theory incorporates different parts of \ac{ML} to assess feature, sample, or element importance. This solution is an adaptation of~\cite{ghorbani2020neuron} to find the error division of each neuron. 
Let $\Phi$ be the set of all neurons in the network, and let $\Phi_s \subseteq \Phi$.
Define the operator $\calM(\Phi_s)$ as the network in which neurons in $\Phi \setminus \Phi_s$ have been ``zeroed out''. Zeroing out a neuron means removing the neuron while keeping the residual connection $r$.
The Shapley value $\varphi_{l,j,i}$ for neuron $\phi_{l,j,i}$ is then computed as:
{\small
\begin{align}    
    \varphi_{l,j,i} &= \sum_{\Phi_s \subseteq \Phi \setminus \{\phi_{l,j,i}\}} \frac{|\Phi_s|! \, (|\Phi|-|\Phi_s|-1|)!}{|\Phi|!} \, \Delta_{l,j,i}(\Phi_s) \label{eq:error_shapley} \\
    \Delta_{l,j,i}&(\Phi_s) = \calV(\calM(\Phi_s \cup \{\phi_{l,j,i}\})) - \calV(\calM(\Phi_s)) \nonumber,
\end{align}}%
which $|.|$ is cardinality operator, $\Delta_i(\Phi_s)$ is marginal contribution. The value function $\calV(\calM)$ measures the error made by the network $\calM$, defined as: $\calV(\calM) = \frac{1}{|\calX|} \sum_{x,y \in \calX} |y - \calM(x)|$.
Please check Appendix~\ref{sec:shapley_example} for a more detailed example of how the Shapley value works.

Although Shapley approach leads to a fair allocation, computing it exactly is intractable for the large number of permutations when the number of players (neurons) exceeds 30~\cite{ghorbani2020neuron}.
To address this, we use an \ac{MC} sampling-based algorithm that estimates Shapley values by averaging marginal error changes across randomly sampled permutations of neurons.

\textbf{Case Studies to evaluate error division:}
We evaluate the proposed error division methods on the same functions defined in Section~\eqref{lbl:test_lipdivision}.
The network structure is identical to that used in the Lipschitz division experiment, with \textit{equal Lipschitz division} applied. Table~\ref{tbl:Error-division-summary} summarizes the number of violations for each method across experiments. 

\begin{table}[!ht]
    \centering
    \caption{Comparison of error bounds of different error division methods.}
    \label{tbl:Error-division-summary}
    \rowcolors{4}{lightgray}{white}  
    \begin{tabular}{lcccc} 
        \hline
        \rowcolor{headerblue}     
        
        & \multicolumn{4}{c}{\textbf{Violations' rate}}\\ 
        \hhline{>{\arrayrulecolor{headerblue}}->{\arrayrulecolor{black}}----}
        \rowcolor{headerblue}     
        \textbf{Methods} & {$f_1$} & {$f_2$} & {$f_3$} & {$f_4$}\\              
        \hline

        Equal division     & 0.007 & 0.013 & 0.002 & 0.037 \\
        Last layer error & 0.005 & 0.001 & 0.0   & 0.002 \\
        Shapley value   & 0.007 & 0.000 & N/A   & 0.005 \\
        Appx. Shapley & 0.007 & 0.001 & 0.0   & 0.005 \\
        
        \hline
    \end{tabular}
\end{table}

As reported in Table~\ref{tbl:Error-division-summary}, the violation rate of all methods decreased significantly compared to Table~\ref{tbl:Lipschitz-division-summary}, where error division was assumed to be zero. 
For the cosine function, all the error division approaches eliminate violations, as illustrated in Fig.~\ref{fig:Lipschitz-division-and-Error-division}~{(b)}. 
The Shapley method for experiment $f_3$ was infeasible since $|\Phi|=30$ which results in $2^{30}$ subsets. Even with an optimized model evaluation that takes~$1$ms, the full computation would require weeks on a single machine.

\section{CASE STUDIES}
\label{sec:experiments}
In this section, we evaluate the proposed method, \ac{DAREK}, through several experiments. 
We compared \ac{DAREK}'s error bounds with probabilistic uncertainty bounds, such as \acp{GP} and Ensemble, and evaluated the computational complexity of \ac{DAREK} with \acp{GP} and Ensemble.  
We also evaluate \ac{DAREK} for estimating an object's shape from a single laser scan, its distance-awareness in a high-dimensional face-recognition experiment, and its performance in a multi-agent navigation experiment.
All experiments use cubic splines ($k=3$) with first and $k$th derivative Lipschitz constants set to one, unless otherwise reported. 

\begin{figure}
\begin{overpic}[width=0.25\textwidth,trim=12pt 15pt 0pt 12pt, clip]{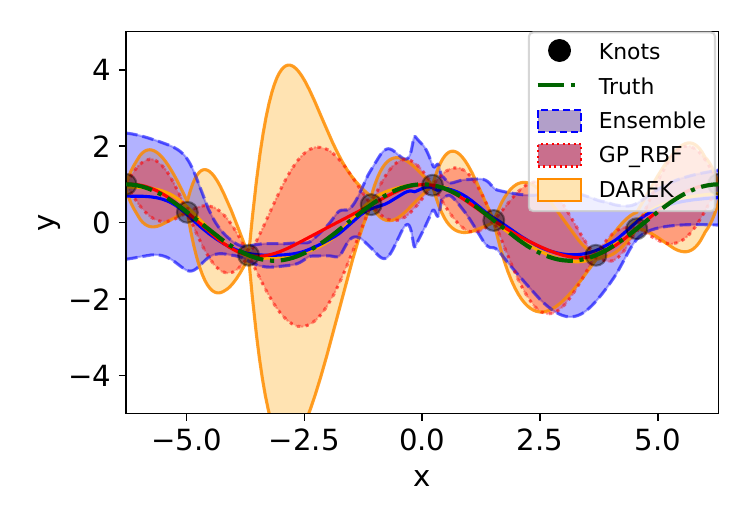}
        \put(-1,60){\textbf{(a)}} 
\end{overpic}%
\begin{overpic}[width=0.25\textwidth,trim=12pt 15pt 0pt 12pt, clip]{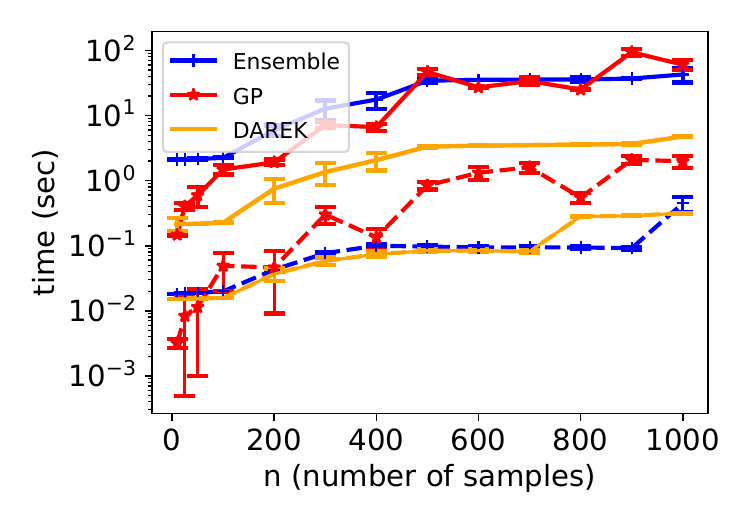}
        \put(-1,60){\textbf{(b)}} 
\end{overpic}
\vspace{-3mm}
\caption{\textbf{(a)} The comparison of error bounds of \ac{DAREK} on cosine function with Ensemble and \ac{GP}. Ensemble and \ac{GP}'s uncertainty bounds are shown within the $\pm3\sigma$ range.
\textbf{(b)} Computation complexity comparison of {\color{orange}\ac{DAREK} (orange)}, {\color{red}\ac{GP} (red)}, and {\color{blue}Ensemble (blue)} models. Training and inference times are plotted with solid and dashed lines, respectively.
    \label{fig:gantt_chart}}
\vspace{-2mm}
\label{fig:Err-comparison}
\end{figure}

\textbf{Comparison with Baselines:}
To the best of our knowledge, this is the first work to introduce distance-aware worst-case error bounds under higher-order Lipschitz smoothness assumptions.
In any case, Fig.~\ref{fig:Err-comparison} presents a comparison between the proposed two-layer \ac{DAREK}, Ensemble, and \ac{GP} in terms of both uncertainty estimation and computational complexity.
The Ensemble method employs an \ac{MC} approach with ten \ac{KAN} models initialized independently.
For the \ac{GP}, we use a \ac{RBF} kernel with fixed hyperparameters: unit length scale and unit variance. As shown in image~{(a)}, the uncertainty estimated by the Ensemble does not follow a consistent or interpretable pattern, whereas the \ac{GP} inherently captures the distance awareness through its interpolation mechanism. The comparison with \ac{GP} is \textit{unfair} without explicitly relating Lipschitz constants to \ac{RBF} kernel parameters, which is left for future work. 
Fig.~\ref{fig:Err-comparison}~{(b)} presents the computational performance of the three models for different numbers of training set sizes, demonstrating trends consistent with the theoretical complexity discussed earlier in Section~\ref{sec:overview}. 

\begin{figure}
\includegraphics[width=0.48\textwidth,trim=0pt 0pt 0pt 0pt, clip]{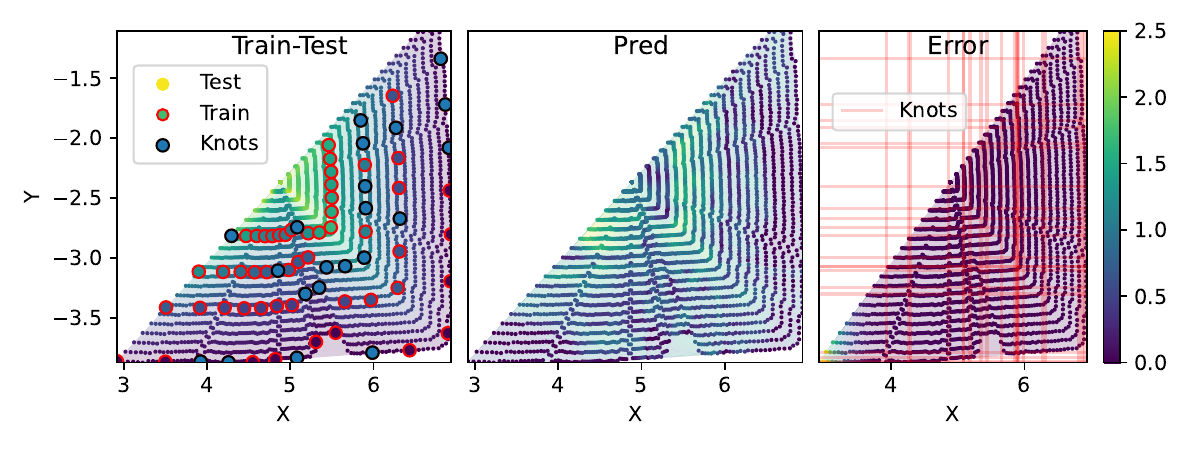}
\includegraphics[width=0.48\textwidth,trim=0pt 0pt 0pt 0pt, clip]{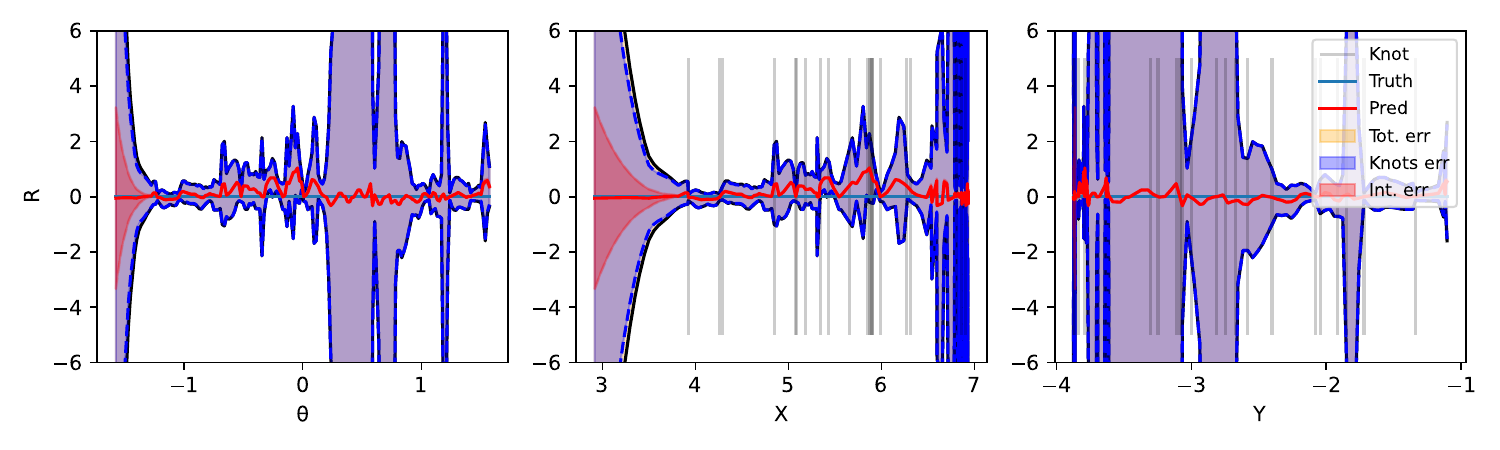}
\caption{In the first row, the plots from left to right show the train-test dataset, the trained model's predictions, and the interpolation error of each test point. The second row shows the predicted distance from the object boundary ($R$) versus the laser scanner angle ($\theta$), the $x$ location, and the $y$ location of the boundary point.}
\label{fig:Laser-Scan}
\end{figure}

\textbf{Object Shape Estimation:}
To evaluate the generalization of \ac{DAREK} to real-world mapping problem, we consider the object shape estimation from a laser scan in 2D. 
We train a two-layer KAN with units [2,20,1] and 20 knots to learn the Signed Distance Function (SDF) profile of an object, using laser scan data from~\cite{howard2003robotics}. 
The goal is to estimate the object's shape from scan measurements. 
The top-left plot of Fig.~\ref{fig:Laser-Scan} shows the training and test samples with selected knot locations. The top-middle plot presents the model's predictions at test points, while the top-right plot depicts the corresponding prediction errors. Red lines indicate the positions of knots in the input layer. The bottom row of plots shows the prediction error near the object's boundary as a function of the laser scan angle $\theta$, and the Cartesian coordinates $x$ and $y$, respectively. 

\begin{figure}
    \centering
    \includegraphics[width=0.7\linewidth]{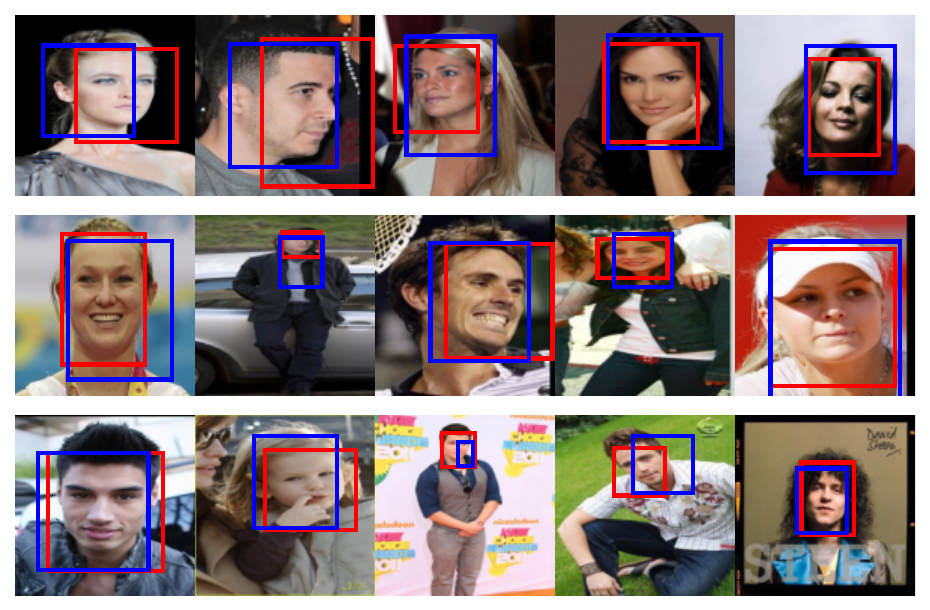}
    \caption{Face detection on sample test images from CelebA dataset~\cite{liu2015deep} using \ac{DAREK}. {\color{red}Red boxes} indicate ground-truth bounding boxes, and {\color{blue}blue boxes} indicate \ac{DAREK} predictions. Note that it uses only 100 projected dimensions out of a 25k-dimensional feature space of ResNet18. }
    \label{fig:face_recognition}
\end{figure}

\textbf{Face detection:}
\label{sec:face-dataset}
To demonstrate the effectiveness of \ac{DAREK} on high dimension inputs, we 
evaluate \ac{DAREK} on the Celebrity Attribute face detection dataset~\cite{liu2015deep}. 
The task is to predict the bounding box of the face in the given image.
To reduce the dataset size, we randomly selected 8k images for training and 1k images for testing.
We reduced feature dimensionality to 2, 20, and 100 using two feature extraction strategies: PCA-based projection and deep Res-Net18 feature projection. 
These reduced dimensionality inputs were used to train uncertainty-aware methods to predict the location of the bounding box along with corresponding uncertainty. 
We compared the following methods: one-layer \ac{DAREK} (DK1), two-layer \ac{DAREK} with 10 hidden units (DK2), an ensemble of one-layer KANs (ENS1), an ensemble of two-layer KANs with 10 hidden units (ENS2), an exact GP (ExGP) that uses the same knots as the \ac{DAREK} as its training data, and a variational sparse GP (ApxGP)~\cite{hensman2015scalable} that uses \ac{DAREK} knots as inducing points and learns a variational posterior.
Networks were trained for 1000 iterations with a learning rate of 0.1 and \acp{GP} for 2000 iterations with a learning rate of 0.5.
All models used a decay factor of 0.9 every 200 iterations.
Performance was evaluated using normalized root mean squared error (RMSE), intersection over union (IOU), and SDA~\eqref{eq:sampled-distance-awareness}.
The results are reported in Table~\ref{tab:facerecognition_main}.
Additional results on dimensional sizes (5, 10, 50, 200) are shown in Appendix~\ref{sec:appendx-additional-experiments-face}. 
RMSE and IOU remain consistent across models and features, with ResNet-18 features yielding lower RMSE and higher IOU.
\Ac{DAREK}'s SDA~\eqref{eq:sampled-distance-awareness} increases from approximately 60\% at two dimensions to around 95\% at 100 dimensions, outperforming ENS1 and ENS2, which do not follow a consistent pattern. 
\Ac{GP} achieves near-perfect SDA due to its kernel-based formulation. 
Example predictions are illustrated in Fig.~\ref{fig:face_recognition}. 

\begin{table}[t]
\centering
\caption{Results for high-dimensional experiments across two feature extraction methods (PCA and ResNet-18). 
}
\label{tab:facerecognition_main}
\setlength{\tabcolsep}{4pt}
\renewcommand{\arraystretch}{1.05}

\resizebox{\linewidth}{!}{
\rowcolors{4}{lightgray}{white}  
\begin{tabular}{l | ccc | ccc | ccc}
\hline
\rowcolor{headerblue}
& \multicolumn{3}{c|}{RMSE $\downarrow$} & \multicolumn{3}{c|}{IOU $\uparrow$} & \multicolumn{3}{c}{SDA $\uparrow$} \\
\hhline{>{\arrayrulecolor{headerblue}}->{\arrayrulecolor{black}}---------}
\rowcolor{headerblue}
Model & 2      & 20    & 100   &     2 &    20 &   100 &      2 &    20 &   100  \\
\hline
\multicolumn{10}{l}{\textbf{(a)} PCA
}\\
DK1   & 0.136  & 0.133 & \textbf{0.129} & \textbf{0.451}  & 0.452 & \textbf{0.466} & 0.676 & 0.967 & 0.952 \\
DK2   & 0.137  & 0.190 & 0.184 & 0.448  & 0.375 & 0.339 & 0.677 & 0.972 & 0.950 \\
ENS1  & 0.136  & 0.133 & 0.130 & 0.450  & 0.451 & 0.464 & 0.571 & 0.563 & 0.431 \\
ENS2  & \textbf{0.135}  & \textbf{0.128} & 0.265 & 0.449  & \textbf{0.469} & 0.145 & 0.527 & 0.530 & 0.564 \\
ExGP  & 0.138  & 0.161 & 0.137 & 0.435  & 0.359 & 0.448 & 1.000 & 1.000 & 1.000 \\
ApxGP & 0.138  & 0.137 & 0.137 & 0.427  & 0.447 & 0.447 & 0.977 & 1.000 & 1.000 \\
\midrule
\end{tabular}}

\rowcolors{2}{white}{lightgray}  
\resizebox{\linewidth}{!}{ 
\begin{tabular}{l | ccc | ccc | ccc} 
\multicolumn{10}{l}{\textbf{(b)} ResNet-18 }\\

DK1    & 0.100 & 0.088 & 0.074  & 0.535 & 0.557 & 0.606  & 0.704 & 0.892 & 0.837 \\
DK2    & 0.100 & 0.103 & 0.073  & 0.533 & 0.470 & 0.608  & 0.691 & 0.918 & 0.882 \\
ENS1   & 0.100 & 0.089 & 0.075  & 0.535 & 0.557 & 0.603  & 0.566 & 0.577 & 0.468 \\ 
ENS2   & 0.100 & 0.084 & 0.069  & 0.530 & 0.575 & 0.628  & 0.515 & 0.562 & 0.550 \\
ExGP   & 0.139 & 0.140 & 0.137  & 0.431 & 0.427 & 0.447  & 0.542 & 1.000 & 1.000  \\
ApxGP  & 0.133 & 0.137 & 0.137  & 0.445 & 0.447 & 0.447  & 0.729 & 1.000 & 1.000 \\
\bottomrule
\end{tabular}}
\end{table}

\textbf{Safe Multi-Agent Experiment:} 
We set up a multi-agent environment based on Cheng et al.~\cite{cheng2020safe} to evaluate \ac{DAREK} in a safe control application. 
Each agent follows double-integrator dynamics in 2D space, with state $\bfx = (p_t,v_t)$ containing positions and velocities, of all agents in $x$ and $y$ directions, and control input $\bfa$ contains the acceleration in $x$ and $y$ directions of all agents.
The system dynamics are $\bfx_{t+1} = \bff(\bfx_t, \bfa_t) + \bfd(\bfx_t)$, where $\bfd$ is disturbance. Each agent has an initial state, a goal state, and a decentralized controller. 
The blue agent uses a robust controller (Equation~\eqref{eq:robust-controller}), while the red agents use a primal controller without the polytope constraint~\eqref{eq:polytop}.
The control framework first solves a \ac{MPC} to generate the optimal trajectory, $\bfx_{\text{des}}$, and then solves \ac{CBF} to compute the safe control inputs that keep the robot close to the optimal trajectory.
Assuming bounded disturbances, a robust \ac{CBF} can be formulated to jointly optimize safety and optimality. 
The nonlinear uncertainty bound at a specified confidence level can be approximated by the polytope:
\begin{align}
    \{\bfd \in \mathbf{R}^n | G \bfd \le \bfg\},
    \label{eq:polytop}
\end{align}
where $\bfd$ is disturbance within the polytope, $G$ defines its orientation, and $\bfg$ sets the bounds along those directions.
\begin{align}
    \min_{\bfa \in \calA, \bfd \in D} & \| \bff(\bfx_t,\bfa) - \bfx_{\text{des}} \| \notag\\
    \text{s.t.} \quad & \forall \bfd \in \{\bfd \in \bbR^n | G \bfd \le \bfg \} \notag \\
    & H_1(\bfx_t) \bfd + \bfa^\top H_2 (\bfx_t) \bfd + H_3(\bfx_t) \bfa \le \bfk_c (\bfx_t) \notag \\
    & \| \bfa \|_2 \le \bfa_{\max} \qquad \text{(actuation limits),}
    \label{eq:robust-controller}
\end{align}
where $H_1$, $H_2$, and $H_3$ are projecting the effect of disturbance, control-disturbance coupling, and control input in safety, and $k_c$ defines the safety bounds. 
These functions are defined in~\cite{cheng2020safe}.
We used a pretrained \ac{DAREK} model with architecture $[2,5,4]$ and $100$ knots for cubic spline, trained offline in a low-noise environment (one percent noise on acceleration), then fine-tuned online.
During execution, the model buffers $(\Delta \bfx, \bfd)$ pairs and retrains every 10 steps for 10 epochs. 
To ensure distance-awareness, we use the error bound with linear fit (EBL)~\eqref{eq:error-at-knots-ebl} for error at knots in this experiment.
\ac{DAREK} produces uncorrelated error bounds, which are linearly independent, and the polytope is directly obtained: ${-\bfu_f \le \bfd \le \bfu_f}$. 

The system dynamics is defined as follows:
\begin{align}
    \bfx_{t+1} &= 
    \begin{bmatrix}
        p_{t} + v_t dt\\
        v_{t} - k_v |v_t|^2
    \end{bmatrix}
    +
    \begin{bmatrix}
        0 \\
        k_d (|v_t| + 1) dt
    \end{bmatrix} a
    +    \notag \\
    &\quad \underbrace{
    \begin{bmatrix}
        0 \\
        k_d (|v_t| + 1) dt
    \end{bmatrix} \bfd_a
    +
    \begin{bmatrix}
    \bfd_p\\
    \bfd_v
    \end{bmatrix} 
    }_{\bfd(\bfx)},
\end{align}
where $dt$ is the step time, $k_v$ is the drag coefficient, and $k_d$ is the acceleration scaling factor. The terms $\bfd_p$, $\bfd_v$, and $\bfd_a$ represent disturbances in position, velocity, and acceleration, respectively.  
Uncertainty model learns the disturbance function $\bfd(\bfx)$ and Lipschitz constants $L_{\partial p_{+1}/\partial v} = 0$ and 
$L_{\partial v_{+1}/\partial v} = k_d \bfd_{a,\text{max}} dt$.
These Lipschitz constants cannot capture $\bfd_p$ and $\bfd_v$, which have to be captured with error
at knots. 

For higher order Lipschitz, assuming $\calL^{(k)}_f$ is the $k$th order Lipschitz and bounds $k$th derivative of $f$, Assumption~\ref{assm:k-Lips}.
Now, the $(k+1)$th Lipschitz using the trivial error bound is:
\begin{align}
    \calL_f^{(k+1)} \le 2 \calL_f^{(k)} |\Delta \bfx|.
    \label{eq:highorder-trivial-bound}
\end{align}

\begin{figure}
    \centering
    \begin{overpic}[width=0.98\linewidth,trim = 20 15px 20 0 ,clip]{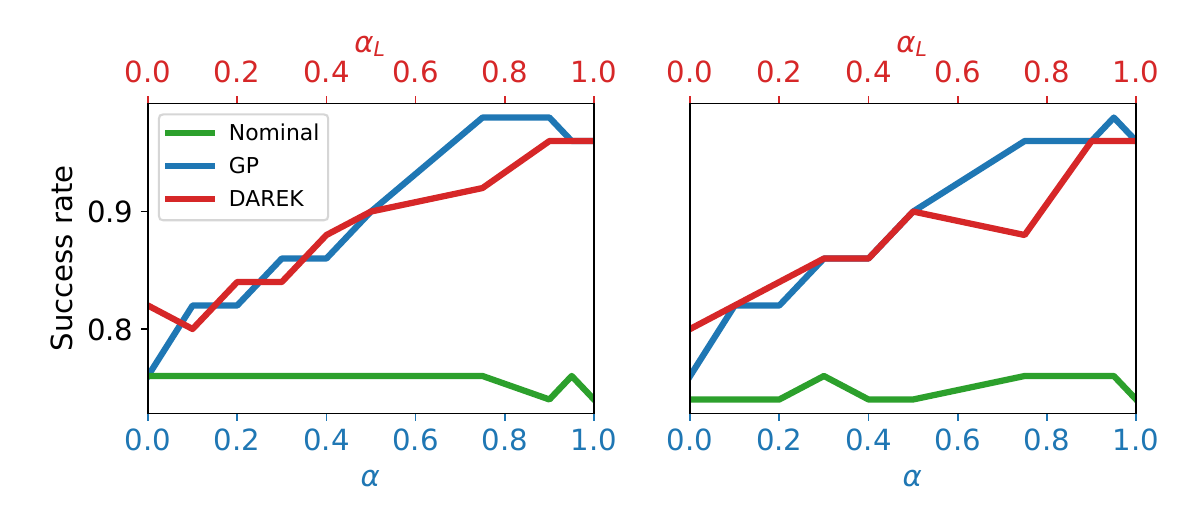}
    \put(3,33){\textbf{(a)}}
    \put(51,33){\textbf{(b)}}
    \end{overpic}
    \caption{Success rate in multi-agent simulations under different safety levels. The GP controller uses $\alpha$ to represent the percentage of the uncertainty coverage, while \ac{DAREK} uses $\alpha_L$, denoting the percentage of the true Lipschitz constant applied during simulation. \textbf{(a)} Uniform noise, \textbf{(b)} Gaussian noise.}
    \label{fig:Success-rate}
\end{figure}

\begin{figure}
    \centering
    \includegraphics[width=0.98\linewidth]{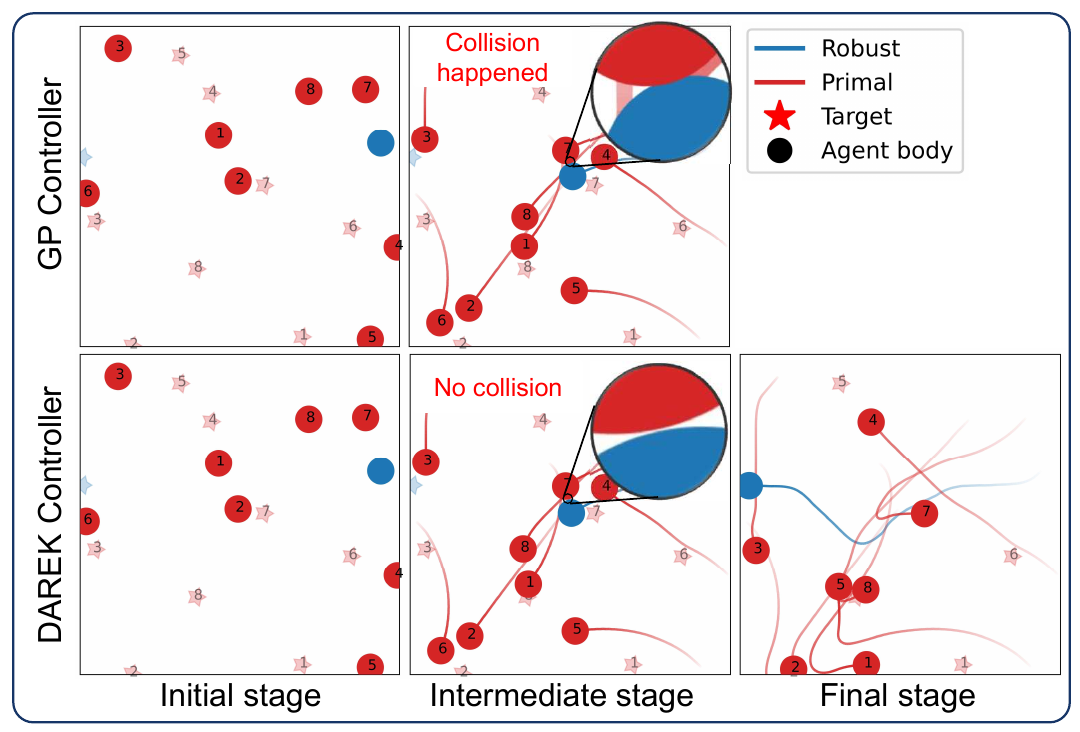}
    \caption{Illustration of a multi-agent simulation. The plot compares a test scenario where the \ac{GP} controller failed to complete the trajectory, whereas the \ac{DAREK} controller successfully reached the goal. It shows the initial positions, the intermediate stage where the \ac{GP} controller collided and the \ac{DAREK} controller avoided it, and the final stage where the \ac{DAREK} controller reached the goal. Agent trajectories are shown as lines, their bodies at the current state as circles, and goals as stars. 
    }
    \label{fig:Multiagent-simulation}
\end{figure}

All simulations are conducted with a fixed noise level of 10\% and are repeated over 50 randomized environments. A trial is successful if the robust agent reaches its goal without collisions. Fig.~\ref{fig:Success-rate} depicts the robust controllers' plots, \ac{GP} and \ac{DAREK}, which outperform the primal controller and reach up to 95 percent success rate for different confidence levels. 
Fig.~\ref{fig:Multiagent-simulation} illustrates a failure case for \ac{GP} due to a collision, while \ac{DAREK} successfully navigated to the goal.
The trivial bound in Equation~\eqref{eq:highorder-trivial-bound} is used on average 82 percent of the time.

\section{Conclusion}
We introduced \ac{DAREK}, distance-aware error analysis for \ac{SNN} that provides theoretically tight and interpretable worst-case error bounds. 
We evaluate the correctness and quality of these bounds on various function approximation tasks, consistently showing that the proposed bound safely overestimates the true function. 
We further validate \ac{DAREK} on an object shape estimation problem. 
Additionally, we integrate \ac{DAREK} into a multi-agent safe control framework, demonstrating its utility in a simulated multi-agent environment.
Across experiments, \ac{DAREK} was competitive with \ac{GP} while being more computationally efficient. 

\ac{DAREK} offers several \textbf{advantages}.
It combines the strengths of both parametric and non-parametric models, offering flexibility and analytical tractability.
The ability to localize errors to the nearest knots makes the model interpretable and facilitates explanation and debugging. 
Distance awareness is inherently embedded in the error formulation, making the model more conservative as test points move farther from training samples. 
Moreover, \ac{DAREK} is computationally efficient, making it suitable for real-time systems.
Additionally, \ac{DAREK} provides worst-case error guarantees, making it highly suitable for safety-critical \textbf{applications} such as robotics, autonomous control systems, and medical decision-making, where conservative behavior is essential. It also has potential utility in signal processing, risk management, and other domains that require robust, distance-aware regression.
\\ 
Despite its strengths, \ac{DAREK} has several \textbf{limitations}.
The proposed Lipschitz and error division methods are not guaranteed to match the Lipschitz and errors of the true system. 
Fixing the spline knots using a subset of the training data can limit generalization if the sample is unrepresentative, and it also requires access to training data. 
Furthermore, aggregating worst-case bounds, especially across dimensions, can grow quickly and lead to excessive conservatism. 
Additionally, poor knot placement, either too far or too close, can cause high interpolation or unwanted oscillations, known as Runge’s phenomenon~\cite{chen2014solving}.
These limitations are not fundamental to the approach and can be addressed in future research.

\section*{References}
\bibliographystyle{IEEEtran}
\bibliography{bib/main-journal}

@inproceedings{ataei2025darek,
  title={DAREK-Distance Aware Error for Kolmogorov Networks},
  author={Ataei, Masoud and Khojasteh, Mohammad Javad and Dhiman, Vikas},
  booktitle={ICASSP},
  pages={1--5},
  year={2025},
  organization0={IEEE}
}

@article{ataei2024dadee,
  title={{DADEE}: Well-calibrated uncertainty quantification in neural networks for barriers-based robot safety},
  author={Ataei, Masoud and Dhiman, Vikas},
  journal={arXiv:2407.00616},
  year={2024}
}

@article{friedman1990adaptive,
  title={Adaptive spline networks},
  author={Friedman, Jerome},
  journal={NeurIPS},
  volume={3},
  year={1990}
}

@inproceedings{uncini1997adaptive,
  title={Adaptive spline neural networks for signal processing applications},
  author={Uncini, Aurelio and Piazza, Francesco},
  booktitle={INTERNATIONAL SYMPOSIUM on INTELLIGENT SYSTEMS AMSEISIS},
  volume={97},
  pages={11--13},
  year={1997}
}

@article{gunther2021spline,
  title={Spline parameterization of neural network controls for deep learning},
  author={G{\"u}nther, Stefanie and Pazner, Will and Qi, Dongping},
  journal={arXiv:2103.00301},
  year={2021}
}

@article{hong2011modeling,
  title={Modeling of complex-valued Wiener systems using B-spline neural network},
  author={Hong, Xia and Chen, Sheng},
  journal={IEEE Transactions on Neural Networks},
  volume={22},
  number={5},
  pages={818--825},
  year={2011},
  publisher={IEEE}
}

@article{fakhoury2022exsplinet,
  title={ExSpliNet: An interpretable and expressive spline-based neural network},
  author={Fakhoury, Daniele and Fakhoury, Emanuele and Speleers, Hendrik},
  journal={Neural Networks},
  volume={152},
  pages={332--346},
  year={2022},
  publisher={Elsevier}
}

@article{keskin2018splinenets,
  title={Splinenets: Continuous neural decision graphs},
  author={Keskin, Cem and Izadi, Shahram},
  journal={NeurIPS},
  volume={31},
  year={2018}
}

@inproceedings{diamant2024conformalized,
  title={Conformalized deep splines for optimal and efficient prediction sets},
  author={Diamant, Nathaniel and Hajiramezanali, Ehsan and Biancalani, Tommaso and Scalia, Gabriele},
  booktitle={International Conference on Artificial Intelligence and Statistics},
  pages={1657--1665},
  year={2024},
  organization0={PMLR}
}

@inproceedings{balestriero2018spline,
  title={A spline theory of deep learning},
  author={Balestriero, Randall and others},
  booktitle={ICML},
  pages={374--383},
  year={2018},
  organization0={PMLR}
}

@inproceedings{campolucci1996neural,
  title={Neural networks with adaptive spline activation function},
  author={Campolucci, P and Capperelli, F and Guarnieri, S and Piazza, Francesco and Uncini, Aurelio},
  booktitle={Proceedings of 8th Mediterranean Electrotechnical Conference on Industrial Applications in Power Systems, Computer Science and Telecommunications (MELECON 96)},
  volume={3},
  pages={1442--1445},
  year={1996},
  organization0={IEEE}
}

@book{harris1993intelligent,
  title={Intelligent control: aspects of fuzzy logic and neural nets},
  author={Harris, Christopher J and Moore, Chris G and Brown, Martin},
  volume={6},
  year={1993},
  publisher={World Scientific}
}

@article{montanelli2020error,
  title={Error bounds for deep ReLU networks using the Kolmogorov--Arnold superposition theorem},
  author={Montanelli, Hadrien and Yang, Haizhao},
  journal={Neural Networks},
  volume={129},
  pages={1--6},
  year={2020},
  publisher={Elsevier}
}

@article{takacs2016approximation,
  title={Approximation error estimates and inverse inequalities for B-splines of maximum smoothness},
  author={Takacs, Stefan and Takacs, Thomas},
  journal={Mathematical Models and Methods in Applied Sciences},
  volume={26},
  number={07},
  pages={1411--1445},
  year={2016},
  publisher={World Scientific}
}

@article{sande2020explicit,
  title={Explicit error estimates for spline approximation of arbitrary smoothness in isogeometric analysis},
  author={Sande, Espen and Manni, Carla and Speleers, Hendrik},
  journal={Numerische Mathematik},
  volume={144},
  number={4},
  pages={889--929},
  year={2020},
  publisher={Springer}
}

@article{beltran2024b,
  title={B-Spline Artificial Neural Networks in Robust Induction Motor Control},
  author={Beltran-Carbajal, Francisco and Ya{\~n}ez-Badillo, Hugo and Galvan-Perez, Daniel and Rivas-Cambero, Ivan and Sotelo, David and Sotelo, Carlos},
  journal={IEEE Access},
  year={2024},
  publisher={IEEE}
}

@inproceedings{qian2024investigating,
  title={Investigating KAN-Based Physics-Informed Neural Networks for EMI/EMC Simulations},
  author={Qian, Kun and Kheir, Mohamed},
  booktitle={International Conference on Intelligent Systems, Blockchain, and Communication Technologies},
  pages={40--48},
  year={2024},
  organization0={Springer}
}

@inproceedings{kolmogorov1957representation,
  title={On the representation of continuous functions of many variables by superposition of continuous functions of one variable and addition},
  author={Kolmogorov, Andrei Nikolaevich},
  booktitle={Doklady Akademii Nauk},
  volume={114},
  number={5},
  pages={953--956},
  year={1957},
  organization0={Russian Academy of Sciences}
}

@inproceedings{hecht1987kolmogorov,
  title={Kolmogorov’s mapping neural network existence theorem},
  author={Hecht-Nielsen, Robert},
  booktitle={Proceedings of the international conference on Neural Networks},
  volume={3},
  pages={11--14},
  year={1987},
  organization0={IEEE press New York, NY, USA}
}

@article{ismailov2023three,
  title={A three layer neural network can represent any multivariate function},
  author={Ismailov, Vugar E},
  journal={Journal of Mathematical Analysis and Applications},
  volume={523},
  number={1},
  pages={127096},
  year={2023},
  publisher={Elsevier}
}

@article{bozorgasl2024wav,
  title={Wav-kan: Wavelet kolmogorov-arnold networks},
  author={Bozorgasl, Zavareh and Chen, Hao},
  journal={arXiv:2405.12832},
  year={2024}
}

@article{koenig2024kan,
  title={KAN-ODEs: Kolmogorov--Arnold network ordinary differential equations for learning dynamical systems and hidden physics},
  author={Koenig, Benjamin C and Kim, Suyong and Deng, Sili},
  journal={Computer Methods in Applied Mechanics and Engineering},
  volume={432},
  pages={117397},
  year={2024},
  publisher={Elsevier}
}

@article{bresson2024kagnns,
  title={Kagnns: Kolmogorov-arnold networks meet graph learning},
  author={Bresson, Roman and Nikolentzos, Giannis and Panagopoulos, George and Chatzianastasis, Michail and Pang, Jun and Vazirgiannis, Michalis},
  journal={arXiv:2406.18380},
  year={2024}
}

@article{abueidda2025deepokan,
  title={Deepokan: Deep operator network based on kolmogorov arnold networks for mechanics problems},
  author={Abueidda, Diab W and Pantidis, Panos and Mobasher, Mostafa E},
  journal={Computer Methods in Applied Mechanics and Engineering},
  volume={436},
  pages={117699},
  year={2025},
  publisher={Elsevier}
}

@article{bertsimas2021probabilistic,
  title={Probabilistic guarantees in robust optimization},
  author={Bertsimas, Dimitris and Den Hertog, Dick and Pauphilet, Jean},
  journal={SIAM Journal on Optimization},
  volume={31},
  number={4},
  pages={2893--2920},
  year={2021},
  publisher={SIAM}
}

@inproceedings{vaca2024kolmogorov,
  title={Kolmogorov-arnold networks (kans) for time series analysis},
  author={Vaca-Rubio, Cristian J and Blanco, Luis and Pereira, Roberto and Caus, M{\`a}rius},
  booktitle={2024 IEEE Globecom Workshops (GC Wkshps)},
  pages={1--6},
  year={2024},
  organization={IEEE}
}

@article{genet2024tkan,
  title={Tkan: Temporal kolmogorov-arnold networks},
  author={Genet, Remi and Inzirillo, Hugo},
  journal={arXiv:2405.07344},
  year={2024}
}

@article{schmidt2021kolmogorov,
  title={The Kolmogorov--Arnold representation theorem revisited},
  author={Schmidt-Hieber, Johannes},
  journal={Neural networks},
  volume={137},
  pages={119--126},
  year={2021},
  publisher={Elsevier}
}

@article{shi2022efficiently,
  title={Efficiently computing local lipschitz constants of neural networks via bound propagation},
  author={Shi, Zhouxing and Wang, Yihan and Zhang, Huan and Kolter, J Zico and Hsieh, Cho-Jui},
  journal={NeurIPS},
  volume={35},
  pages={2350--2364},
  year={2022}
}

@article{dhiman2021control,
  title={Control barriers in bayesian learning of system dynamics},
  author={Dhiman, Vikas and Khojasteh, Mohammad Javad and Franceschetti, Massimo and Atanasov, Nikolay},
  journal={IEEE transactions on automatic control},
  volume={68},
  number={1},
  pages={214--229},
  year={2021},
  publisher={IEEE}
}

@article{seaman2012hidden,
  title={Hidden dangers of specifying noninformative priors},
  author={Seaman III, John W and Seaman Jr, John W and Stamey, James D},
  journal={The American Statistician},
  volume={66},
  number={2},
  pages={77--84},
  year={2012},
  publisher={Taylor \& Francis}
}

@inproceedings{chakraborty2024scalable,
  title={Scalable Model-Based Gaussian Process Clustering},
  author={Chakraborty, Anirban and Chakraborty, Abhisek},
  booktitle={ICASSP},
  pages={5730--5734},
  year={2024},
  organization0={IEEE}
}

@article{liu2020gaussian,
  title={When Gaussian process meets big data: A review of scalable GPs},
  author={Liu, Haitao and Ong, Yew-Soon and Shen, Xiaobo and Cai, Jianfei},
  journal={IEEE Transactions on Neural Networks and Learning Systems},
  volume={31},
  number={11},
  pages={4405--4423},
  year={2020},
  publisher={IEEE}
}

@article{nguyen2009model,
  title={Model learning with local gaussian process regression},
  author={Nguyen-Tuong, Duy and Seeger, Matthias and Peters, Jan},
  journal={Advanced Robotics},
  volume={23},
  number={15},
  pages={2015--2034},
  year={2009},
  publisher={Taylor \& Francis}
}

@article{snelson2005sparse,
  title={Sparse Gaussian processes using pseudo-inputs},
  author={Snelson, Edward and Ghahramani, Zoubin},
  journal={NeurIPS},
  volume={18},
  year={2005}
}

@inproceedings{song2024novel,
  title={Novel Architecture of Deep Feature-Based Gaussian Processes with an Ensemble of Kernels},
  author={Song, Yuanqing and Liu, Yuhao and Djuri{\'c}, Petar M},
  booktitle={ICASSP},
  pages={6750--6754},
  year={2024},
  organization0={IEEE}
}

@inproceedings{jantre2024learning,
  title={Learning active subspaces for effective and scalable uncertainty quantification in deep neural networks},
  author={Jantre, Sanket and Urban, Nathan M and Qian, Xiaoning and Yoon, Byung-Jun},
  booktitle={ICASSP},
  pages={5330--5334},
  year={2024},
  organization0={IEEE}
}

@article{lakshminarayanan2017simple,
  title={Simple and scalable predictive uncertainty estimation using deep ensembles},
  author={Lakshminarayanan, Balaji and Pritzel, Alexander and Blundell, Charles},
  journal={{NeurIPS}},
  volume={30},
  year={2017}
}

@inproceedings{blundell2015weight,
  title={Weight uncertainty in neural network},
  author={Blundell, Charles and Cornebise, Julien and Kavukcuoglu, Koray and Wierstra, Daan},
  booktitle={ICML},
  pages={1613--1622},
  year={2015},
  organization0={PMLR}
}

@inproceedings{welling2011bayesian,
  title={Bayesian learning via stochastic gradient Langevin dynamics},
  author={Welling, Max and Teh, Yee W},
  booktitle={ICML},
  pages={681--688},
  year={2011},
  organization0={Citeseer}
}

@article{lu2017ensemble,
  title={Ensemble sampling},
  author={Lu, Xiuyuan and Van Roy, Benjamin},
  journal={NeurIPS},
  volume={30},
  year={2017}
}

@inproceedings{ritter2018scalable,
  title={A scalable laplace approximation for neural networks},
  author={Ritter, Hippolyt and Botev, Aleksandar and Barber, David},
  booktitle={ICLR},
  volume={6},
  year={2018},
  organization0={International Conference on Representation Learning}
}

@InProceedings{gal2016dropout,
  title = 	 {Dropout as a Bayesian Approximation: Representing Model Uncertainty in Deep Learning},
  author = 	 {Gal, Yarin and Ghahramani, Zoubin},
  booktitle = 	 {ICML},
  pages = 	 {1050--1059},
  year = 	 {2016},
  volume = 	 {48},
  month = 	 {20--22 Jun},
  publisher =    {PMLR},
}

@article{maddox2019simple,
  title={A simple baseline for bayesian uncertainty in deep learning},
  author={Maddox, Wesley J and Izmailov, Pavel and Garipov, Timur and Vetrov, Dmitry P and Wilson, Andrew Gordon},
  journal={{NeurIPS}},
  volume={32},
  year={2019}
}

@inproceedings{van2020uncertainty,
  title={Uncertainty estimation using a single deep deterministic neural network},
  author={Van Amersfoort, Joost and Smith, Lewis and Teh, Yee Whye and Gal, Yarin},
  booktitle={ICML},
  pages={9690--9700},
  year={2020},
  organization0={PMLR}
}

@inproceedings{mukhoti2023deep,
  title={Deep deterministic uncertainty: A new simple baseline},
  author={Mukhoti, Jishnu and Kirsch, Andreas and van Amersfoort, Joost and Torr, Philip HS and Gal, Yarin},
  booktitle={CVPR},
  pages={24384--24394},
  year={2023}
}

@article{van2021feature,
  title={On feature collapse and deep kernel learning for single forward pass uncertainty},
  author={Van Amersfoort, Joost and Smith, Lewis and Jesson, Andrew and Key, Oscar and Gal, Yarin},
  journal={arXiv:2102.11409},
  year={2021}
}

@article{liu2020simple,
  title={Simple and principled uncertainty estimation with deterministic deep learning via distance awareness},
  author={Liu, Jeremiah and Lin, Zi and Padhy, Shreyas and Tran, Dustin and Bedrax Weiss, Tania and Lakshminarayanan, Balaji},
  journal={NeurIPS},
  volume={33},
  pages={7498--7512},
  year={2020}
}

@article{gawlikowski2023survey,
  title={A survey of uncertainty in deep neural networks},
  author={Gawlikowski, Jakob and Tassi, Cedrique Rovile Njieutcheu and Ali, Mohsin and Lee, Jongseok and Humt, Matthias and Feng, Jianxiang and Kruspe, Anna and Triebel, Rudolph and Jung, Peter and Roscher, Ribana and others},
  journal={Artificial Intelligence Review},
  volume={56},
  number={Suppl 1},
  pages={1513--1589},
  year={2023},
  publisher={Springer}
}

@inproceedings{guo2017calibration,
  title={On calibration of modern neural networks},
  author={Guo, Chuan and Pleiss, Geoff and Sun, Yu and Weinberger, Kilian Q},
  booktitle={ICML},
  pages={1321--1330},
  year={2017},
  organization0={PMLR}
}

@inproceedings{liu2020robust,
  title={Robust regression for safe exploration in control},
  author={Liu, Anqi and Shi, Guanya and Chung, Soon-Jo and Anandkumar, Anima and Yue, Yisong},
  booktitle={Learning for Dynamics and Control},
  pages={608--619},
  year={2020},
  organization0={PMLR}
}

@inproceedings{cheng2020safe,
  title={Safe multi-agent interaction through robust control barrier functions with learned uncertainties},
  author={Cheng, Richard and Khojasteh, Mohammad Javad and Ames, Aaron D and Burdick, Joel W},
  booktitle={IEEE CDC},
  pages={777--783},
  year={2020},
  organization0={IEEE}
}

@book{de1978practical,
  title={A practical guide to splines},
  author={De Boor, Carl},
  volume={27},
  year={1978},
  publisher={springer New York}
}

@article{berkenkamp2017safe,
  title={Safe model-based reinforcement learning with stability guarantees},
  author={Berkenkamp, Felix and Turchetta, Matteo and Schoellig, Angela and Krause, Andreas},
  journal={NeurIPS},
  volume={30},
  year={2017}
}

@article{howard2003robotics,
  title={The robotics data set repository (radish)},
  author={Howard, Andrew},
  journal={http://radish. sourceforge. net/},
  year={2003}
}

@book{phillips2003interpolation,
  title={Interpolation and approximation by polynomials},
  author={Phillips, George M},
  volume={14},
  year={2003},
  publisher={Springer Science \& Business Media}
}

@article{igelnik2003kolmogorov,
  title={Kolmogorov's spline network},
  author={Igelnik, Boris and Parikh, Neel},
  journal={IEEE Transactions on Neural Networks},
  volume={14},
  number={4},
  pages={725--733},
  year={2003},
  publisher={IEEE}
}

@article{aziznejad2020deep,
  title={Deep neural networks with trainable activations and controlled Lipschitz constant},
  author={Aziznejad, Shayan and Gupta, Harshit and Campos, Joaquim and Unser, Michael},
  journal={IEEE Transactions on Signal Processing},
  volume={68},
  pages={4688--4699},
  year={2020},
  publisher={IEEE}
}

@article{bohra2020learning,
  title={Learning activation functions in deep (spline) neural networks},
  author={Bohra, Pakshal and Campos, Joaquim and Gupta, Harshit and Aziznejad, Shayan and Unser, Michael},
  journal={IEEE Open Journal of Signal Processing},
  volume={1},
  pages={295--309},
  year={2020},
  publisher={IEEE}
}

@inproceedings{
liu2025kan,
title={{KAN}: Kolmogorov{\textendash}Arnold Networks},
author={Ziming Liu and Yixuan Wang and Sachin Vaidya and Fabian Ruehle and James Halverson and Marin Soljacic and Thomas Y. Hou and Max Tegmark},
booktitle={ICLR},
year={2025},
url0={https://openreview.net/forum?id=Ozo7qJ5vZi}
}

@book{apostol1967calculus,
  title={Calculus, Volume 1},
  author={Apostol, Tom M},
  year={1967},
  publisher={John Wiley \& Sons}
}

@inproceedings{damianou2013deep,
  title={Deep gaussian processes},
  author={Damianou, Andreas and Lawrence, Neil D},
  booktitle={Artificial intelligence and statistics},
  pages={207--215},
  year={2013},
  organization0={PMLR}
}

@book{jaulin2001interval,
  title={Interval analysis},
  author={Jaulin, Luc and Kieffer, Michel and Didrit, Olivier and Walter, Eric},
  year={2001},
  publisher={Springer}
}

@book{rasmussen2006gaussian,
  title={Gaussian processes for machine learning},
  author={Williams, Christopher KI and Rasmussen, Carl Edward},
  volume={1},
  number={1},
  year={2006},
  publisher={The MIT Press}
}

@article{berkenkamp2019no,
  title={No-regret Bayesian optimization with unknown hyperparameters},
  author={Berkenkamp, Felix and Schoellig, Angela P and Krause, Andreas},
  journal={Journal of Machine Learning Research},
  volume={20},
  number={50},
  pages={1--24},
  year={2019}
}

@inproceedings{cheng2021limits,
  title={Limits of probabilistic safety guarantees when considering human uncertainty},
  author={Cheng, Richard and Murray, Richard M and Burdick, Joel W},
  booktitle={ICRA},
  pages={3182--3189},
  year={2021},
  organization0={IEEE}
}

@article{shapley1953value,
  title={A value for n-person games},
  author={Shapley, Lloyd S},
  journal={Contribution to the Theory of Games},
  volume={2},
  year={1953}
}

@article{ghorbani2020neuron,
  title={Neuron shapley: Discovering the responsible neurons},
  author={Ghorbani, Amirata and Zou, James Y},
  journal={NeurIPS},
  volume={33},
  pages={5922--5932},
  year={2020}
}

@inbook{wahba2006spline,
    author = {Wahba, Grace},
    publisher = {John Wiley \& Sons, Ltd},
    isbn = {9780471667193},
    title = {Spline Functions},
    booktitle = {Encyclopedia of Statistical Sciences},
    chapter = {},
    pages = {},
    doi = {https://doi.org/10.1002/0471667196.ess3095.pub2},
    url0 = {https://onlinelibrary.wiley.com/doi/abs/10.1002/0471667196.ess3095.pub2},
    eprint = {https://onlinelibrary.wiley.com/doi/pdf/10.1002/0471667196.ess3095.pub2},
    year = {2006}
}

@article{scikit-learn,
  title={Scikit-learn: Machine Learning in {P}ython},
  author={Pedregosa, F. and Varoquaux, G. and Gramfort, A. and Michel, V.
          and Thirion, B. and Grisel, O. and Blondel, M. and Prettenhofer, P.
          and Weiss, R. and Dubourg, V. and Vanderplas, J. and Passos, A. and
          Cournapeau, D. and Brucher, M. and Perrot, M. and Duchesnay, E.},
  journal={Journal of Machine Learning Research},
  volume={12},
  pages={2825--2830},
  year={2011}
}

@book{boyd2004convex,
  title={Convex optimization},
  author={Boyd, Stephen P and Vandenberghe, Lieven},
  year={2004},
  publisher={Cambridge university press}
}

@article{lipton2018mythos,
  title={The mythos of model interpretability: In machine learning, the concept of interpretability is both important and slippery.},
  author={Lipton, Zachary C},
  journal={Queue},
  volume={16},
  number={3},
  pages={31--57},
  year={2018},
  publisher={ACM New York, NY, USA}
}

@article{nanda2023progress,
  title={Progress measures for grokking via mechanistic interpretability},
  author={Nanda, Neel and Chan, Lawrence and Lieberum, Tom and Smith, Jess and Steinhardt, Jacob},
  journal={arXiv:2301.05217},
  year={2023}
}

@article{tohme2024isr,
  title={{ISR}: Invertible Symbolic Regression},
  author={Tohme, Tony and Khojasteh, Mohammad Javad and Sadr, Mohsen and Meyer, Florian and Youcef-Toumi, Kamal},
  journal={arXiv:2405.06848},
  year={2024}
}

@article{pleiss2021limitations,
  title={The limitations of large width in neural networks: A deep Gaussian process perspective},
  author={Pleiss, Geoff and Cunningham, John P},
  journal={NeurIPS},
  volume={34},
  pages={3349--3363},
  year={2021}
}

@inproceedings{chen2014solving,
  title={Solving the problem of Runge phenomenon by pseudoinverse cubic spline},
  author={Chen, Dechao and Qiao, Tianjian and Tan, Hongzhou and Li, Mingming and Zhang, Yunong},
  booktitle={2014 IEEE 17th International Conference on Computational Science and Engineering},
  pages={1226--1231},
  year={2014},
  organization0={IEEE}
}

@article{wu2018deterministic,
  title={Deterministic variational inference for robust bayesian neural networks},
  author={Wu, Anqi and Nowozin, Sebastian and Meeds, Edward and Turner, Richard E and Hernandez-Lobato, Jose Miguel and Gaunt, Alexander L},
  journal={arXiv:1810.03958},
  year={2018}
}

@inproceedings{hernandez2015probabilistic,
  title={Probabilistic backpropagation for scalable learning of bayesian neural networks},
  author={Hern{\'a}ndez-Lobato, Jos{\'e} Miguel and Adams, Ryan},
  booktitle={ICML},
  pages={1861--1869},
  year={2015},
  organization0={PMLR}
}

@inproceedings{lee2018deep,
  title={Deep Neural Networks as Gaussian Processes},
  author={Lee, Jaehoon and Bahri, Yasaman and Novak, Roman and Schoenholz, Samuel S and Pennington, Jeffrey and Sohl-Dickstein, Jascha},
  booktitle={ICLR},
  year={2018}
}

@article{jacot2018neural,
  title={Neural tangent kernel: Convergence and generalization in neural networks},
  author={Jacot, Arthur and Gabriel, Franck and Hongler, Cl{\'e}ment},
  journal={NeurIPS},
  volume={31},
  year={2018}
}

@article{lederer2019uniform,
  title={Uniform error bounds for Gaussian process regression with application to safe control},
  author={Lederer, Armin and Umlauft, Jonas and Hirche, Sandra},
  journal={NeurIPS},
  volume={32},
  year={2019}
}

@book{egerstedt2009control,
  title={Control theoretic splines: optimal control, statistics, and path planning},
  author={Egerstedt, Magnus and Martin, Clyde},
  year={2009},
  publisher={Princeton University Press}
}

@article{unser2002splines,
  title={Splines: A perfect fit for signal and image processing},
  author={Unser, Michael},
  journal={IEEE Signal processing magazine},
  volume={16},
  number={6},
  pages={22--38},
  year={2002},
  publisher={IEEE}
}

@inproceedings{fey2018splinecnn,
  title={Splinecnn: Fast geometric deep learning with continuous b-spline kernels},
  author={Fey, Matthias and Lenssen, Jan Eric and Weichert, Frank and M{\"u}ller, Heinrich},
  booktitle={CVPR},
  pages={869--877},
  year={2018}
}

@article{morris2021hilbert,
  title={Hilbert 13: Are there any genuine continuous multivariate real-valued functions?},
  author={Morris, Sidney},
  journal={Bulletin of the American Mathematical Society},
  volume={58},
  number={1},
  pages={107--118},
  year={2021}
}

@article{kuurkova1992kolmogorov,
  title={Kolmogorov's theorem and multilayer neural networks},
  author={K\.urkov\`a\hspace{0pt}, V{\v{e}}ra},
  journal={Neural networks},
  volume={5},
  number={3},
  pages={501--506},
  year={1992},
  publisher={Elsevier}
}

@article{epperson1987runge,
  title={On the Runge example},
  author={Epperson, James F},
  journal={The American Mathematical Monthly},
  volume={94},
  number={4},
  pages={329--341},
  year={1987},
  publisher={Taylor \& Francis}
}

@inproceedings{liu2015deep,
  title={Deep learning face attributes in the wild},
  author={Liu, Ziwei and Luo, Ping and Wang, Xiaogang and Tang, Xiaoou},
  booktitle={Proceedings of the IEEE international conference on computer vision},
  pages={3730--3738},
  year={2015}
}

@inproceedings{hensman2015scalable,
  title={Scalable variational Gaussian process classification},
  author={Hensman, James and Matthews, Alexander and Ghahramani, Zoubin},
  booktitle={Artificial intelligence and statistics},
  pages={351--360},
  year={2015},
  organization0={PMLR}
}

\begin{IEEEbiography}[{\includegraphics[width=1in,height=1.25in, clip,keepaspectratio]{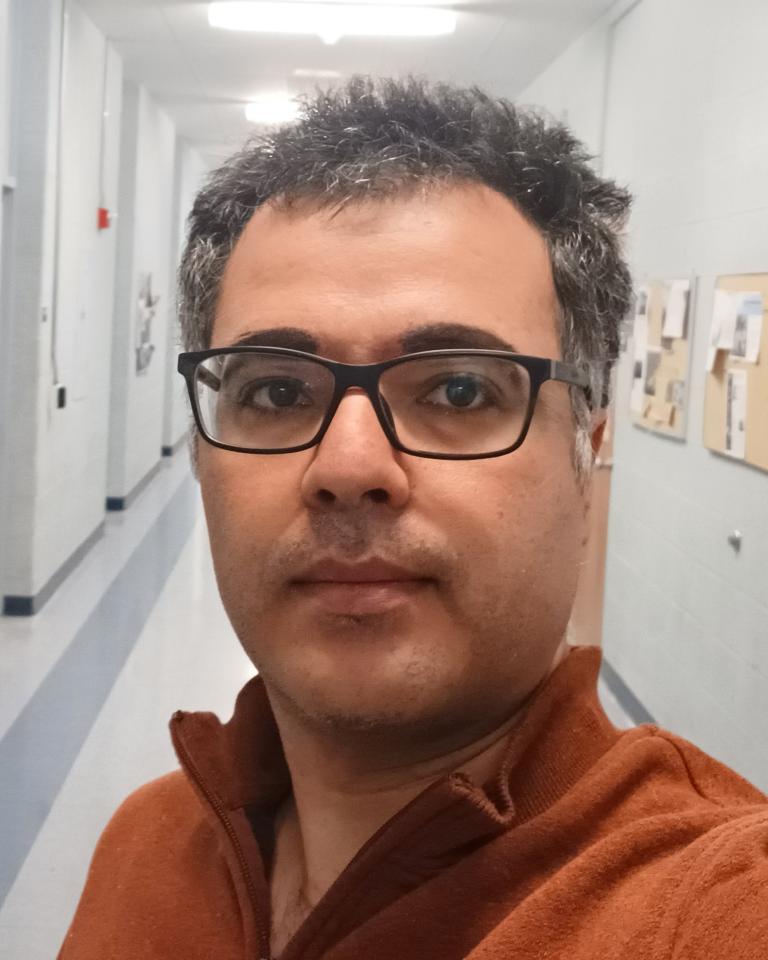}}]{Masoud Ataei} received his M.Sc. degree in Electrical Engineering from Amirkabir University of Technology and is currently pursuing his Ph.D. in Electrical Engineering at the University of Maine. He serves as a Research Assistant in the Computer Vision and Autonomous Robotics lab at the University of Maine and is passionate about advancing safe control systems, particularly in autonomous robotics. His current research focuses on developing a risk-aware safe control framework for autonomous vehicles, combining model-based techniques with machine learning approaches to enable reliable navigation in uncertain environments.
\end{IEEEbiography}

\begin{IEEEbiography}[{\includegraphics[width=1in,height=1.25in,clip,keepaspectratio]{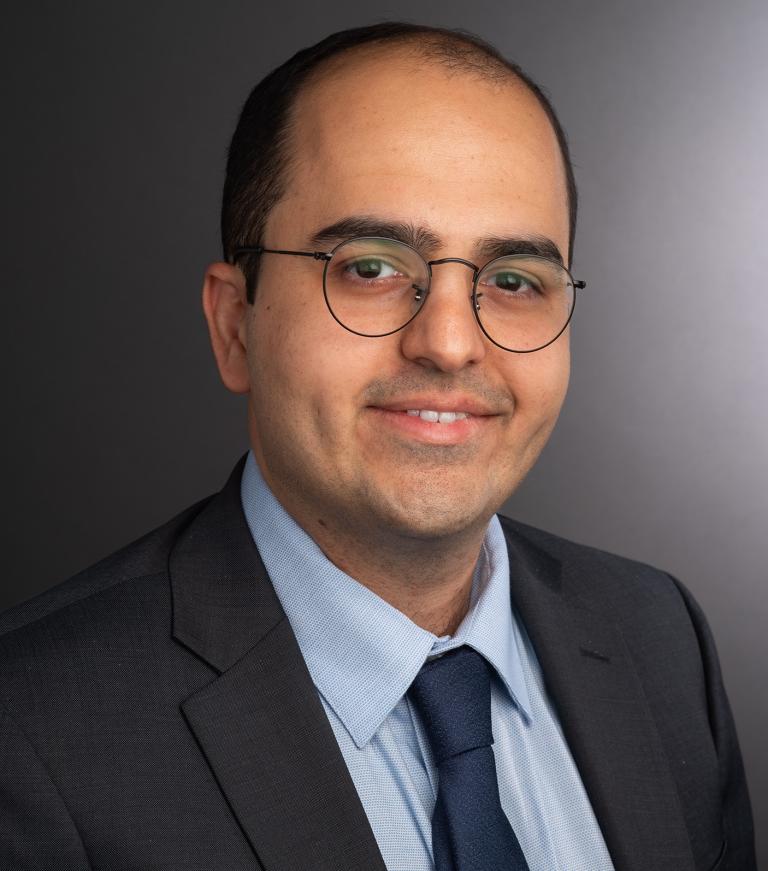}}]{Mohammad Javad Khojasteh} (Member,
IEEE) received the Ph.D. and M.Sc. degrees in electrical and computer engineering from the University of California, San Diego in 2019 and 2017, respectively. 
He is a Gleason Endowed Assistant Professor with the Rochester Institute of Technology (RIT). Before joining RIT, he held postdoctoral positions at Marine Physical Laboratory (MPL) at Scripps Institution of Oceanography (SIO), Department of Mechanical Engineering and Laboratory for Information and Decision Systems (LIDS) at Massachusetts Institute of Technology (MIT), and Center for Autonomous Systems and Technologies (CAST) at California Institute of Technology (Caltech), where he worked with Team CoSTAR as visitor at NASA Jet Propulsion Laboratory. He received the Gleason Chair in 2024 and the 2025 Provost’s Learning Innovation Grant at RIT. Dr. Khojasteh's publications, co-authored with colleagues and students, have received awards, including Tammy L. Blair Student Paper Award (second place) from the International Society of Information Fusion.
\end{IEEEbiography}

\begin{IEEEbiography}[{\includegraphics[width=1in,height=1.25in,clip,keepaspectratio]{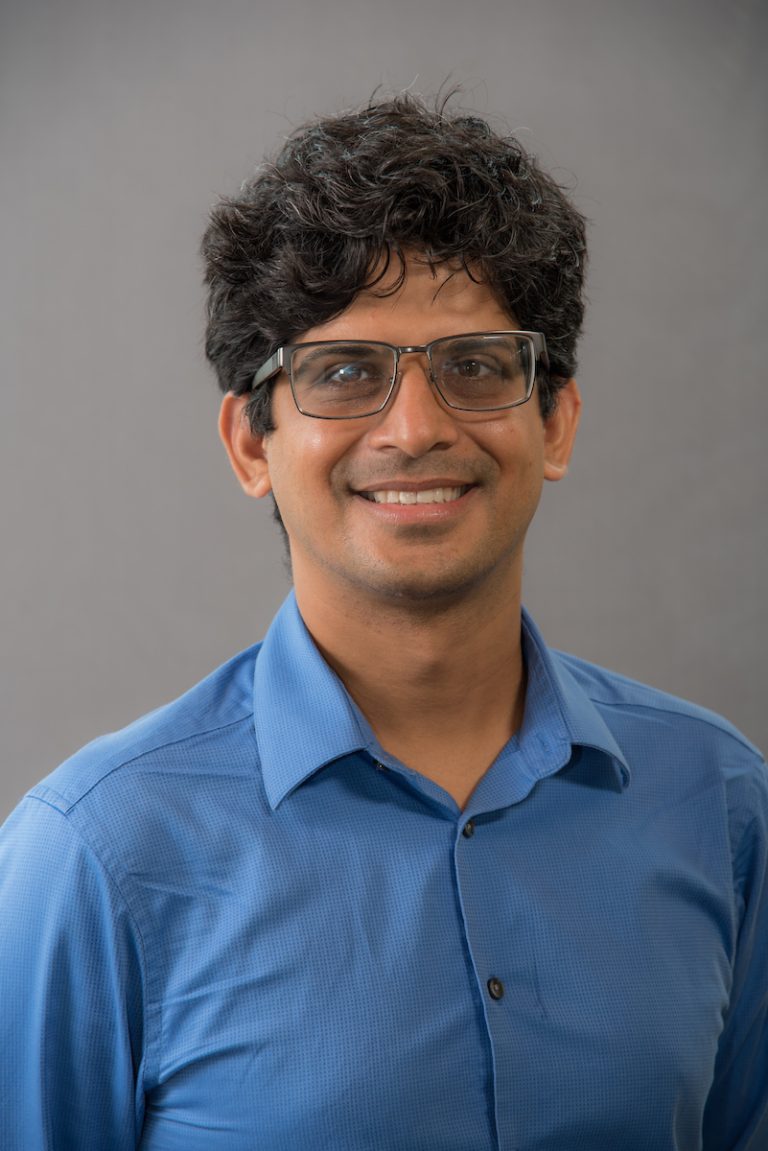}}]{Vikas Dhiman} (Member, IEEE) received the
Electrical Engineering degree from the Indian Institute of Technology, Roorkee, in 2008, the M.S. degree from the University at Buffalo, in 2014, and the Ph.D. degree from the University of Michigan, Ann Arbor, in 2019. His works lie in the localization, mapping, and control algorithms for applications in robotics. He was a Postdoctoral Researcher with the University of California, San Diego, from 2019 to 2021. He is currently an Assistant Professor with the ECE Department, the University of Maine.
\end{IEEEbiography}

\ifarxiv

\renewcommand{\thesection}{\Alph{section}}  
\renewcommand{\thesubsection}{\Alph{section}-\Roman{subsection}} 
\renewcommand{\thesectiondis}{\Alph{section}.}  
\renewcommand{\thesubsectiondis}{\Roman{subsection}.}
\setcounter{section}{0}                     

\label{fn:sup-material}
\onecolumn
\twocolumn[{
\begin{center}
    \Large 
   \textbf{DAREK - DISTANCE AWARE ERROR FOR KOLMOGOROV NETWORKS (Supplementary Material)} \\[0.5em]
    \normalsize 
    Masoud Ataei$^{\star}$ \qquad  Mohammad Javad Khojasteh$^{\dagger}$ \qquad Vikas Dhiman$^{\star}$ \\
$^{\star}$Electrical and Computer Engg. Dept., University of Maine, Orono, ME, USA \\
\small $^{\dagger}$Electrical and Microelectronic Engg. Dept.,  Rochester Institute of Technology, Rochester, NY, USA \\[1em]
\end{center}
}]

\section{Background and Proofs}
\label{sec:background-proofs}
Here, we review the background and definitions used in the paper. We also provide proof for the theories used in this paper. 

\subsection{Mean Value Theorem:}
\label{sec:background-proofs-meanvalue}
The result of following Theorem establishes the relation between divided differences and derivatives.

\begin{theorem}[Mean value theorem~\cite{de1978practical}]
\label{thm:divided-diff-mean-value}
    For a $k-1$ times differentiable function $f:\bbR\to\bbR$ and
    a set of $k$ points $\calT = \{\tau_1, \dots, \tau_k\}$, there exists $\zeta \in \calR(\calT)$ such that:
    \begin{align}
    [\tau_1,\hdots,\tau_{k}]f &=\frac{1}{(k-1)!} f^{(k-1)}(\zeta) \quad \exists \zeta \in \calR(\calT).
\end{align}
Here, $f^{(k-1)}(.)$ denotes the $k-1$th derivative of the function and $\calR(\calT)= [\min(\calT),\max(\calT)]$ the interval spanned by the knots $\calT$.
\end{theorem}%

\begin{proof}
    Suppose we approximate a function $f$ with a $k-1$th order newton's polynomial $\hat{f} = \calP_{k-1}$, since $\hat{f}$ passes through $k$ points at $f(\tau_{1:k})$ then we know $f(\tau_i) = \hat{f}(\tau_i) \; \forall i \in {1,\dots, k}$. It means $e(\tau_i) = f(\tau_i)-\hat{f}(\tau_i) = 0$. Since, $f - \hat{f}$ is a continuous function then between every two continuous knots there exists at least one point that derivative of $f-\hat{f}$ will be zero, having $k$ point will make $k-1$ point of $e'$ to be zero. Continuing this process results that there are one zero on $k-1$th derivative of $e$ function which means $(f-\hat{f})^{(k-1)}(\zeta) = 0$ at one point, lets call it $\zeta$. Then we have:
    \begin{align}
        f^{(k-1)} &= (\calP_k)^{(k-1)} \notag\\
        &= ([\tau_1]f + \sum_{i=1}^{k-1}[\tau_{1:i+1}]f \prod_{j=1}^{i}(x-\tau_j))^{(k-1)} \notag\\
        &= (k-1)! \quad [\tau_1,\hdots,\tau_{k}]f \notag \\
        \Rightarrow [\tau_1,\hdots,\tau_{k}]f &= \frac{1}{(k-1)!}f^{(k-1)}(\zeta).
     \end{align}
\end{proof}

\subsection{Linearity of Divided Difference:}
\label{sec:background-proofs-linearity-divided-diff}
Divided differences are linear with respect to the function values, i.e., $[\bftau](\alpha f + \beta g) = \alpha [\bftau] f + \beta [\bftau] g$ for a set of $k$ distinct knots $\bftau = \{\tau_1, \dots, \tau_k\}$.
We use this property to establish the linearity of Newton's polynomial; the proof is as follows.

\begin{proof}
Let $f$ and $g$ be two functions, and let $\alpha$ and $\beta$ be two scalars. We will show that the divided difference~\eqref{eq:divided-differences} operator satisfies linearity. For the base case, consider the zeroth-order divided differences: 

\begin{align}
    [\tau_1](\alpha f+\beta g) &= (\alpha f+\beta g)(\tau_1) \notag\\
    &=\alpha  f(\tau_1)+\beta g(\tau_1) \notag\\
    &= \alpha [\tau_1]f+\beta [\tau_1]g.
\end{align}

We proceed to the first-order divided differences to confirm linearity:
\begin{align}
    [\tau_1,\tau_2]&(\alpha f +\beta g) = \frac{[\tau_2](\alpha f+\beta g)-[\tau_1](\alpha f+\beta g)}{\tau_2-\tau_1} \notag\\
    &= \frac{\left (\alpha f(\tau_2)+\beta g(\tau_2) \right )-\left (\alpha f(\tau_1)+\beta g(\tau_1) \right )}{\tau_2-\tau_1} \notag\\
    &= \alpha \frac{ f(\tau_2) - f(\tau_1)}{\tau_2-\tau_1}+\beta \frac{ g(\tau_2)- g(\tau_1)}{\tau_2-\tau_1} \notag\\
    &= \alpha [\tau_1,\tau_2]f + \beta [\tau_1,\tau_2]g.
\end{align}

For higher-order divided differences, we use induction. Assume that for $k-1$th order divided differences, the linearity holds:
\begin{align}
    [\tau_1,\dots,\tau_{k-1}]&(\alpha f+\beta g) = \notag \\
    & \alpha [\tau_1,\dots,\tau_{k-1}]f+ 
    \beta [\tau_1,\dots,\tau_{k-1}]g.
\end{align}

Now, for the $k$th order divided differences, we have:
\begin{align}
    [\tau_1&,\dots,\tau_k](\alpha f+\beta g) \notag\\
    &= \frac{[\tau_2,\dots,\tau_k](\alpha f+\beta g)-[\tau_1,\dots,\tau_{k-1}](\alpha f+\beta g)}{\tau_k-\tau_1} \notag\\
    &= \alpha [\tau_1,\dots,\tau_k]f + \beta [\tau_1,\dots,\tau_k]g.
\end{align}

Thus, the divided difference operator is linear for all orders.
\end{proof}

\subsection{Linearity of Newton's polynomial operator:}
\label{lemma:Linearity-Newton's-poly}
Lemma~\ref{thm:error-at-knots} implicitly depends on this property, and it has also been used in Equation~\eqref{eq:linearity-newt-poly}, and its proof is as follows.

\begin{proof}
From definition of Newton's polynomial~\eqref{eq:newton-form-poly} for a function $\alpha f + \beta g$, we have:
\begin{align}
    \calP_{k,n}&[\alpha f + \beta g](x) = 
    [\tau_n](\alpha f+ \beta g) \notag\\
    &+ \sum_{i=1}^{k} [\tau_{n},\dots,\tau_{n+i}](\alpha f + \beta g) \prod_{j=n}^{n+i-1} (x-\tau_j).
\end{align}

Using the linearity properties of divided differences, we can rewrite this as:
\begin{align}
    \calP&_{k,n}[\alpha f + \beta g](x) = 
    (\alpha [\tau_n]f+ \beta [\tau_n]g) \notag\\ 
    &+ \sum_{i=1}^{k} (\alpha [\tau_{n},\dots,\tau_{n+i}]f + \beta [\tau_{n},\dots,\tau_{n+i}]g) \prod_{j=n}^{n+i-1} (x-\tau_j) \notag\\
    &= \alpha \calP_{k,n}[f](x) + \beta \calP_{k,n}[g](x).
\end{align}
Therefore, $\calP_{k,n}$ is a linear operator.
\end{proof}

\subsection{Proof of Lemma~\ref{thm:error-at-knots}:}
\label{ptf:thm-error-at-knot}
Note that $\calP_{k,j}[\calD_{\hat{f}_{[j]}}(\bftau_{1:m})](x) = \hat{f}_{[j]}(x)$ because $\hat{f}_{[j]}(x)$ is a $k$th-order polynomial that generated through the $k+1$ nearby knots $\{(\tau_i, \hat{f}_{[j]}(\tau_i))\}_{i=j}^{j+k}$ and  Newton's polynomial is a unique $k$th-order polynomial that is generated through the same data $k+1$ knots~\cite{de1978practical}.
From the linearity of Newton's polynomial operator at the same knots we have,
\begin{align}
    \calP_{k,j}[\calD_{\hat{f}_{[j]}}](x) =
    \calP_{k,j}[\calD_f](x) - \calP_{k,j}[e_j^f](x).
    \label{eq:linearity-newt-poly}
\end{align}
To simplify notation, we drop the knots $\bftau_{1:m}$ when clear from the context.
Combining these observations, for $x \in [\tau_j, \tau_{j+1})$ and $j \in \{1, \dots, m - 1\}$, we get,
\begin{align}
    f(x)&-\hat{f}_j(x) = f(x)-\calP_{k,j}[\calD_{\hat{f}_{[j]}}](x)
    \notag\\
    &\overset{\eqref{eq:linearity-newt-poly}}{=}
    f(x) - \calP_{k,j}[\calD_f](x) + \calP_{k,j}[\calD_{e^f_j}](x) 
    \notag\\
    \text{or } |f(x)&-\hat{f}_{[j]}(x)|
    \le  |f(x) - \calP_{k,j}[\calD_f](x)| + |\calP_{k,j}[\calD_{e^f_j}](x)| \notag\\
    &\overset{Thm.~\ref{thm:poly-interp-bound}}{\le} \bar{u}_f(x) + |\calP_{k,j}[\calD_{e^f_j}](x)|.
\end{align}
The resulting inequality is tight when inequality from Theorem~\ref{thm:poly-interp-bound} is tight and $\sgn(f(x) -\calP_{k,j}[\calD_f](x)) = \sgn(\calP_{k,j}[\calD_{e^f_j}](x))$.
\hfill$\blacksquare$

\subsection{Proof of Theorem~\ref{thm:error-kan-layer}:}
\label{ptf:thm-error-kan-layer}
    Consider a function $f$ which is approximated with the residual structure of $\hat{f}_{[j]} = \hat{r} + \phi_{k}$ on interval $[\tau_j,\tau_{j+1})$. Since Newton's polynomial is uniquely determined by $k+1$ points, we can accordingly construct Newton's polynomial of $\phi_{k}$. 
    Also, $\calP_{k,j}[\calD_{\phi_{k}}] = \phi_{k}$ on interval $[\tau_j,\tau_{j+1})$. Note that the non-zero error of neuron can be defined as $e_{[j]}^{f-\hat{r}} = f - \hat{f}_{[j]} = f - \hat{r} - \phi_{k}$, and since $\phi_k$ in $j$th interval is a polynomial, and polynomial interpolation is a linear operation, we can define $\phi_k = \calP_{k,j}[\calD_{\phi_{k}}] = \calP_{k,j}[\calD_{f - \hat{r}}] - \calP_{k,j}[\calD_{e_{[j]}^{f-\hat{r}}}]$. Substituting this expression into following equation obtains:
\begin{align}
    f - \hat{f}_{[j]} 
    &= (f-\hat{r}) - \calP_{k,j}[\calD_{f - \hat{r}}] + \calP_{k,j}[\calD_{e_{[j]}^{f-\hat{r}}}] .
\end{align}
Then the error bound of the $f-\hat{r}$ would be: 
\begin{align}
    |f- \hat{f}_{[j]}| &\le |(f-\hat{r}) - \calP_{k,j}[\calD_{f - \hat{r}}] | + |\calP_{k,j}[\calD_{e_{[j]}^{f-\hat{r}}}] |\notag\\
    &= u_{f-\hat{r}}(x; \bftau_{1:m}).
\end{align}
\hfill$\blacksquare$

\subsection{Interpolation Error from Theorem~\ref{thm:poly-interp-bound} is Distance-aware.}
\label{sec:background-proofs-DA_of_IntErr}
Interpolation error~\eqref{eq:int-error} is distance aware because it satisfies distance awareness condition~\eqref{eq:dist-aware-cond}.
Here, we first show that the interpolation error has a unique maximum between consecutive knots and is monotonically increasing until it reaches the maximum, then decreasing, and show how this satisfies distance-awareness for differentiable uncertainty~\eqref{eq:dist-aware-cond}.

For a set of distinct and consecutive knots $\bftau_{1:k+1}$, the interpolation error~\eqref{eq:int-error}, within the interval $\tau_j \le x < \tau_{j+1}$, is given by 
\begin{align}
\bar{u}_{[j]}(x,\bftau^{(j)}) &\defeq \alpha \left|\prod_{i=1}^{k+1}(x-\tau_{i}^{(j)})\right|\\
&=\alpha \prod_{i=1}^{j}(x-\tau_{i}^{(j)}) \prod_{i=j+1}^{k+1}(\tau_{i}^{(j)}-x),
\end{align}
where $\alpha \defeq \frac{\calL^{k+1}_f}{(k+1)!}$.
We shorten  $\bar{u}_{[j]}(x,\bftau^{(j)})$ to $\bar{u}_{[j]}(x)$ when knots are clear from the context. 
Take the derivative of $\bar{u}_j(x)$ with respect to $x$, and write it as $\bar{u}'_{[j]}(x) = \bar{u}_{[j]}(x) g_{[j]}(x)$, where $g_{[j]}(x) =\sum_{i=1}^{j}\frac{1}{x-\tau_i}-\sum_{i=j+1}^{k+1}\frac{1}{\tau_i-x}$.
The function $g_{[j]}(x)$ is strictly decreasing since its derivative is negative $g'_{[j]}(x)=-\sum_{i=1}^{k+1}{(x-\tau_i)^{-2}}<0$.
In addition, $g_{[j]}(x) \to +\infty$, $x \to \tau_j^+$, $g_{[j]}(x) \to -\infty$ as $x\to \tau_{j+1}^-$ and $g_{[j]}(x^*) = 0$ when $\sum_{i=1}^{j}\frac{1}{x-\tau_i} = \sum_{i=j+1}^{k+1}\frac{1}{\tau_i-x}$.
Since, $g_{[j]}(x)$ is a continuous and strictly decreasing function in interval $(\tau_j,\tau_{j+1})$, therefore by the intermediate value theorem then there exists a unique 
$x^* \in (\tau_j,\tau_{j+1})$ such that $g_{[j]}(x^*)=0$.
Since $\bar{u}'_{[j]}(x) = \bar{u}_{[j]}(x) g_{[j]}(x)$ and $\bar{u}_{[j]}(x) > 0$, the sign of $\bar{u}'_{[j]}(x)$ is completely determined by the sign of $g_{[j]}(x)$.
Consequently, $\bar{u}'_{[j]}(x) > 0$ for $x \in (\tau_j,x^*)$ and $\bar{u}'_{[j]}(x) < 0$ for $x \in (x^*,\tau_{j+1})$.
Choose an inducing distance function $d_u(x, \tau^*)$ for distance-awareness such that, the nearest knot is $\tau^* = \tau_j$ for $x \in [\tau_j, x^*)$ and $\tau^* = \tau_{j+1}$ for $x \in [x^*, \tau_{j+1})$, 
\begin{align}
    d_u(x,\tau^*) \defeq \begin{cases}
        x - \tau_j & \text{ if } x \in [\tau_j, x^*)\\
        \tau_{j+1} - x & \text{ if } x \in [x^*, \tau_{j+1})
    \end{cases}.
\end{align} 
Substituting $\bar{u}_{[j]}(x)$ and $d_u(x, \tau^*)$ into the requirement of distance-awareness~\eqref{eq:dist-aware-cond}:
\begin{align}
\bar{u}'_{[j]}(x) d'_u(x, \tau^*) = \begin{cases}
    \bar{u}'_{[j]}(x) \frac{d}{dx}(x-\tau_j) \ge 0 & \text{ if } x \in [\tau_j,x^*) \\
    \bar{u}'_{[j]}(x) \frac{d}{dx}(\tau_{j+1}-x) \ge 0 &\text{ otherwise. }
\end{cases}
\end{align}
\hfill$\blacksquare$

\subsection{Distance-Awareness of Spline Error (Lemma~\ref{thm:error-at-knots}).}
\label{sec:background-proofs-DA_of_Spline}
The error at knot $\calP_{k,j}(x)$ does not change the number of unique maxima of the interpolation error, and therefore preserves distance awareness.

From \eqref{eq:error-at-knots-ebl}, the neuron uncertainty consists of two terms: an interpolation error term $\bar{u}_j(x)$ and a correction term $|\calP_{1,j}(x)|$ capturing the error at the knots. 
If we use a linear function for error at knots, 
$|\calP_{1,j}(x)|= |e_j^f| + c_j (x - \tau_j)$,
where $c_j = (|e_{j+1}^f| - |e_j^f|)  / (\tau_{j+1}-\tau_j)$ is a constant and finite because the knots are distinct. The derivative of~\eqref{eq:error-at-knots-ebl} becomes $u_j'(x)=\bar{u}_j'(x)+c_j$.
From Appendix~\ref{sec:background-proofs-DA_of_IntErr}, $\bar{u}_j'$ is continuous $(\tau_j, \tau_{j+1})$ and changes from $+\infty$ to $-\infty$ with a unique zero.
Adding a finite, constant term to it will preserve continuity, monotonicity, and the boundary limits $\pm\infty$ of $u_j$.
Consequently, $u_j(x)$ remains strictly increasing on $x \in (\tau_j,\tilde{x}^*)$ and strictly decreasing on $x \in (\tilde{x}^*,\tau_{j+1})$ and retains the distance awareness property by the substitution into~\eqref{eq:dist-aware-cond} as shown in Appendix~\ref{sec:background-proofs-DA_of_IntErr}, where $\tilde{x}^*$ is the unique maxima of $u_{[j]}(x)$ within the knots $[tau_j, \tau_{j+1})$.

\section{Shapley Example}
\label{sec:shapley_example}
Here we explain how we Shapley value for error division in a simple example. 
Consider the network $\hat{f}(x) = h(g_1(x)+g_2(x))$. This has the set of neurons $\Phi = \{h, g_1, g_2\}$ and which can produce eight different architectures including $\{x, g_1(x), g_2(x), h(x), h(g_1(x)), h(g_2(x)), h(g_1(x)+g_2(x)) \}$. 
The operator $\calM$ can construct each instance by passing the set of neurons to it, for instance $\calM(\{\emptyset\}) = x$ or $\calM(\{h, g_2\}) = h(g_2(x))$. 
Also, note that the permutation in this problem does not affect the output, $\calM(\{h, g_1\}) = \calM(\{g_1, h\})$.

Assume we want to calculate the contribution of $g_1$. All possible sets without $g_1$ would be $\Phi_s \subseteq \Phi \setminus \{g_1\} = \{  \emptyset, \{ g_2\}, \{h \}, \{g_2, h \} \}$ and we can calculate the error each of eight networks over the entire dataset is made then substitute values in~\eqref{eq:error_shapley}. And, we assume we only have access to knots data $f(\tau_{1:m})$ (however it can be any data in the dataset), then the evaluation for one of eight sub-networks would be:
\[
\calV(\calM(\{h, g2\})) = \calV(h(g_2)) = \frac{1}{m}\sum_{i=1}^{m}(y_i - h(g_2(\tau_i))).
\]

Then the complete equation would be:
{\small
\begin{align}
\varphi_{g_1}
&= \frac{0!2!}{3!} \frac{1}{m}\sum_{i=1}^{m}
\big(g_1(\tau_i) - \tau_i\big) &  _{\Phi_s=\emptyset} \notag\\
&+ \frac{1!1!}{3!} \frac{1}{m}\sum_{i=1}^{m}
\big((g_2(\tau_i)+g_1(\tau_i)) - g_2(\tau_i)\big) & {\tiny _{\Phi_s=\{g_2\}}} \notag\\
&+ \frac{1!1!}{3!} \frac{1}{m}\sum_{i=1}^{m}
\big(h(g_1(\tau_i)) - h(\tau_i)\big) & {\tiny _{\Phi_s=\{h\}}} \notag\\
&+ \frac{2!0!}{3!} \frac{1}{m}\sum_{i=1}^{m}
\big(h(g_2(\tau_i)+g_1(\tau_i)) - h(g_2(\tau_i))\big) & {\tiny _{\Phi_s=\{g_2,h\}}}
\end{align}
}

\section{Additional Experiments}
\label{sec:appendx-additional-experiments}
In this section, we present additional experiments for the paper: studying the error of a single neuron and a two-layer DAREK, providing details on the Lipschitz division experiment, and a high-dimensional face recognition experiment. 
\subsection{Spline Error Bound}
\label{sec:exp-spline}
We trained a one-layer DAREK on a $\cos$ function. We used 20 equally spaced points from $[-2\pi,2\pi]$ as training data, and selected $9$ knots from these samples. The model was trained for 200 epochs with $lr=1.0$.
As Fig.~\ref{fig:Spline-Err}~{(a)} depicts, while the fitted model does not perfectly follow the actual function, the overall error bound completely covers the true function. The error bounds between knots grow from one knot and shrink when meeting another knot. 
In this study, we use the error bound with spline fit (EBS)~\eqref{eq:error-at-knots-ebs} for the error at knots.

\begin{figure}

\begin{overpic}[width=0.24\textwidth,trim=0pt 15pt 12pt 12pt, clip]{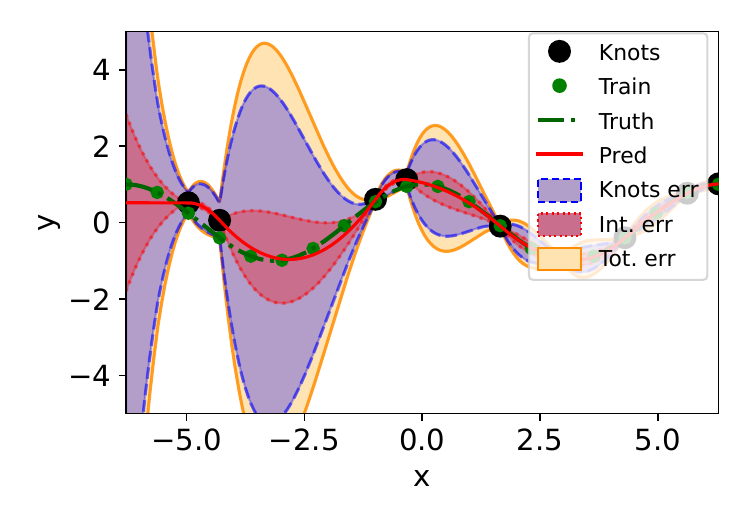}
        \put(-1,60){\textbf{(a)}} 
\end{overpic}%
\begin{overpic}[width=0.24\textwidth,trim=15pt 15pt 0pt 15pt, clip]{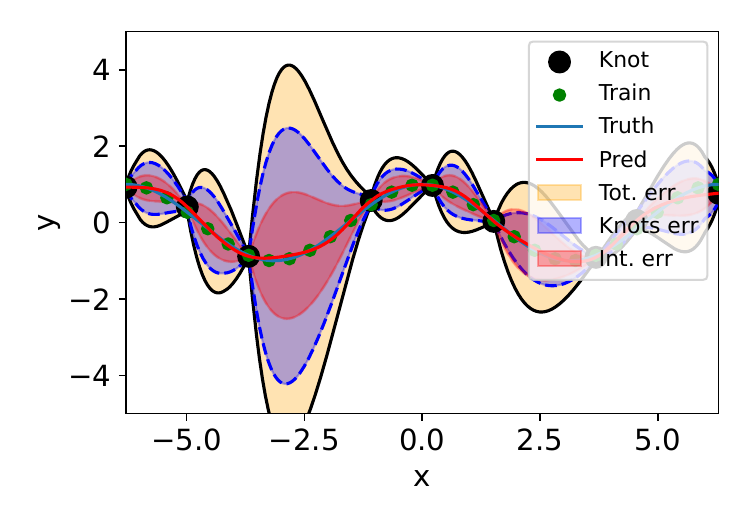}
        \put(-1,60){\textbf{(b)}} 
\end{overpic}
\caption{The error bounds of DAREK model on $\cos$ function. \textbf{(a)} one-layer model,  
\textbf{(b)} two-layer DAREK model. 
}
\label{fig:Spline-Err}
\end{figure}

\subsection{Two-layer DAREK Error Bound}
\label{sec:appendx-additional-experiments-simple}
In this experiment, we trained a two-layer spline network with two hidden units to approximate the $cos$ function. 
The network was trained on 30 equally spaced points from $[-2\pi,2\pi]$, using $9$ knots selected from these points. Training was conducted over 200 epochs with a learning rate of 0.05. Fig.~\ref{fig:Spline-Err}~{(b)} shows the overall output error. 
To analyze layer-wise behavior, we provide visualizations of the intermediate neuron error in Fig.~\ref{fig:KAN-2Layer-Err}, showing each input-output pair separately. 
Since the output of the first layer serves as the input domain for the second layer, any extrema in the first layer can reverse the direction of the input to the second layer, making the composite input non-functional unless sorted. Although the interpolation error of the second layer remains continuous with respect to its input, the propagated error from earlier layers may appear discontinuous due to this sorting process. To better illustrate this phenomenon, we color-coded different monotonic segments of the first layer's output. 
In this study, we use error bound for spline (EBS)~\eqref{eq:error-at-knots-ebs} for the error of individual neurons.

\begin{figure*}[ht]
\includegraphics[width=0.45\textwidth,trim=0pt 10pt 15pt 15pt, clip]{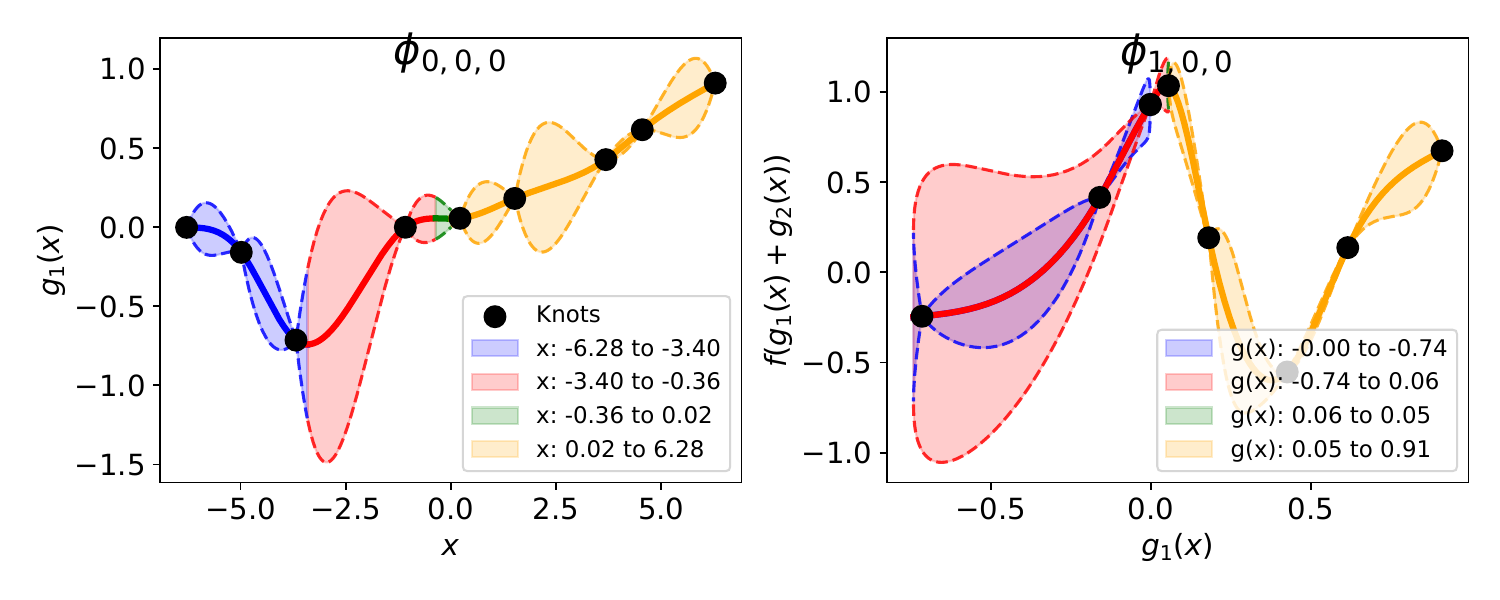}%
\includegraphics[width=0.45\textwidth,trim=0pt 10pt 15pt 15pt, clip]{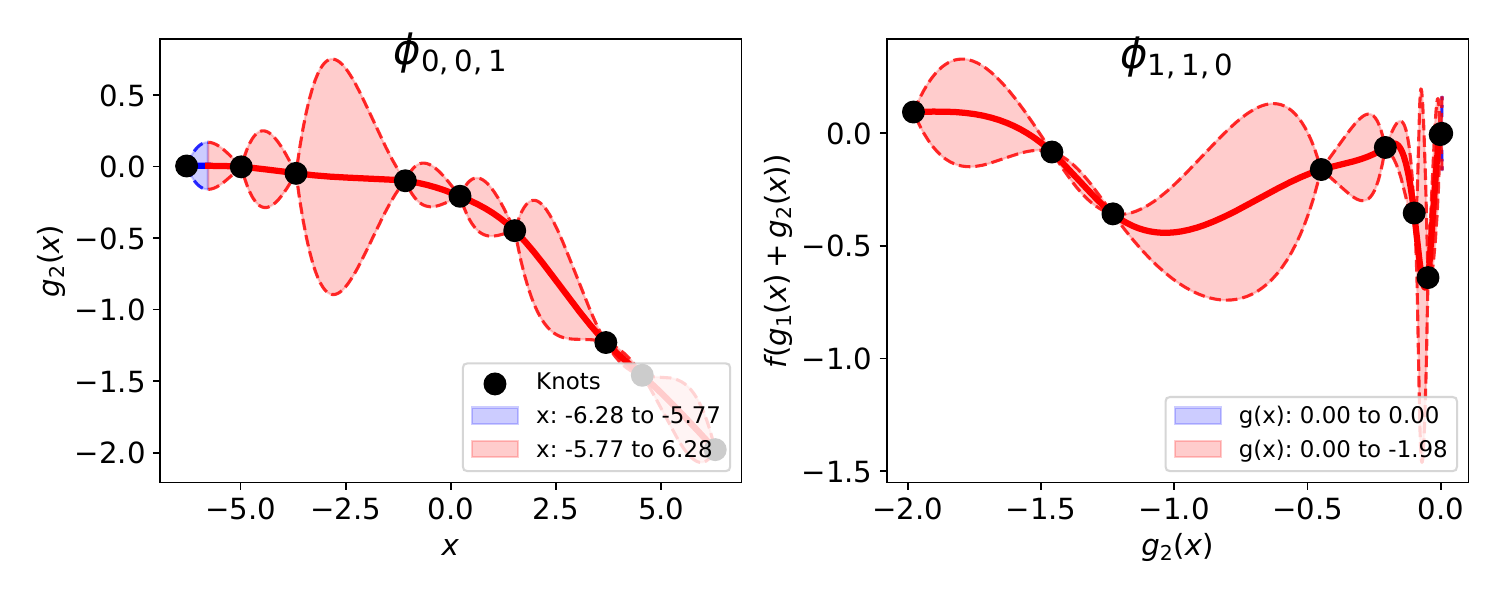}
\caption{Layer-wise interpolation errors of two-layer DAREK model on $cos$ function; color highlights monotonic segments of the first layer affecting the second layer's input. 
}
\label{fig:KAN-2Layer-Err}
\end{figure*}

\subsection{Lipschitz Division Results}
\label{sec:appdx_Lipschitz_division_results}
In this section, we provide details of the computed Lipschitz constants from section~\ref{sec:Lip-division} and Table~\ref{tbl:Lipschitz-division-summary}. The parameters are provided in Table~\ref{tbl:Lipschitz-division-summary2}. 
For function $f_1$, we used 50 training data equally spaced from $-2 \pi$ to $2 \pi$ and we used a two-layer DAREK with two hidden units, cubic splines, and 9 knots. We only trained the model for 50 epochs with lr=0.1 to leave some error at knots for error study as well.  

\begin{table*}[!ht]
    \centering
    \caption{Lipschitz constants parameters estimated using different methods on different datasets in section~\ref{sec:Lip-division}.}
    \label{tbl:Lipschitz-division-summary2}
    \rowcolors{3}{white}{lightgray}  
    \begin{tabular}{l|ccc|ccc|ccc|ccc|}         
        \hline
        \rowcolor{headerblue}     
        
        & \multicolumn{3}{c|}{\textbf{$f_1$}}
        & \multicolumn{3}{c|}{\textbf{$f_2$}}
        & \multicolumn{3}{c|}{\textbf{$f_3$}}
        & \multicolumn{3}{c|}{\textbf{$f_4$}} \\ 
        \hhline{>{\arrayrulecolor{headerblue}}->{\arrayrulecolor{black}}------------}        
        \rowcolor{headerblue}     
        \textbf{Methods} & {$\calL_g^k$} & {$\calL_h^1$} & {$\calL_h^k$}
                         & {$\calL_g^k$} & {$\calL_h^1$} & {$\calL_h^k$}
                         & {$\calL_g^k$} & {$\calL_h^1$} & {$\calL_h^k$}
                         & {$\calL_g^k$} & {$\calL_h^1$} & {$\calL_h^k$}\\
        \hline
        Equal       & 0.500 & 0.500 & 0.500 & 1.414 & 0.548 & 1.414 & 0.577 & 0.258 & 0.577 & 0.816 & 0.365 & 0.816 \\ 
        Linear   & 0.000 & 0.345 & 0.500 & 0.001 & 0.030 & 3.999 & 0.500 & 0.001 & 0.000 & 1.000 & 0.001 & 0.000 \\ 
        Logarithmic & 0.500 & 0.500 & 0.500 & 0.397 & 0.297 & 1.008 & 0.175 & 0.120 & 0.285 & 0.381 & 0.218 & 0.524 \\ 
        Worst-case  & 1.000 & 1.000 & 1.000 & 20.00 & 3.000 & 20.00 & 10.0  & 2.0   & 10.00 & 10.00 & 2.000 & 10.00 \\ 
        Optimization           & 1.000 & 1.000 & 0.368 & 6.439 & 1.645 & 20.00 & 10.0  & 1.406 & 10.00 & 10.00 & 2.000 & 10.00 \\
        \hline
    \end{tabular}
\end{table*}

For function $f_2$, we used 2500 training pairs, a two-layer DAREK with two inputs and five hidden units, cubic splines, and 20 knots for each neuron. We trained the model until its MSE loss reached $0.256$.

\begin{figure}
    \includegraphics[width=0.45\textwidth,trim=0pt 0pt 0pt 0pt, clip]{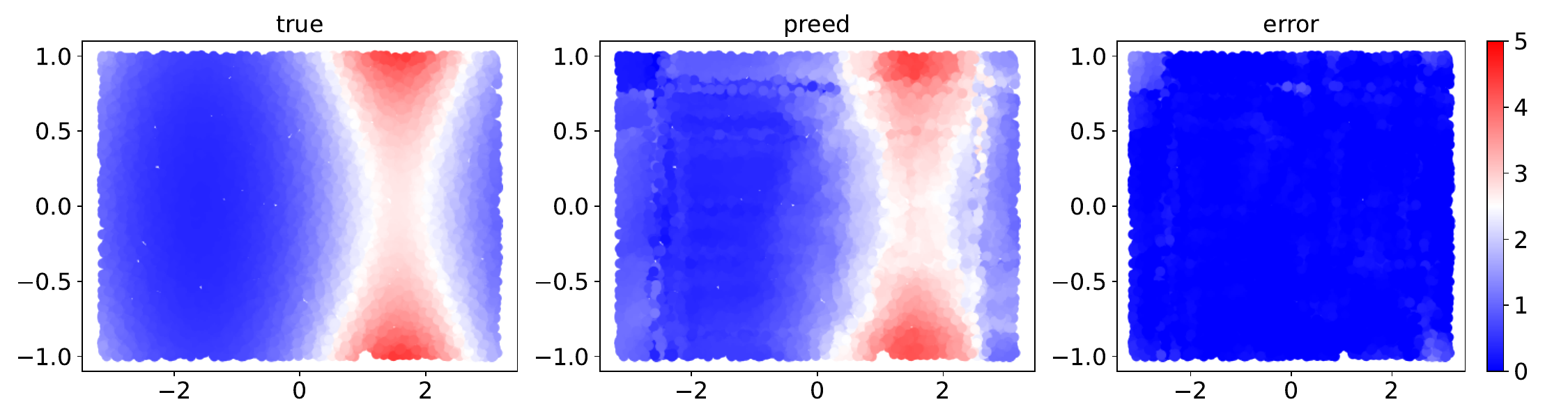}
    \caption{The true dataset, mean prediction, and error estimation of function $f_2(x,y) = \exp(\sin(x)+y^2/2)$. }
\end{figure}

In the experiment for function $f_3$, we used 2500 training pairs, a two-layer DAREK with two inputs and ten hidden units, cubic splines, and 30 knots. We trained the model until its MSE loss reached $0.592$.

\begin{figure}
    \includegraphics[width=0.45\textwidth,trim=0pt 0pt 30pt 0pt, clip]{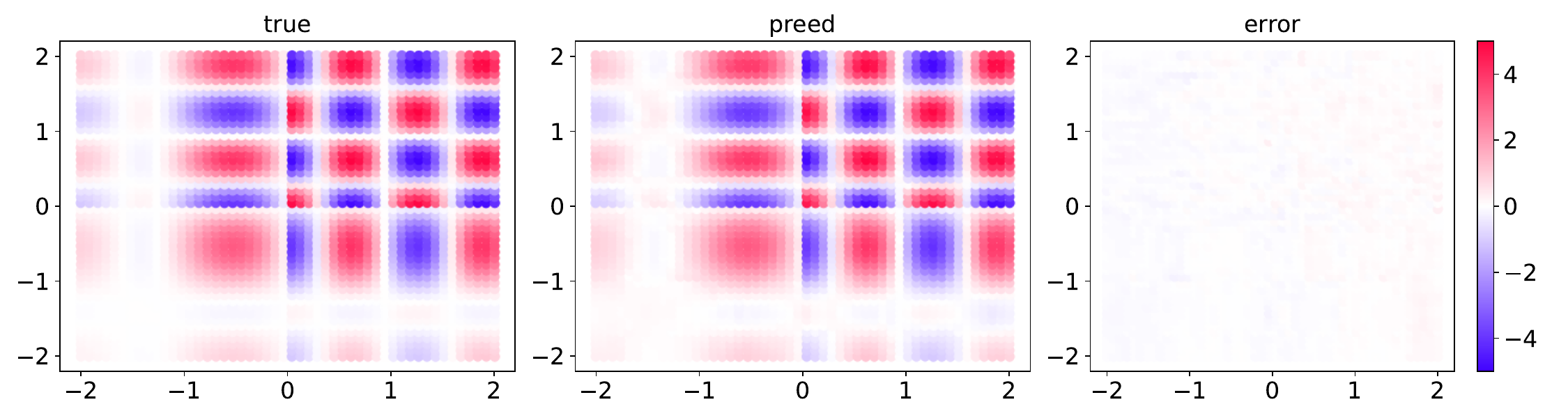}
    \caption{The true dataset, mean prediction, and error estimation of function $f_3$ as defined in~\eqref{eq:discont_eq}.}
\end{figure}

For the Twomoon dataset, $f_4$, we used two moon datasets from the sklearn package, each with 500 training samples, and a model similar to the one we used for $f_2$, and trained the model until its MSE loss reached $0.735$. 
In this study, we use EBL~\eqref{eq:error-at-knots-ebl} for the error of individual neurons.

\begin{figure}
    \includegraphics[width=0.45\textwidth,trim=0pt 0pt 0pt 0pt, clip]{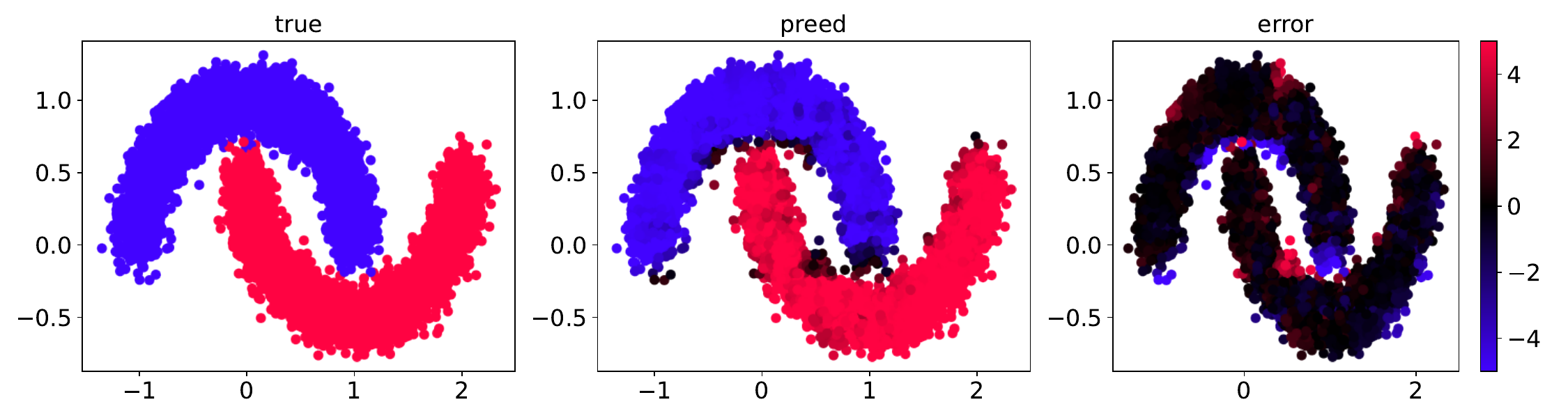}
    \caption{The true dataset, mean prediction, and error estimation of the twomoon dataset.}
\end{figure}

\subsection{Face detection}
\label{sec:appendx-additional-experiments-face}
We provide the full results for face recognition experiments in Table~\ref{tab:highdim}, extending the \textit{High-dimensional experiment} in Section~\ref{sec:experiments} to additional feature dimensions (2, 5, 10, 20, 50, 100, and 200). 
These results are consistent with the main findings: DAREK (DK1 and DK2) maintains stable RMSE and IoU, and SDA improves as the feature dimension increases.
ENS1 and ENS2 do not provide consistent SDA performance, and GP-based methods (ExGP and ApxGP) achieve near-perfect SDA. 

\begin{table*}[t]
\centering
\caption{Results for high-dimensional experiments across four feature extraction methods (PCA and ResNet-18). 
}
\label{tab:highdim}
\setlength{\tabcolsep}{4pt}
\renewcommand{\arraystretch}{1.05}

\rowcolors{1}{white}{lightgray}  
\textbf{(a) PCA: resize to $224 \times 224 \times 3$ then select top $n$ PCA dimensions}\\[4pt]
\resizebox{\textwidth}{!}{%
\begin{tabular}{l|ccccccc|ccccccc|ccccccc}
\arrayrulecolor{black}\cline{1-22}
\rowcolor{headerblue}
& \multicolumn{7}{c|}{RMSE $\downarrow$} 
& \multicolumn{7}{c|}{IoU $\uparrow$} 
& \multicolumn{7}{c}{SDA $\uparrow$} \\

\arrayrulecolor{black}\cline{2-22}

\rowcolor{headerblue}
Model 
& 2 & 5 & 10 & 20 & 50 & 100 & 200
& 2 & 5 & 10 & 20 & 50 & 100 & 200
& 2 & 5 & 10 & 20 & 50 & 100 & 200 \\

\arrayrulecolor{black}\cline{1-22}
DK1
  & 0.136 & 0.133 & 0.133 & 0.133 & 0.126 & 0.129 & 0.149
  & 0.451 & 0.456 & 0.454 & 0.452 & 0.469 & 0.466 & 0.410
  & 0.676 & 0.856 & 0.947 & 0.967 & 0.974 & 0.952 & 0.969 \\
 
DK2
  & 0.137 & 0.139 & 0.134 & 0.190 & 0.123 & 0.184 & 0.135
  & 0.448 & 0.443 & 0.454 & 0.375 & 0.486 & 0.339 & 0.451
  & 0.677 & 0.857 & 0.947 & 0.972 & 0.974 & 0.950 & 0.971 \\
  
ENS1
  & 0.136 & 0.133 & 0.133 & 0.133 & 0.126 & 0.130 & 0.151
  & 0.450 & 0.457 & 0.455 & 0.451 & 0.467 & 0.464 & 0.407
  & 0.571 & 0.554 & 0.568 & 0.563 & 0.501 & 0.431 & 0.378 \\
 
ENS2
  & 0.135 & 0.131 & 0.131 & 0.128 & 0.183 & 0.265 & 0.129
  & 0.449 & 0.456 & 0.446 & 0.469 & 0.311 & 0.145 & 0.431
  & 0.527 & 0.542 & 0.545 & 0.530 & 0.588 & 0.564 & 0.525 \\

ExGP 
 & 0.138 & 0.149 & 0.137 & 0.161 & 0.138 & 0.137 & 0.143
 & 0.435 & 0.400 & 0.450 & 0.359 & 0.437 & 0.448 & 0.420
 & 1.000 & 1.000 & 1.000 & 1.000 & 1.000 & 1.000 & 1.000 \\
 
ApxGP
  & 0.138 & 0.137 & 0.137 & 0.137 & 0.137 & 0.137 & 0.137
  & 0.427 & 0.447 & 0.447 & 0.447 & 0.447 & 0.447 & 0.447
  & 0.977 & 1.000 & 1.000 & 1.000 & 1.000 & 1.000 & 1.000 \\
\arrayrulecolor{black}\cline{1-22}
\end{tabular}}

\vspace{8pt}
 
\textbf{(b) ResNet-18: resize to $224{\times}224{\times}3$, extract ResNet-18 features, select top $n$ via PCA}\\[4pt]
\resizebox{\textwidth}{!}{%
\rowcolors{1}{white}{lightgray}  
\begin{tabular}{l|ccccccc|ccccccc|ccccccc}
\arrayrulecolor{black}\cline{1-22}
\rowcolor{headerblue}
& \multicolumn{7}{c|}{RMSE $\downarrow$} 
& \multicolumn{7}{c|}{IoU $\uparrow$} 
& \multicolumn{7}{c}{SDA $\uparrow$} \\

\arrayrulecolor{black}\cline{2-22}

\rowcolor{headerblue}
Model 
& 2 & 5 & 10 & 20 & 50 & 100 & 200
& 2 & 5 & 10 & 20 & 50 & 100 & 200
& 2 & 5 & 10 & 20 & 50 & 100 & 200 \\

\arrayrulecolor{black}\cline{1-22}
DK1
  & 0.100 & 0.097 & 0.090 & 0.088 & 0.079 & 0.074 & 0.079
  & 0.535 & 0.538 & 0.562 & 0.557 & 0.586 & 0.606 & 0.589
  & 0.704 & 0.686 & 0.875 & 0.892 & 0.788 & 0.837 & 0.935 \\
 
DK2
  & 0.100 & 0.100 & 0.093 & 0.103 & 0.083 & 0.073 & 0.072
  & 0.533 & 0.535 & 0.553 & 0.470 & 0.573 & 0.608 & 0.626
  & 0.691 & 0.707 & 0.899 & 0.918 & 0.858 & 0.882 & 0.902 \\

ENS1 
  & 0.100 & 0.097 & 0.089 & 0.089 & 0.080 & 0.075 & 0.081 
  & 0.535 & 0.540 & 0.565 & 0.557 & 0.585 & 0.603 & 0.583 
  & 0.578 & 0.551 & 0.566 & 0.562 & 0.364 & 0.222 & 0.408 \\

ENS2
  & 0.100 & 0.094 & 0.087 & 0.084 & 0.174 & 0.069 & 0.065
  & 0.530 & 0.551 & 0.569 & 0.575 & 0.161 & 0.628 & 0.646
  & 0.515 & 0.533 & 0.573 & 0.562 & 0.461 & 0.550 & 0.542\\

ExGP
 & 0.139 & 0.140 & 0.139 & 0.140 & 0.140 & 0.137 & 0.138
 & 0.431 & 0.447 & 0.434 & 0.427 & 0.426 & 0.447 & 0.444
 & 0.542 & 1.000 & 1.000 & 1.000 & 1.000 & 1.000 & 1.000 \\

ApxGP
  & 0.133 & 0.137 & 0.137 & 0.137 & 0.137 & 0.137 & 0.137
  & 0.445 & 0.447 & 0.447 & 0.447 & 0.447 & 0.447 & 0.447
  & 0.729 & 1.000 & 1.000 & 1.000 & 1.000 & 1.000 & 1.000 \\
\arrayrulecolor{black}\cline{1-22}
\end{tabular}}

\end{table*}

\fi
\end{document}